# Elementary methods provide more replicable results in microbial differential abundance analysis

Juho Pelto[1,2], Kari Auranen[2,3], Janne V. Kujala[2], Leo Lahti[1]

[1] Department of Computing, University of Turku, Turku, Finland

[2] Department of Mathematics and Statistics, University of Turku, Turku, Finland

[3] Department of Clinical Medicine, University of Turku, Turku, Finland

Differential abundance analysis is a key component of microbiome studies. Although dozens of methods exist there is currently no consensus on the preferred methods. While the correctness of results in differential abundance analysis is an ambiguous concept and cannot be fully evaluated without setting the ground truth and employing simulated data, we argue that a well-performing method should be effective in producing highly reproducible results.

We compared the performance of 14 differential abundance analysis methods by employing datasets from 53 taxonomic profiling studies based on 16S rRNA gene or shotgun metagenomic sequencing. For each method, we examined how the results replicated between random partitions of each dataset and between datasets from separate studies. While certain methods showed good consistency, some widely used methods were observed to produce a substantial number of conflicting findings. Overall, when considering consistency together with sensitivity, the best performance was attained by analyzing relative abundances with a non-parametric method (Wilcoxon test or ordinal regression model) or linear regression/t-test. Moreover, a comparable performance was obtained by analyzing presence/absence of taxa with logistic regression.

**Corresponding author**

Juho Pelto, jepelt@utu.fi

## Introduction

Studying associations between microbial taxa and external variables, such as disease status or environmental exposure, is central to microbiome research. These associations can be investigated with differential abundance analysis (DAA), which in its simplest form compares the abundances of microbial taxa between two experimental groups, e.g. subjects with and without the disease in question. Despite this seemingly simple goal, performing DAA reliably has proven to be challenging due to some peculiar statistical properties of microbiome data. Indicative of those challenges, numerous DAA methods developed in recent years tend to yield remarkably differing results [1], [2]. For instance, while one method may detect hundreds of differentially abundant taxa in a particular dataset, another method may detect none [1]. Until now, no consensus has emerged on the best performing DAA method.

The discrepancy in results obtained from different DAA methods naturally raises the question about which methods provide the most *correct* results. Answering such a question would evidently require knowing the *ground truth,* namely, the true values of taxon-wise differential abundances, against which the DAA results could be compared. As the ground truth behind any real microbiome dataset is very rarely known, evaluating the correctness of DAA results in practice requires the use of a preset ground truth and *simulated* data.

There is, however, an obvious problem in relying on simulations in that there is no guarantee that the set ground truth or the simulated data would correspond to their real-world equivalents in all relevant aspects. For instance, the assumed abundance differences may not realistically mimic the effects of the studied condition. Furthermore, simulated data may not correspond to count data that would emerge in a real-world experiment under the assumed ground truth. To simulate realistic data, one would need to accurately model different biases that the experimental workflow, consisting of e.g. DNA extraction and sequencing, introduces to the observed counts. Such biases should include, for instance, taxon-wise biases, addressing the fact that different taxa are detected with varying sensitivities in the experimental workflow [3]. Indicative of the problems in evaluating DAA methods by employing simulated data are the discordant conclusions on methods obtained under different simulation approaches [2], [4], [5], [6], [7], [8], [9].

Furthermore, different methods are designed to estimate different types of differential abundance (DA) and, consequently, different ground truths. In particular, most currently available taxonomic profiling data are *compositional* [10]. This problem must be accounted for in DAA by what we here call a *normalization strategy*. However, the normalization strategies incorporated in the DAA methods vary. While certain methods *aim* to estimate DA in terms of absolute abundances through employing advanced normalization strategies [11], [12], [13], others merely aim to estimate DA in terms of *relative abundances* by employing an equivalent of the simple TSS normalization (counts divided by the library sizes) [14], [15]. Further variation in the targeted ground truths is introduced by the choice between using untransformed or log transformed counts, i.e., between estimating DA with respect to arithmetic or geometric means. Therefore, due to the ambiguities in the definition of DA, a specific DAA result may not be unambiguously classified as correct or incorrect.

Because of the many problems in evaluating the correctness of DAA results, we decided to focus on the *consistency* and *replicability* of the results. In particular, we study how different methods can replicate statistical significance and the direction (sign) of taxon-wise DA between random partitions of datasets and between different studies. Our goal was to identify methods that can declare a high number of taxa as being associated with the studied condition while being able to accurately reproduce these results in another dataset. Such methods can be considered capable of effectively detecting robust signals in microbiome data. In addition, our approach reveals methods that are not even internally consistent and should therefore be avoided in general. Importantly, we perform all our evaluations by employing datasets from (N = 53) real human gut microbiome studies.

Although a few benchmarking studies have examined certain aspects of consistency or replicability on a limited number of datasets [1], [16], [17], a comprehensive benchmarking study focusing on consistency and replicability and employing a large number of datasets is still missing. Furthermore, while we concentrate on evaluating how different DAA methods perform in the basic two-group comparison, we also evaluate the methods in the presence of covariates. Although including covariates is often essential in practice, only a few previous studies have addressed this aspect of DAA, and mostly on simulated datasets [2], [7], [18]. Our study is also the first to evaluate the coverage of the confidence intervals that different DAA methods provide. Lastly, our comparison includes logistic regression for presence/absence of taxa, which has not been included in previous benchmarking studies.

## Materials and methods

### Analysis frameworks

We extracted 69 datasets from 53 real *human gut microbiome* studies employing 16S rRNA gene sequencing (16S, 23 studies) or shotgun metagenomic sequencing (shotgun, 30 studies) [19], [20]. We included studies that compared groups consisting of healthy and non-healthy individuals (e.g. colorectal cancer, CRC). We call these groups *control* and *case* groups, respectively. If a single study compared more than two groups (e.g. CRC, adenoma and healthy), multiple datasets from the study were extracted (e.g. CRC vs. healthy and adenoma vs. healthy). The most common conditions examined in the included studies were CRC (16 studies), inflammatory bowel disease (IBD, 9), adenoma (7), obesity (OB, 4), type 2 diabetes (T2D, 4), Clostridioides difficile infection (CDI, 3), type 1 diabetes (T1D, 3), autism spectrum disorder (ASD, 2) and overweight (OW, 2).

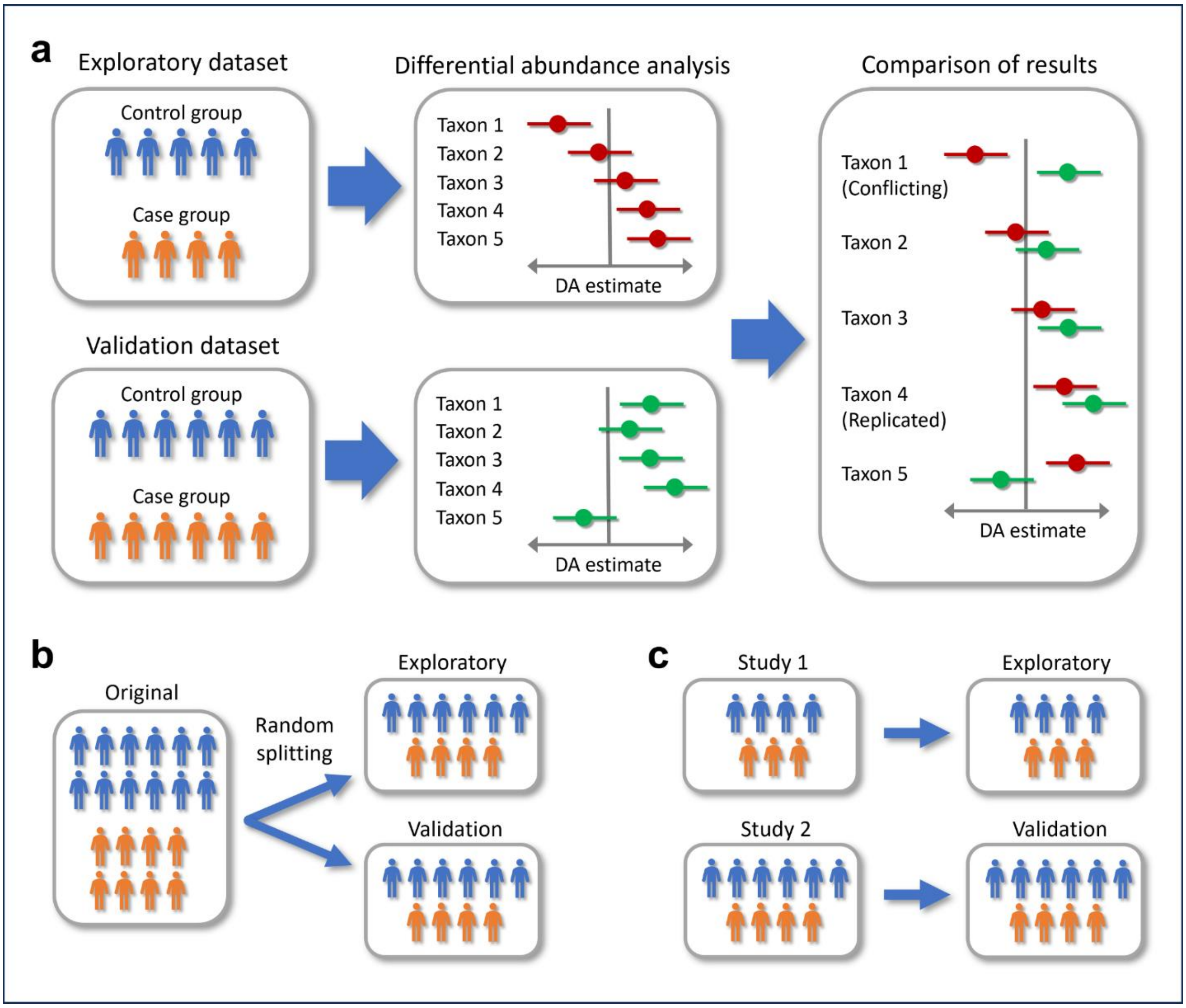

**Figure 1** a) The basic workflow. DAA was performed on exploratory and validation datasets and the results were compared between them. If the result for a taxon was significant in both exploratory and validation datasets but the directions were opposite, the results were considered conflicting (Taxon 1). The result for a taxon was considered replicated if it was significant and had the same direction in exploratory and validation datasets (Taxon 4). b) In the split-data analyses each exploratory/validation pair of datasets was constructed by randomly splitting an original dataset. c) In the separate study analyses, datasets from separate studies were used as exploratory and validation datasets. In all subfigures, the individuals belonging to the control and case groups are indicated with blue and orange, respectively.

In all analyses, DAA was first performed in an *exploratory dataset* to search for statistically significant differences in taxon-wise abundances and, subsequently, in the *validation dataset(s)* to validate the results (Figure 1a). We used two analysis frameworks to study the replication of DAA results both within studies and between studies. In the *split-data analyses* (within study), 285 exploratory-validation pairs of datasets were constructed by randomly splitting 57 original datasets five times into two equal sized halves (Figure 1b). These analyses were used to evaluate whether DAA methods perform as consistently as they theoretically should. In the *separate study analyses* (between studies), 50 datasets from separate studies were used as exploratory and/or validation datasets (Figure 1c). Here the goal was to investigate how the results provided by each DAA method replicate across studies in practice. In both types of analyses, the required minimum number of subjects in each group (case or control) in each exploratory or validation dataset was 10. A summary of the datasets used in each type of analysis is given in Table 1. Additional details on constructing the pairs of exploratory and validation datasets are given in the Appendix.

The analyses on 16S and shotgun datasets were performed on the *genus* and *species* level, respectively. Furthermore, following the standard practice in the field, we excluded taxa with prevalence < 10%, separately from each exploratory and validation dataset.

| Analysis / sequencing type | Number of datasets | Sample size median (min-max) | Number of taxa median (min-max) | Significant taxa median (min-max) |
|---|---|---|---|---|
| **Split-data analyses** | | | | |
| **16S** | 24 × 5 × 2 | 57 (21-374) | 79 (36-188) | 48 (13-104)[1] |
| **Shotgun** | 33 × 5 × 2 | 53 (25-255) | 201 (91-276) | 110 (33-233)[1] |
| **Separate study analysis** | | | | |
| **16S** | 23 | 112 (32-747) | 80 (45-188) | 27 (5-81) |
| **Shotgun** | 27 | 96 (20-509) | 203 (96-260) | 44 (9-125) |

**Table 1** Summary of the datasets used in the split-data analyses and the separate study analyses by sequencing type. In the split-data analyses each original dataset was randomly split five times to form five pairs consisting of an exploratory and a validation dataset. Sample size refers to the number of subjects in a single exploratory or validation dataset. Number of taxa refers to the number of genera and species in the 16S and shotgun datasets, respectively, after filtering out the taxa with prevalence < 10%. Significant taxa refers to the number of statistically significant taxa (FDR adjusted p < .05) detected in an exploratory dataset by *at least one* of the 14 DAA methods. For the details for each dataset, see Tables A1.1 and A1.2.

[1] In at least one of the five random splits

**Evaluation metrics**

A DAA result for each taxon was defined to be (statistically) significant in the exploratory datasets if the false discovery rate (FDR) adjusted p-value was below a chosen *nominal* FDR level α. We note that the true FDR is at (or below) the nominal FDR level α only if a method provides valid p-values. We used the standard choice $\alpha = .05$ but repeated the analyses by using $\alpha = .01$, $\alpha = .10$ and $\alpha = .20$. FDR adjusted p-values were calculated by the Benjamini-Hochberg method [21] or by a method integrated into the DAA method (ANCOM-BC2, DESeq2, LDM, ZicoSeq). For each DAA method and within each pair of exploratory and validation datasets, a taxon was called a *candidate taxon* if it was significant in the exploratory dataset and it was present (after the 10% prevalence filtering) in the validation dataset(s) (Taxa 1, 4, and 5 in Figure 1a). As there were typically under ten candidate taxa, we considered the validation process as testing a few well-formed hypotheses. Therefore, the statistical significance in the validation datasets was defined as *unadjusted* p-value < .05. Lastly, the *direction* of a taxon was defined as the sign of its estimated DA (positive, if estimated to be more abundant in the case group). The following metrics were calculated to evaluate the performance of the DAA methods.

*Percentage of conflicting results*

The result for a candidate taxon was defined as *conflicting* if the taxon was significant also in the validation dataset, but the direction was *opposite* to what it was in the exploratory dataset (Taxon 1 in Figure 1a). A conflicting result was considered a serious error as it would lead to conflicting inferences. Therefore, the *percentage of conflicting results (Conflict%*), namely, the percentage of candidate taxa for which a conflicting result was observed, was used as our first main metric for consistency. In the split-data analyses, a conflicting result signifies a false result in either the exploratory or the validation dataset and we derived an approximate upper limit of $\alpha \times .50 \times .05$ for an ideal Conflict% (for derivation, see the Appendix).

*Replication percentage*

The result for a candidate taxon was defined as *replicated* if the taxon was significant also in the validation dataset and it had the *same* direction as in the exploratory dataset (Taxon 4 in Figure 1a). *Replication percentage* (*Replication%*), namely, the percentage of candidate taxa whose result was replicated in the validation dataset, was our second main metric for consistency. Especially in the split-data analyses, if a method detects any *truly* differentially abundant taxon in an exploratory dataset, it should often be able to detect the same taxon in the validation dataset. Consequently, a low replication percentage may indicate a high FDR.

*Number of significant taxa*

The sensitivity of a given DAA method was measured by the total *number of significant taxa* (number of "hits", *NHits*) in the exploratory datasets. We note that a high NHits does not necessarily indicate a high power to detect *true* effects as it may also be due to a high number of false positive findings. Nevertheless, when accompanied with low Conflict% and high Replication%, higher NHits can be considered an indication of higher statistical power.

*Additional metrics: Correlation of estimates and overlap percentage of confidence intervals*

In addition to the above metrics based on the statistical significance and the direction of DA, we evaluated metrics related to the estimation of the magnitude of DA (i.e. the effect size). The consistency of DA estimates was measured by the Spearman *correlation of the estimates* between each exploratory and the corresponding validation dataset.

Furthermore, we evaluated the *overlap percentage of confidence intervals (CI%)*. For a given taxon, confidence intervals were considered overlapping, if there was any overlap between the confidence intervals calculated in the exploratory and the corresponding validation datasets (e.g. the confidence intervals for Taxa 2, 3 and 4 in Figure 1 were considered overlapping). We examined 83.4% confidence intervals that should ideally overlap (at least) 95% of times in the split-data analyses if the sampling distribution of each DA estimate is approximately normal [22]. Confidence intervals were evaluated only for the candidate taxa (with FDR adjusted p < .05) with at least 10% prevalence in *both* experimental groups (case and control). LDM, edgeR and ZicoSeq were excluded from this analysis as they do not provide confidence intervals or standard errors in their output.

**Calculation of the overall values of the evaluation metrics**

The overall values of Conflict%, Replication% and CI% reported in Results were calculated over all candidate taxa in all exploratory datasets. For instance, if 5 out of 8 candidate taxa found in one exploratory dataset replicated and 2 out of 2 candidate taxa found in another exploratory dataset replicated, $Replication\% = \frac{5+2}{8+2} = 0.70 = 70\%$ (for the details for the separate study analyses, see the Appendix). The number of significant taxa (NHits) was the sum of the numbers of significant taxa in all exploratory datasets. The average correlation of DA estimates was calculated as the hyperbolic tangent transformed mean of the inverse hyperbolic tangent transformed Spearman correlation coefficients.

**Evaluation and ranking of differential abundance analysis methods**

The evaluation and ranking of the DAA methods proceeded as follows. In the split-data analyses, we first examined whether the methods performed adequately according to our criterion for Conflict% and had a high Replication% compared to the other methods. In the separate study analyses, we first considered which methods were among the most consistent methods according to Conflict% and Replication%. Lastly, the methods with adequate/high consistency were ranked by the total number of significant taxa found in the exploratory datasets.

Additionally, we evaluated whether the low consistency of some methods could be justified by their high sensitivity. This was done by comparing the consistency of all methods under constant high sensitivity, that is, when the nominal FDR levels (α) were chosen so that each method detected the same high number of taxa in the exploratory datasets.

**Included differential abundance analysis methods**

We included 14 DAA methods that provided the DA estimate and p-value for each taxon and had an up-to-date R implementation available. The collection of methods included recent methods designed specifically for microbial DAA, namely, ANCOM-BC2 [11], corncob [14], fastANCOM [13], LDM [23], LinDA [12] and ZicoSeq [2]. We also included methods that were originally designed for differential expression analysis for RNA-Seq data, but which have also been used for microbial DAA, namely, ALDEx2 [24], DESeq2 [25], edgeR [26], limma voom [27], [28] and metagenomeSeq [29].

Additionally, we included two general statistical methods commonly used for DAA in practice [30]. These are the analysis of log transformed TSS normalized counts with a linear regression model, i.e. *the t-test* when no covariates are included, as implemented in the MaAsLin2 R package [31], and the analysis of (untransformed) TSS normalized counts with a non-parametric method. As the non-parametric method we chose ordinal regression model (ORM/Wilcoxon), which can be seen as a generalization of the familiar Wilcoxon test in the sense that it can incorporate covariates while giving basically the same p-values as the Wilcoxon test in the two-group comparison [32]. Lastly, we included logistic regression analysis of the observed presence/absence (non-zero/zero count) of taxa (LogR).

Except for the LogR approach, the methods can be roughly described in terms of two factors. First, the methods address the compositionality of microbiome data using different normalization strategies. For instance, corncob, LDM, MaAsLin2/t-test and ORM/Wilcoxon (effectively) employ the simple TSS normalization while ALDEx2 (CLR), DESeq2 (RLE), edgeR (TMM), limma-voom (TMM) and metagenomeSeq (CSS) apply more elaborate normalizations to count data. Furthermore, ANCOM-BC2 and LinDA employ an estimated bias correction to the DA estimates, while fastANCOM and ZicoSeq search for non-differentially abundant taxa and use those as the reference taxa in the analysis.

Second, the methods employ different transformations and statistical models to analyze the (possibly normalized) count data. A linear model/t-test with log (or CLR) transformation is employed in ALDEx2, ANCOM-BC2, fastANCOM, limma-voom, LinDA, MaAsLin2/t-test and metagenomeSeq. A linear model is also used in ZicoSeq (with power transformation) and in LDM (with no transformation *and* arcsine-root transformation). Untransformed counts are analyzed with a negative binomial model in DESeq2 and edgeR, and with a beta-binomial model in corncob. Lastly, the ranks of the (TSS normalized) counts are analyzed in ORM/Wilcoxon.

| Method | Normalization strategy | Log | Model | Covariates | CI | Typical run-time |
|---|---|---|---|---|---|---|
| ALDEx2 | CLR | x | Linear | x | x[1] | 30 s |
| ANCOM-BC2 | Bias correction | x | Linear | x | x | 20 s |
| corncob | TSS |  | Beta-binomial | x | x[1] | 10 s |
| DESeq2 | RLE |  | Negative binomial | x | x | 7 s |
| edgeR | TMM |  | Negative binomial | x |  | 0.5 s |
| fastANCOM | Reference taxa | x | Linear | x | x | 0.01 s |
| LDM | TSS |  | Linear | x |  | 30 s |
| limma voom | TMM | x | Linear | x | x | 0.07 s |
| LinDA | Bias correction | x | Linear | x | x | 0.2 s |
| LogR | - |  | Binary logistic | x | x | 0.6 s |
| MaAsLin2/t-test | TSS | x | Linear | x | x | 2 s |
| metagenomeSeq | CSS | x | Linear |  | x | 0.8 s |
| ORM/Wilcoxon | TSS |  | Ordinal (proportional odds)[2] | x | x[1] | 3 s |
| ZicoSeq | Reference taxa | [3] | Linear | x |  | 20 s |

**Table 2** DAA methods included in this study. Normalization strategy: method to address compositionality of count data. Log: log transformation was performed for (normalized) counts. Model: Statistical model fit to (transformed/normalized) counts. Covariates: Can incorporate covariates. CI: Provides confidence intervals (or standard errors) for DA. Typical run-time: Average run-time when run on different sized datasets on a standard laptop (for details, see figure A2). CLR: Centered log ratio, TSS: Total sum scaling, RLE: Relative log expression, TMM: Trimmed mean of M values, CSS: Cumulative sum scaling.

[1] Confidence intervals do not match exactly the p-values given by the method.

[2] Essentially non-parametric

[3] Power transformation

While most of the tested methods had several parameters that could be adjusted by the user, we primarily employed the default settings as this is likely how these methods are applied in practice. For the performance obtained with some other settings, see Figures A14.1 – A14.4 in the Appendix. The list of the compared DAA methods is given in Table 2 and the details of running the methods are presented in the Appendix.

# Results

## Split-data analyses

The overall values of the evaluation metrics are presented in the text and in Figure 2 (stratified by sample size and sequencing type, see Figures A17.1 and A19.1, respectively; for the distributions, see Figures A4.1 - A4.3, A5 and A6 in the Appendix). The number of conflicting and replicated results on each randomly split original dataset are given in Figure 3.

*Percentage of conflicting results*

Seven methods, namely, ALDEx2, fastANCOM, LinDA, LogR, MaAsLin2/t-test, ORM/Wilcoxon and ZicoSeq, performed properly as their Conflict% was mostly below the thresholds for ideal performance, namely, 0.025%, 0.125%, 0.25% and 0.50% with $\alpha = .01, .05, .10$ and $.20$, respectively. The performances of corncob (Conflict% = 0.3% at $\alpha = .05$), limma-voom (0.3%) and metagenomeSeq (0.3%) were also tolerable, but for LDM the Conflict% (0.8%) was too high. Most notably, however, ANCOM-BC2 (3.0%) and especially DESeq2 (7.2%) and edgeR (6.3%) had their Conflict% order(s) of magnitude above the acceptable level.

*Replication percentage*

Replication percentages (Replication%) ranged from 35% to almost 80% at α = .05. The highest Replication% was observed for ALDEx2 (79%), LogR (78%), fastANCOM (77%), MaAsLin2/t-test (77%) and ORM/Wilcoxon (76%). They were followed by LinDA (72%), limma-voom (70%), corncob (69%), ZicoSeq (68%), metagenomeSeq (64%) and LDM (64%). The lowest Replication% was observed for ANCOM-BC2 (35%), DESeq2 (45%) and edgeR (42%).

The high Conflict% and low Replication% observed for ANCOM-BC2, DESeq2 and edgeR indicated that these methods likely provide systematically too low p-values. Additional evidence for this was gathered by examining the performance of the methods on 500 (= 50 x 10) datasets with randomly permuted group labels (corresponding to a scenario where there are no truly differentially abundant taxa). For instance, instead of the nominal 5%, the percentage of (not multiplicity adjusted) p-values < .05 was 18.4% for edgeR and 10.5% for DESeq2 (Figure A7). However, for ANCOM-BC2 this percentage was 3.1%, indicating other reasons for its low consistency.

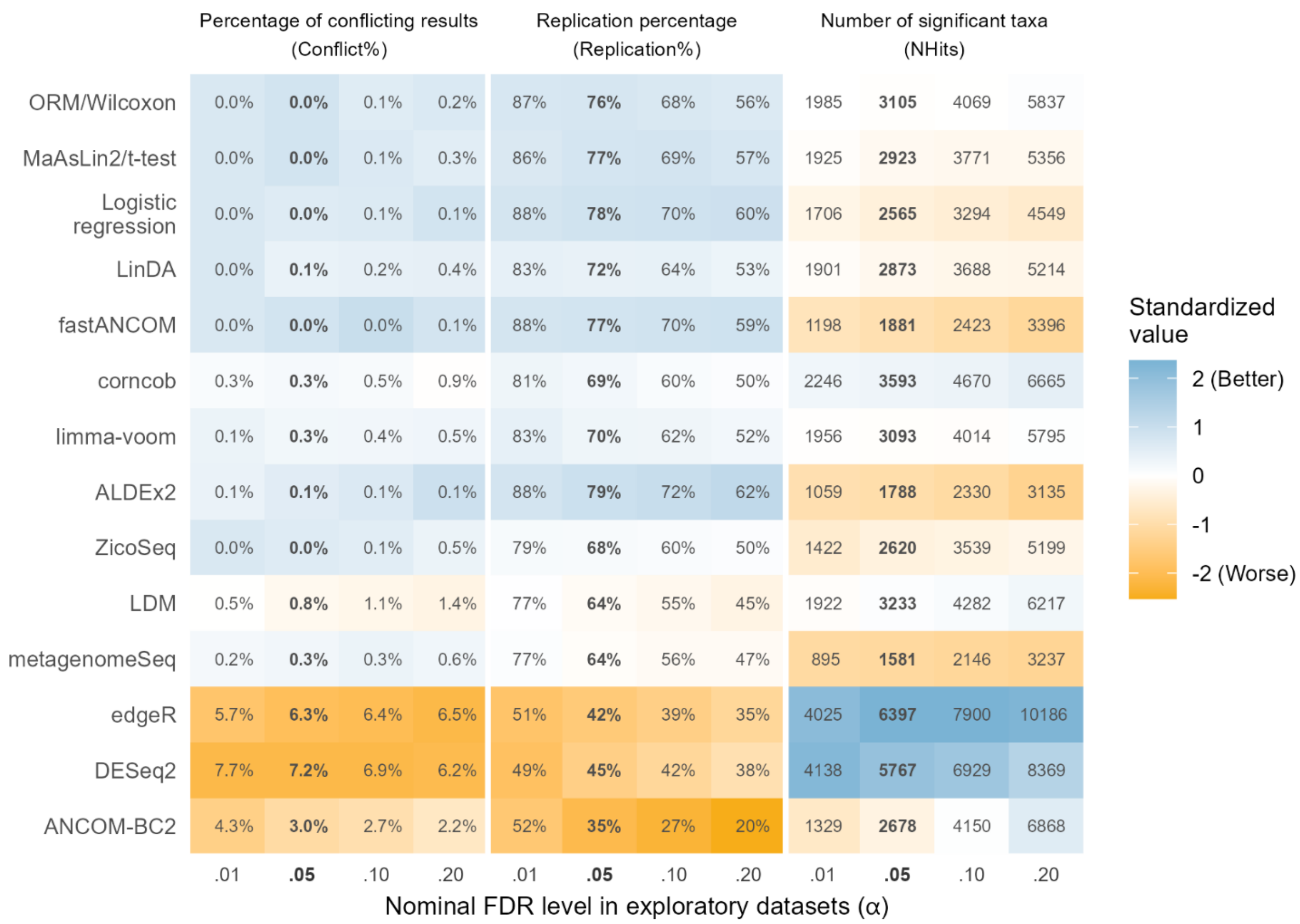

**Figure 2** The performance of 14 DAA methods in terms of consistency and sensitivity on 57 randomly split real microbiome datasets. The methods are in rank order based on the mean of the standardized values of the metrics. (Conflict% was square root transformed before the standardization.) Values based on the nominal FDR level α = .05 are shown in bold. Each original dataset was split five times to form pairs consisting of an exploratory and a validation dataset, thus totaling 285 pairs of datasets. Candidate taxon = A taxon that was significant (FDR adjusted $p < \alpha$) in an exploratory dataset and present in the validation dataset. Conflict% = The percentage of candidate taxa that were significant ($p < .05$) in the validation dataset, but in the opposite direction to that in the exploratory dataset. Replication% = The percentage of candidate taxa that were significant ($p < .05$) in the validation dataset in the same direction as in the exploratory dataset. NHits = The total number of significant (FDR adjusted $p < \alpha$) taxa found in the 285 exploratory datasets. A higher NHits can be considered better when it is accompanied by low Conflict% and high Replication%.

*Number of significant taxa*

By a large margin the most sensitive methods were edgeR and DESeq2. With the nominal FDR level α = .05, they identified a total of 6397 and 5767 significant taxa in the 285 (= 57 × 5) exploratory datasets, respectively. They were followed by corncob (NHits = 3593), LDM (3233), ORM/Wilcoxon (3105), limma-voom (3093), MaAsLin2/t-test (2923) and LinDA (2873). A little behind these methods were ANCOM-BC2 (2678), ZicoSeq (2620) and LogR (2565). The least sensitive methods were fastANCOM (1881), ALDEx2 (1788) and metagenomeSeq (1581).

While DESeq2 and edgeR identified a high number of taxa (NHits = 5767 – 6397) with the nominal FDR level α = .05, LDM, ZicoSeq, limma-voom, corncob, LinDA, MaAsLin2/t-test and ORM/Wilcoxon detected a number of similar magnitude (NHits around 5000) when α = .20 was used. In the latter cases, however, Conflict% was mostly at least an order of magnitude lower and Replication% was somewhat higher than in the former cases. Moreover, when the nominal FDR levels were chosen so that each method detected exactly 6000 taxa, the highest consistency was achieved by ORM/Wilcoxon, MaAsLin2/t-test and LogR while the lowest was demonstrated by ANCOM-BC2, DESeq2 and edgeR (Figure A3).

*Correlation of estimates and overlap percentage of confidence intervals*

The average Spearman correlation between DA estimates in the exploratory and validation datasets was highest for fastANCOM (.43), limma-voom (.43), LinDA (.43), ALDEx2 (.43) and MaAsLin2/t-test (.42), followed by ORM/Wilcoxon (.40), corncob (.38), ZicoSeq (.36), LogR (.36) (Figure A5). The lowes values were observed for metagenomeSeq (.32), DESeq2 (.30), edgeR (.29), LDM (.28) and ANCOM-BC2 (.22).

The overlap percentage of confidence intervals (CI%) for candidate taxa was highest for ALDEx2 (93%), LogR (93%), fastANCOM (92%), ORM/Wilcoxon (91%) and MaAsLin2/t-test (89%) (Figure A6). It was especially low for DESeq2 (65%) and ANCOM-BC2 (62%).

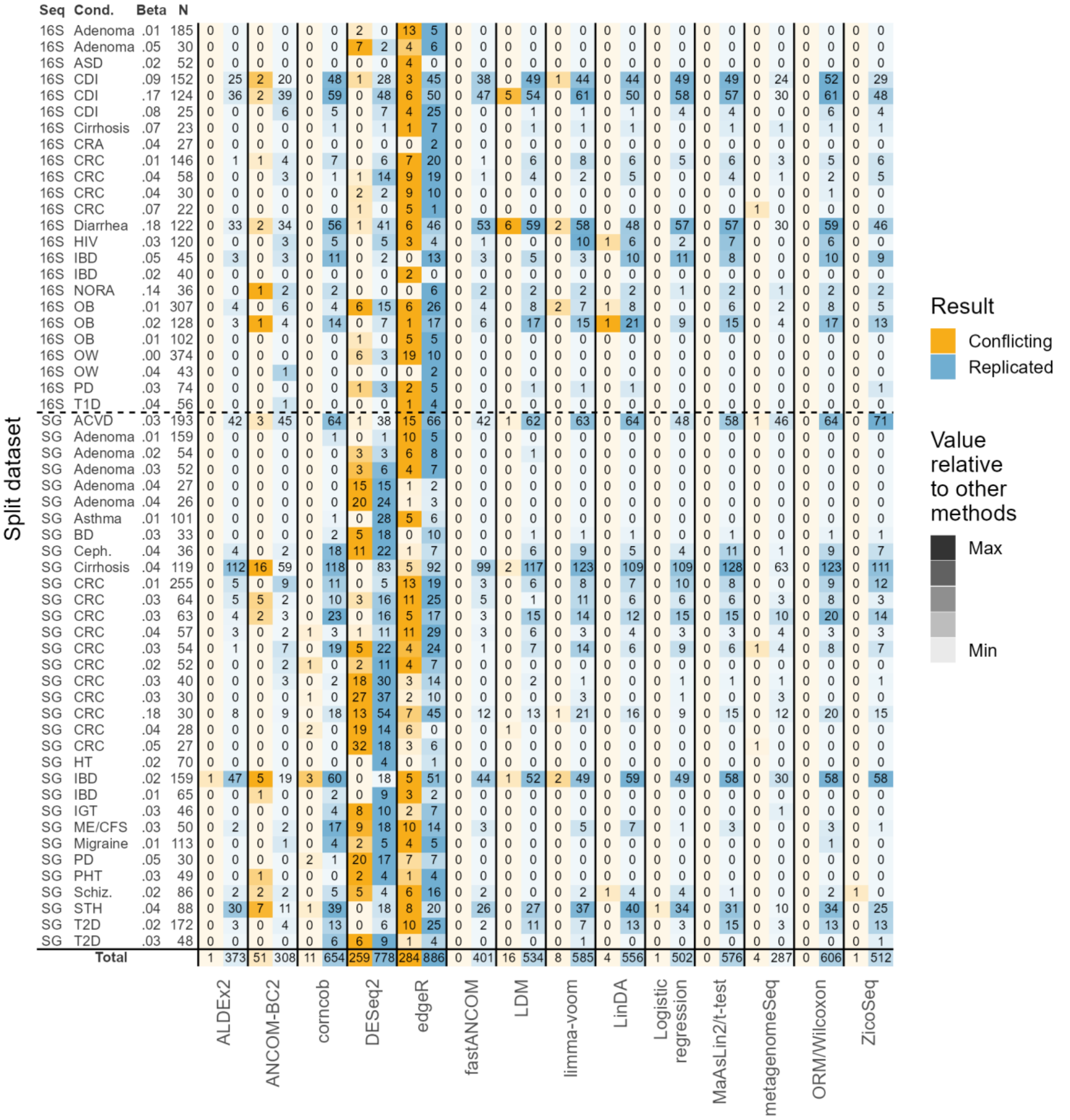

**Figure 3** The number of conflicting and replicated results found by 14 DAA methods on 57 randomly split real microbiome datasets. Each original dataset was split to form a pair consisting of an exploratory and a validation dataset. The splitting was performed five times for each original dataset. In each slot is the number of taxa for which a conflicting or replicated result was found in *at least one* of such pair. Conflicting result = the result for a taxon was significant in the exploratory datasets (FDR adjusted p < .05) and validation datasets (p < .05) but in opposite directions. Replicated result = the result for a taxon was significant in the exploratory dataset and validation datasets in the same direction. Seq. = sequencing type (16S or SG = shotgun); Cond. = the studied condition; Beta = Beta diversity explained by the experimental group (case/control); N = the sample size in a single exploratory or validation dataset. ACVD: Atherosclerotic cardiovascular disease; BD: Behcet's disease; Ceph.: Cephalosporins; CRA: Chronic, treated rheumatoid arthritis; HIV: Human immunodeficiency virus; HT: Hypertension; IGT: Impaired glucose tolerance; ME/CFS: Myalgic encephalomyelitis/chronic fatigue syndrome; NASH: Nonalcoholic steatohepatitis; NORA: New-onset untreated rheumatoid arthritis; PD: Parkinson's disease; PHT: Pre-hypertension; STH: Soil-transmitted helminths

*Performance with covariates*

The inclusion of 1 - 3 covariates did not notably decrease the performance of most DAA methods (Figure A8). An exception was DESeq2, whose performance dropped substantially. For instance, its Conflict% was as high as 16.0% and Replication% as low as 25%.

**Separate study analyses**

The overall values of the evaluation metrics are presented in the text and in Figure 4 (stratified by sample size and sequencing type, see Figures A17.2 and A19.2, respectively; for the distributions, see Figures A10.1 – A10.3, A11 and A12 in the Appendix). The number of conflicting and replicated results for each exploratory dataset are given in Figure 5.

*Percentage of conflicting results*

The lowest Conflict% at the nominal FDR level α = .05 was achieved by ALDEx2 (1.5%), followed by metagenomeSeq (2.2%), ZicoSeq (2.4%), fastANCOM (3.0%) and LogR (3.1%). The next lowest values were observed for MaAsLin2/t-test (4.0%), limma-voom (4.2%), ORM/Wilcoxon (4.3%), ANCOM-BC2 (4.3%), LinDA (4.8%) and LDM (4.9%). A higher value was observed for corncob (6.5%) and the two highest values for DESeq2 (8.1%) and edgeR (9.2%).

*Replication percentage*

With α = .05, the replication percentages (Replication%) varied little (38% - 44%) between most of the methods. The methods listed by Replication% from highest to lowest were: MaAsLin2/t-test (44%), ORM/Wilcoxon (44%), fastANCOM (43%), LinDA (43%), LDM (42%), ALDEx2 (40%), metagenomeSeq (40%), LogR (39%), limma-voom (39%), ZicoSeq (39%) and corncob (38%). Clearly lower values were observed for edgeR (33%), DESeq2 (25%) and ANCOM-BC2 (21%).

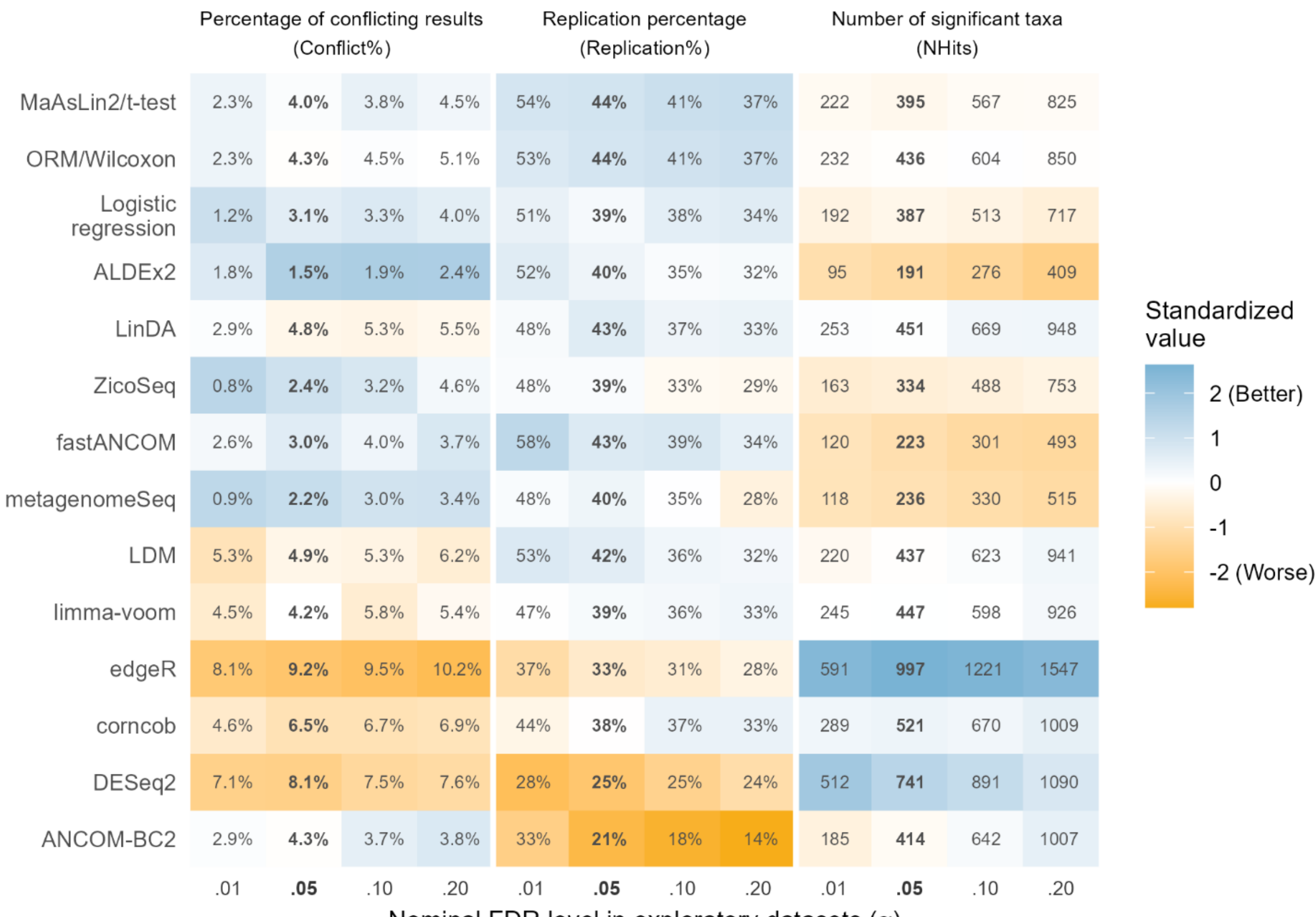

**Figure 4** The performance of 14 DAA methods in terms of sensitivity and consistency of results between separate studies. The methods are in rank order based on the mean of the standardized values of the metrics. (Conflict% was square root transformed before the standardization.) Values based on the nominal FDR level α = .05 are shown in bold. A dataset from one study was used as an exploratory dataset and dataset(s) from other study/studies as the validation dataset(s). Candidate taxon = A taxon that was significant (FDR adjusted p < α) in an exploratory dataset and present in a validation dataset. Conflict% = The percentage of candidate taxa that were significant (p < .05) in the validation dataset, but in the opposite direction to that in the exploratory dataset. Replication% = The percentage of candidate taxa that were significant (p < .05) in the validation dataset in the same direction as in the exploratory dataset. NHits = The total number of significant taxa found in the 37 exploratory datasets. A higher NHits can be considered better when it is accompanied by low Conflict% and high Replication%.

*Number of significant taxa*

The total number significant taxa found in the 37 exploratory datasets (NHits) with α = .05 was clearly highest for edgeR (997), followed by DESeq2 (741). The middle group consisted of corncob (521), LinDA (451), limma-voom (447), LDM (437), ORM/Wilcoxon (436), ANCOM-BC2 (414), MaAsLin2/t-test (395), LogR (387) and ZicoSeq (334). The lowest numbers were observed for metagenomeSeq (236), fastANCOM (223), and ALDEx2 (191).

While edgeR detected the highest number of taxa (997) with Conflict% = 9.2% and Replication% = 33% at the nominal FDR level α = .05, we note that, for instance, LinDA, LDM and limma-voom identified almost the same number of taxa (926 – 948) with lower Conflict% (5.4% - 6.2%) and similar Replication% (32% - 33%) when α was set at .20. Furthermore, when the nominal FDR levels were chosen for each method so that each detected exactly 1000 taxa, the highest consistency was achieved by MaAsLin2/t-test, ORM/Wilcoxon, LogR, limma-voom and LinDA while the lowest consistency was demonstrated by ANCOM-BC2 and DESeq2 (Figure A9).

*Correlation of estimates and overlap percentage of confidence intervals*

The average Spearman correlation between the DA estimates in the exploratory and validation datasets varied little between most methods (Figure A11). It was highest for ORM/Wilcoxon (.25), MaAsLin2/t-test (.24), LinDA (.24), limma-voom (.24), ALDEx2 (.24) and fastANCOM (.24). The lowest values were observed for edgeR (.16), LDM (.15) and ANCOM-BC2 (.13).

The overlap percentage of the 83.4% confidence intervals (CI%) varied between 29% and 47%. The highest values were observed for LogR (47%), MaAsLin2 (46%), and ORM (44%) and metagenomeSeq (42%), and the lowest for fastANCOM (33%), ALDEx2 (32%), ANCOM-BC2 (32%) and DESeq2 (29%) (Figure A12).

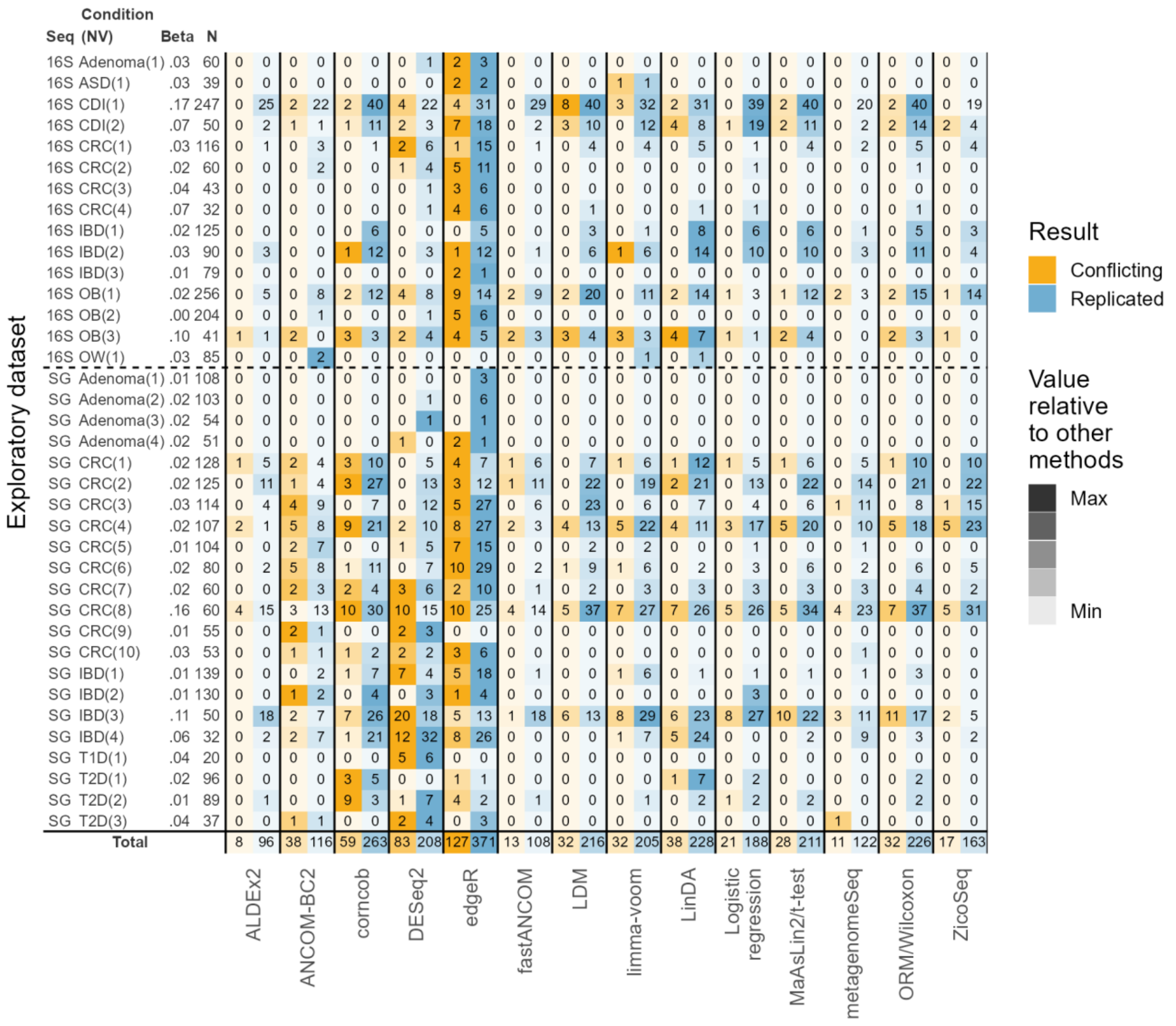

**Figure 5** The number of conflicting and replicated results found by 14 DAA methods when datasets from separate studies were used as exploratory and validation datasets. One exploratory dataset may have had multiple validation datasets (indicated by NV). In each slot is the number of taxa for which a conflicting or replicated result was found in at least one of the validation datasets. Conflicting result = the result for a taxon was significant in the exploratory dataset (FDR adjusted p < .05) and validation dataset(s) (p < .05) but in opposite directions. Replicated result = the result for a taxon was significant in the exploratory dataset and validation dataset(s) in the same direction. Seq. = sequencing type (16S or SG = shotgun); Condition = the studied condition; Beta = Beta diversity explained by the experimental group (case/control); N = the sample size of the exploratory dataset.

## Summary of the results

*Split-data analyses*

Five methods, ALDEx2, fastANCOM, LogR, MaAsLin2/t-test and ORM/Wilcoxon had adequately low Conflict% while also producing the highest replication percentages. Of these methods ORM/Wilcoxon and MaAsLin2/t-test, followed by LogR, were the most sensitive ones. At the opposite end, the most inconsistent methods by far were ANCOM-BC2, DESeq2 and edgeR. For instance, their Conflict% was order(s) of magnitude higher than what was considered ideal. Moreover, while DESeq2 and edgeR were very sensitive, similar sensitivity *with better consistency* could be reached by other methods (especially ORM/Wilcoxon, MaAsLin2/t-test and LogR) when the nominal FDR level was increased (Figure A3).

As most methods detected most of their significant taxa in larger datasets (sample size N > 100) the above overall evaluations were effectively based mainly on the performance on those datasets (Figure A17.1). The most consistent methods did not, however, produce conflicting results even on smaller datasets (N < 40).

*Separate study analyses*

Overall, the most consistent methods were ALDEx2, fastANCOM, LogR, MaAsLin2/t-test, ORM/Wilcoxon and ZicoSeq. ORM/Wilcoxon and MaAsLin2/t-test, followed by LogR and ZicoSeq were more sensitive than ALDEx2 and fastANCOM. The least consistent methods were again ANCOM-BC2, DESeq2 and edgeR. There was, however, a strong negative correlation between consistency and sensitivity (Figure A13), and thus a high sensitivity of some methods compensated for their low consistency to some degree. Nevertheless, when comparing the methods under a constant high sensitivity, ORM/Wilcoxon, MaAsLin2/t-test, LogR, LinDA and limma-voom performed slightly better than the other methods in terms of consistency (Figure A9).

## Discussion

We performed a comprehensive investigation of how well microbial differential abundance analysis results provided by 14 different methods replicate between datasets. We especially studied how different methods can replicate statistical significance and the direction of taxon-wise differential abundance between random partitions of datasets and between different studies. This approach allowed us to identify relatively sensitive methods that perform consistently on datasets from the same study. The identified methods also performed well when considering replication between separate studies. Furthermore, our analyses revealed methods that provide systematically inconsistent results, suggesting caution in their use.

Overall, the best performance was obtained by analyzing TSS normalized counts, i.e. relative abundances, with a non-parametric method (ORM/Wilcoxon), log transformed TSS normalized counts with the t-test/linear regression (MaAsLin2/t-test), or the presence/absence of taxa with logistic regression (LogR). These methods performed adequately according to our criteria in the split-data analyses and were among the most consistent ones also in the separate study analyses. Importantly, they were clearly more sensitive than the other consistent methods. These methods were also the most consistent ones when all methods were forced to detect a high constant number of taxa by adjusting the nominal FDR levels. Moreover, they performed above average when considering the correlation of DA estimates and the overlap percentage of confidence intervals.

At the opposite end, ANCOM-BC2, DESeq2 and edgeR were found to be highly inconsistent. For instance, in the split-data analyses these methods provided at least an order of magnitude more conflicting results than what was considered acceptable. Our finding is in line with some other studies where a correspondingly poor performance of DESeq2 and especially edgeR has been observed in the form of high error rates in group label shuffling on real datasets [1], [2], [16]. Moreover, when covariates were included, the performance of DESeq2 dropped further and the percentages of conflicting and replicated results were almost of the same order of magnitude (Figure A8). Lastly, while DESeq2 and especially edgeR were clearly the most sensitive methods we observed that a similar sensitivity could be achieved more reliably by other methods by using a higher nominal FDR level.

Of the rest of the methods, ALDEx2 and fastANCOM showed good consistency but were notably less sensitive compared to MaAsLin2/t-test and ORM/Wilcoxon. This finding aligns with previous studies that found ALDEx2 to be conservative [1], [2], [16], [33], yet consistently performing [1], [16]. Furthermore,

LinDA and limma-voom had little deficiencies in consistency while being acceptably sensitive. Lastly, corncob, LDM, metagenomeSeq and ZicoSeq compromised a little in consistency and/or sensitivity.

An important observation in our investigation was that the best performance was obtained by the elementary methods that are not specifically designed for (microbial) DAA. It therefore seems that applying advanced strategies to address the compositionality of microbiome data and/or employing complex statistical models reduces either the consistency or sensitivity achieved in DAA. We interpret this to indicate that it may not be generally beneficial to employ complex statistical modeling techniques in DAA. Especially, considering the highly inconsistent performance of DESeq2 and edgeR, the employment of the negative binomial model may not be advisable in DAA (for pure negative binomial model, see Figures A14.1 – A14.4).

Nevertheless, it is well-known that analyzing simple TSS normalized counts can lead to spurious results if the total absolute abundances differ systematically between the experimental groups [5], [9], [34], [35]. Therefore, despite the possible loss in consistency and/or sensitivity, complex normalization strategies may be needed in DAA to avoid such spurious findings. To investigate this, we performed DAA on a real dataset from a study on mice where the absolute microbial abundances were *measured* to be clearly higher in one group [36]. We observed that the methods employing TSS normalization estimated the sign of the "true" DA (based on the measured absolute abundances) comparably to the other methods (Figure A15). Moreover, in our main analyses the consistency of MaAsLin2/t-test and ORM/Wilcoxon was comparable to other methods on all datasets (Figure 3 and Figure 5) and their p-values changed little when TSS normalization was replaced by more advanced normalization methods, such as CSS, GMPR [37], TMM or Wrench [38] (Figure A16). Consequently, while this examination was very limited and our analyses may not include datasets with very large systematic differences in total absolute abundances, we gathered some indirect evidence indicating that sophisticated normalization methods may offer little advantage over simple TSS normalization in many typical studies on human gut microbiome.

While the consistency of several DAA methods was appropriate in the split-data analyses, a general observation was that DAA results were generally substantially less consistent in the separate study analyses. This indicates that any between-study differences in DAA depend largely on actual differences between the study populations and/or differences in the experimental workflows. The possible low replicability of DAA results across studies is thus not necessarily due to any faulty performance of the DAA method. Nevertheless, while the consistency was lower in the separate study analyses, the more consistent methods produced 10 – 20 times more replicated findings than conflicting results.

Confidence intervals have not been investigated in previous benchmarking studies on DAA. We found that at least some methods (ALDEx2, fastANCOM, LogR, MaAsLin2/t-test and ORM/Wilcoxon) provided relatively consistent intervals in the split-data analyses. We note, however, that a high overlap percentage does not necessarily indicate accurate estimates but may also reflect extensively wide confidence intervals. Moreover, apart from DESeq2, adding covariates in the analysis did not essentially alter the performance of the tested DAA methods. Additionally, the run-time of each tested method was mostly under one minute and should thus not be an issue in practice.

An interesting finding was that when simply analyzing the presence/absence of taxa (LogR), the sensitivity was only around 15% lower than that of MaAsLin2/t-test and ORM/Wilcoxon. These observations are in line with two recent studies [39], [40] and they suggest that a substantial amount of relevant information in microbial count data may lie merely in the observed presence/absence of taxa.

It has been reported that the general performance of DAA methods may depend on data characteristics [9] and that the consistency of any method may depend strongly on the dataset [16]. It can therefore be considered a limitation of our study that we merely investigated the methods' overall performance. However, when replication of DAA results is considered, it would be especially relevant to use the same

DAA method in all compared studies. It is therefore important to identify methods that work reasonably well on most datasets.

Due to data availability reasons, we employed only datasets from human gut microbiome studies. Moreover, we always filtered out taxa with prevalence < 10% and did not consider rarefying data. We cannot guarantee how well our findings would apply on other types of data. Furthermore, while we used datasets from a reasonably large number (N = 53) of studies, the total number of candidate taxa in the separate data analyses was relatively low (167 - 785) and most candidate taxa were from a few exploratory datasets. The effect of random variation was thus considerable on our findings in the separate study analyses. Those findings should therefore be seen as only rough indications for the expected replicability of DAA results between studies.

**Key points**

- We benchmarked 14 differential abundance analysis methods by examining the consistency and replicability of their results along with their sensitivity.
- We employed 69 real datasets from 53 microbiome studies.
- The highest consistency with good sensitivity was obtained by analyzing relative abundances or presence/absence with elementary methods (non-parametric test, linear regression/t-test, logistic regression).
- Some widely used methods were found to perform highly inconsistently.
- A higher sensitivity should be aimed for by using a consistent method with a higher significance level.

## Author biographies

**Juho Pelto** is a doctoral researcher in statistics at the Department of Computing, University of Turku. His research focuses on statistical analysis of microbiome data.

**Kari Auranen** is a professor of statistics at the Department of Clinical Medicine and at the Center of Statistics, University of Turku. His primary research interests include the statistical analysis of cohort studies and models of infectious disease transmission.

**Janne V. Kujala** is an associate professor of statistics at the Department of Mathematics and Statistics, University of Turku. His research spans computational statistics and applied probability.

**Leo Lahti** is a professor of data science at the Department of Computing, University of Turku. His research focuses on computational microbiome research.

## Funding

This work was supported by the Research Council of Finland [330887 to J.P., L.L.]; and the European Union’s Horizon 2020 research and innovation programme [952914 to J.P., L.L.].

## Author contributions

JP conceived the idea, carried out the analyses and drafted the manuscript. KA, JK and LL supervised JP and participated in editing and reviewing the manuscript. The authors read and approved the final manuscript.

## Competing interests

The authors declare that they have no competing interests.

## Data availability

The curated datasets and codes supporting the conclusions of this article are available in the https://github.com/jepelt/DAA_replicability repository (Zenodo https://doi.org/10.5281/zenodo.15047338). The original datasets employed in this study are available in the MicrobiomeHD database [19] or in the curatedMetagenomicData (version 3.10.0) R package [20]. All data curation and analyses were performed in R 4.2.3 [41].

## References

[1] J. T. Nearing *et al.*, 'Microbiome differential abundance methods produce different results across 38 datasets', *Nat Commun*, vol. 13, no. 1, p. 342, Jan. 2022, doi: 10.1038/s41467-022-28034-z.

[2] L. Yang and J. Chen, 'A comprehensive evaluation of microbial differential abundance analysis methods: current status and potential solutions', *Microbiome*, vol. 10, no. 1, p. 130, Aug. 2022, doi: 10.1186/s40168-022-01320-0.

[3] M. R. McLaren, A. D. Willis, and B. J. Callahan, 'Consistent and correctable bias in metagenomic sequencing experiments.', *Elife*, vol. 8, Sep. 2019, doi: 10.7554/eLife.46923.

[4] P. J. McMurdie and S. Holmes, 'Waste Not, Want Not: Why Rarefying Microbiome Data Is Inadmissible', *PLoS Comput Biol*, vol. 10, no. 4, p. e1003531, 2014, doi: 10.1371/JOURNAL.PCBI.1003531.

[5] H. Lin and S. Das Peddada, 'Analysis of microbial compositions: a review of normalization and differential abundance analysis.', *NPJ Biofilms Microbiomes*, vol. 6, no. 1, p. 60, Dec. 2020, doi: 10.1038/s41522-020-00160-w.

[6] M. Cappellato, G. Baruzzo, and B. Di Camillo, 'Investigating differential abundance methods in microbiome data: A benchmark study', *PLoS Comput Biol*, vol. 18, no. 9, p. e1010467, Sep. 2022, doi: 10.1371/JOURNAL.PCBI.1010467.

[7] J. Wirbel, M. Essex, S. K. Forslund, and G. Zeller, 'A realistic benchmark for differential abundance testing and confounder adjustment in human microbiome studies', *Genome Biology 2024 25:1*, vol. 25, no. 1, pp. 1–26, Sep. 2024, doi: 10.1186/S13059-024-03390-9.

[8] D. Swift, K. Cresswell, R. Johnson, S. Stilianoudakis, and X. Wei, 'A review of normalization and differential abundance methods for microbiome counts data', *Wiley Interdiscip Rev Comput Stat*, vol. 15, no. 1, p. e1586, Jan. 2023, doi: 10.1002/WICS.1586.

[9] S. Weiss *et al.*, 'Normalization and microbial differential abundance strategies depend upon data characteristics', *Microbiome*, vol. 5, no. 1, pp. 1–18, Mar. 2017, doi: 10.1186/S40168-017-0237-Y/FIGURES/8.

[10] G. B. Gloor, J. M. Macklaim, V. Pawlowsky-Glahn, and J. J. Egozcue, 'Microbiome Datasets Are Compositional: And This Is Not Optional', *Front Microbiol*, vol. 8, no. NOV, Nov. 2017, doi: 10.3389/FMICB.2017.02224.

[11] H. Lin and S. Das Peddada, 'Multigroup analysis of compositions of microbiomes with covariate adjustments and repeated measures', *Nature Methods 2023 21:1*, vol. 21, no. 1, pp. 83–91, Dec. 2023, doi: 10.1038/s41592-023-02092-7.

[12] H. Zhou, K. He, J. Chen, and X. Zhang, 'LinDA: linear models for differential abundance analysis of microbiome compositional data', *Genome Biol*, vol. 23, no. 1, p. 95, Dec. 2022, doi: 10.1186/s13059-022-02655-5.

[13] C. Zhou, H. Wang, H. Zhao, and T. Wang, 'fastANCOM: a fast method for analysis of compositions of microbiomes.', *Bioinformatics*, vol. 38, no. 7, pp. 2039–2041, Mar. 2022, doi: 10.1093/bioinformatics/btac060.

[14] B. D. Martin, D. Witten, and A. D. Willis, 'MODELING MICROBIAL ABUNDANCES AND DYSBIOSIS WITH BETA-BINOMIAL REGRESSION.', *Ann Appl Stat*, vol. 14, no. 1, pp. 94–115, Mar. 2020, doi: 10.1214/19-aoas1283.

[15] Y.-J. Hu and G. A. Satten, 'Compositional analysis of microbiome data using the linear decomposition model (LDM)', *bioRxiv*, p. 2023.05.26.542540, May 2023, doi: 10.1101/2023.05.26.542540.

[16] M. Calgaro, C. Romualdi, L. Waldron, D. Risso, and N. Vitulo, 'Assessment of statistical methods from single cell, bulk RNA-seq, and metagenomics applied to microbiome data', *Genome Biol*, vol. 21, no. 1, pp. 1–31, Aug. 2020, doi: 10.1186/S13059-020-02104-1/FIGURES/7.

[17] Z. D. Wallen, 'Comparison study of differential abundance testing methods using two large Parkinson disease gut microbiome datasets derived from 16S amplicon sequencing', *BMC Bioinformatics*, vol. 22, no. 1, pp. 1–29, Dec. 2021, doi: 10.1186/S12859-021-04193-6/FIGURES/4.

[18] M. Khomich Id, I. Måge, I. Rud, and I. Berget, 'Analysing microbiome intervention design studies: Comparison of alternative multivariate statistical methods', 2021, doi: 10.1371/journal.pone.0259973.

[19] C. Duvallet, S. M. Gibbons, T. Gurry, R. A. Irizarry, and E. J. Alm, 'Meta-analysis of gut microbiome studies identifies disease-specific and shared responses', *Nat Commun*, vol. 8, no. 1, p. 1784, Dec. 2017, doi: 10.1038/s41467-017-01973-8.

[20] E. Pasolli *et al.*, 'Accessible, curated metagenomic data through ExperimentHub', *Nature Methods 2017 14:11*, vol. 14, no. 11, pp. 1023–1024, Oct. 2017, doi: 10.1038/nmeth.4468.

[21] Y. Benjamini and Y. Hochberg, 'Controlling the False Discovery Rate: A Practical and Powerful Approach to Multiple Testing', *Journal of the Royal Statistical Society: Series B (Methodological)*, vol. 57, no. 1, pp. 289–300, Jan. 1995, doi: 10.1111/J.2517-6161.1995.TB02031.X.

[22] M. J. Knol *et al.*, 'The (mis)use of overlap of confidence intervals to assess effect modification', *Eur J Epidemiol*, vol. 26, pp. 253–254, 2011, doi: 10.1007/s10654-011-9563-8.

[23] Y.-J. Hu and G. A. Satten, 'Testing hypotheses about the microbiome using the linear decomposition model (LDM).', *Bioinformatics*, vol. 36, no. 14, pp. 4106–4115, Aug. 2020, doi: 10.1093/bioinformatics/btaa260.

[24] A. D. Fernandes, J. N. Reid, J. M. Macklaim, T. A. McMurrough, D. R. Edgell, and G. B. Gloor, 'Unifying the analysis of high-throughput sequencing datasets: characterizing RNA-seq, 16S rRNA gene sequencing and selective growth experiments by compositional data analysis', *Microbiome*, vol. 2, no. 1, p. 15, Dec. 2014, doi: 10.1186/2049-2618-2-15.

[25] M. I. Love, W. Huber, and S. Anders, 'Moderated estimation of fold change and dispersion for RNA-seq data with DESeq2.', *Genome Biol*, vol. 15, no. 12, p. 550, 2014, doi: 10.1186/s13059-014-0550-8.

[26] M. D. Robinson, D. J. McCarthy, and G. K. Smyth, 'edgeR: a Bioconductor package for differential expression analysis of digital gene expression data.', *Bioinformatics*, vol. 26, no. 1, pp. 139–40, Jan. 2010, doi: 10.1093/bioinformatics/btp616.

[27] M. E. Ritchie *et al.*, 'limma powers differential expression analyses for RNA-sequencing and microarray studies.', *Nucleic Acids Res*, vol. 43, no. 7, p. e47, Apr. 2015, doi: 10.1093/nar/gkv007.

[28] C. W. Law, Y. Chen, W. Shi, and G. K. Smyth, 'voom: Precision weights unlock linear model analysis tools for RNA-seq read counts.', *Genome Biol*, vol. 15, no. 2, p. R29, Feb. 2014, doi: 10.1186/gb-2014-15-2-r29.

[29] J. N. Paulson, O. C. Stine, H. C. Bravo, and M. Pop, 'Differential abundance analysis for microbial marker-gene surveys', *Nat Methods*, vol. 10, no. 12, pp. 1200–1202, Dec. 2013, doi: 10.1038/nmeth.2658.

[30] L. Geistlinger *et al.*, 'BugSigDB captures patterns of differential abundance across a broad range of host-associated microbial signatures', *Nature Biotechnology 2023*, pp. 1–13, Sep. 2023, doi: 10.1038/s41587-023-01872-y.

[31] H. Mallick *et al.*, 'Multivariable association discovery in population-scale meta-omics studies', *PLoS Comput Biol*, vol. 17, no. 11, p. e1009442, Nov. 2021, doi: 10.1371/journal.pcbi.1009442.

[32] Q. Liu, B. E. Shepherd, C. Li, and F. E. Harrell, 'Modeling continuous response variables using ordinal regression', *Stat Med*, vol. 36, no. 27, p. 4316, Nov. 2017, doi: 10.1002/SIM.7433.

[33] S. Hawinkel, F. Mattiello, L. Bijnens, and O. Thas, 'A broken promise: microbiome differential abundance methods do not control the false discovery rate', doi: 10.1093/bib/bbx104.

[34] V. Lloréns-Rico, S. Vieira-Silva, P. J. Gonçalves, G. Falony, and J. Raes, 'Benchmarking microbiome transformations favors experimental quantitative approaches to address compositionality and sampling depth biases', *Nat Commun*, vol. 12, no. 1, p. 3562, Jun. 2021, doi: 10.1038/s41467-021-23821-6.

[35] D. Vandeputte *et al.*, 'Quantitative microbiome profiling links gut community variation to microbial load', *Nature 2017 551:7681*, vol. 551, no. 7681, pp. 507–511, Nov. 2017, doi: 10.1038/nature24460.

[36] J. T. Barlow, S. R. Bogatyrev, and R. F. Ismagilov, 'A quantitative sequencing framework for absolute abundance measurements of mucosal and lumenal microbial communities', *Nature Communications 2020 11:1*, vol. 11, no. 1, pp. 1–13, May 2020, doi: 10.1038/s41467-020-16224-6.

[37] L. Chen, J. Reeve, L. Zhang, S. Huang, X. Wang, and J. Chen, 'GMPR: A robust normalization method for zero-inflated count data with application to microbiome sequencing data', *PeerJ*, vol. 2018, no. 4, p. e4600, Apr. 2018, doi: 10.7717/PEERJ.4600/SUPP-1.

[38] M. S. Kumar, E. V. Slud, K. Okrah, S. C. Hicks, S. Hannenhalli, and H. Corrada Bravo, ‘Analysis and correction of compositional bias in sparse sequencing count data’, *BMC Genomics*, vol. 19, no. 1, pp. 1–23, Nov. 2018, doi: 10.1186/S12864-018-5160-5/FIGURES/10.

[39] Z. Karwowska *et al.*, ‘Effects of data transformation and model selection on feature importance in microbiome classification data’, *Microbiome*, vol. 13, no. 1, pp. 1–14, Dec. 2025, doi: 10.1186/S40168-024-01996-6/FIGURES/4.

[40] R. Giliberti, S. Cavaliere, I. E. Mauriello, D. Ercolini, and E. Pasolli, ‘Host phenotype classification from human microbiome data is mainly driven by the presence of microbial taxa’, *PLoS Comput Biol*, vol. 18, no. 4, Apr. 2022, doi: 10.1371/JOURNAL.PCBI.1010066.

[41] R Core Team, ‘R: A Language and Environment for Statistical Computing’, 2023, *Vienna, Austria*. [Online]. Available: https://www.R-project.org/

# Appendix of “Elementary methods provide more replicable results in microbial differential abundance analysis”

## The datasets employed in the analyses

Condition: the studied condition; Split: included in the split-data analyses; Separate: included in the separate study analyses; Covariates: of age, sex, and BMI, those that were provided with the dataset; N(Control/Case): the sample size (number of participants) of control/case group; Avg. reads: geometric mean of total read counts (i.e. sequencing depths or library sizes).

| Study | Type | Condition | Split | Separate | Covariates | N (Control) | N (Case) | Avg. reads |
|---|---|---|---|---|---|---|---|---|
| [1] | 16S | T1D | x | | Age, Sex | 55 | 57 | 8 706 |
| [2] | 16S | Adenoma | x | x | Age, Sex, BMI | 172 | 198 | 9 552 |
| [2] | 16S | CRC | x | x | Age, Sex, BMI | 172 | 120 | 10 142 |
| [3] | 16S | CRC | x | x | | 22 | 21 | 1 220 |
| [4] | 16S | IBD | | x | | 16 | 146 | 9 282 |
| [5] | 16S | Obesity | x | x | Age, BMI | 428 | 185 | 25 671 |
| [5] | 16S | Overweight | x | x | Age, BMI | 428 | 319 | 25 713 |
| [6] | 16S | ASD | | x | | 20 | 19 | 1 414 |
| [7] | 16S | IBD | | x | Sex | 18 | 107 | 1 045 |
| [8] | 16S | HIV | x | | Sex | 34 | 205 | 10 074 |
| [9] | 16S | IBD | x | x | Age, Sex | 24 | 66 | 1 884 |
| [10] | 16S | Parkinson | x | | | 74 | 74 | 2 656 |
| [11] | 16S | CRA | x | | | 28 | 26 | 2 529 |
| [11] | 16S | NORA | x | | | 28 | 44 | 2 393 |
| [12] | 16S | CDI | x | x | Age, Sex | 154 | 93 | 4 605 |
| [12] | 16S | Diarrhea | x | | Age, Sex | 154 | 89 | 4 687 |
| [13] | 16S | EDD (CDI) | x | x | | 82 | 222 | 2 667 |
| [14] | 16S | ASD | x | x | Sex | 44 | 59 | 5 279 |
| [15] | 16S | Obesity | x | x | | 61 | 195 | 2 159 |
| [15] | 16S | Overweight | x | x | | 61 | 24 | 2 565 |
| [16] | 16S | CDI | x | x | | 25 | 25 | 3 290 |
| [17] | 16S | CRC | | x | | 18 | 14 | 241 |
| [18] | 16S | IBD | x | x | Age, Sex | 35 | 44 | 1 332 |
| [19] | 16S | Adenoma | x | x | Age, Sex | 30 | 30 | 52 774 |
| [19] | 16S | CRC | x | x | Age, Sex | 30 | 30 | 57 504 |
| [20] | 16S | CRC | x | x | Age, Sex, BMI | 75 | 41 | 112 207 |
| [21] | 16S | Cirrhosis | x | | | 23 | 23 | 719 |
| [22] | 16S | Obesity | | x | Age, Sex | 16 | 25 | 9 864 |
| [23] | 16S | Obesity | x | x | Sex | 100 | 104 | 3 548 |

**Table A1.1** The 16S datasets used in this benchmarking study

| Study | Type | Condition | Split | Separate | Covariates | N (Control) | N (Case) | Avg. reads |
|---|---|---|---|---|---|---|---|---|
| [24] | Shotgun | Parkinson | x | | | 28 | 31 | 26 802 551 |
| [25] | Shotgun | Adenoma | x | x | Age, Sex, BMI | 61 | 47 | 52 045 350 |
| [25] | Shotgun | CRC | x | x | Age, Sex, BMI | 61 | 46 | 52 597 788 |
| [26] | Shotgun | CRC | x | x | Age, Sex, BMI | 30 | 30 | 8 636 870 |
| [27] | Shotgun | IBD | | x | | 12 | 20 | 27 617 834 |
| [28] | Shotgun | Adenoma | x | x | Age, Sex, BMI | 28 | 26 | 5 128 558 |
| [28] | Shotgun | CRC | x | x | Age, Sex, BMI | 28 | 27 | 5 559 781 |
| [29] | Shotgun | T1D | | x | Age, Sex, BMI | 10 | 10 | 44 414 623 |
| [30] | Shotgun | IBD | | x | Age | 38 | 12 | 3 714 882 |
| [31] | Shotgun | ACVD | x | | Sex | 171 | 214 | 52 656 199 |
| [32] | Shotgun | IGT | x | | Age, BMI | 43 | 49 | 26 104 542 |
| [32] | Shotgun | T2D | x | x | Age, BMI | 43 | 53 | 27 820 530 |
| [33] | Shotgun | IBD | | x | BMI | 10 | 129 | 69 215 316 |
| [33] | Shotgun | T1D | | x | | 10 | 31 | 53 985 803 |
| [33] | Shotgun | T2D | | x | | 10 | 79 | 55 798 662 |
| [34] | Shotgun | Hypertension | x | | | 41 | 99 | 44 498 422 |
| [34] | Shotgun | Pre-hypertension | x | | | 41 | 56 | 44 178 298 |
| [35] | Shotgun | ME/CFS | x | | Sex, BMI | 50 | 50 | 54 649 832 |
| [36] | Shotgun | IBD | x | x | Sex, BMI | 236 | 81 | 51 814 434 |
| [37] | Shotgun | T2D | x | x | Sex, BMI | 174 | 170 | 38 298 897 |
| [38] | Shotgun | Cirrhosis | x | | Age, Sex, BMI | 114 | 123 | 44 482 671 |
| [39] | Shotgun | Cephalosporins | x | | Age, Sex, BMI | 36 | 36 | 125 324 066 |
| [40] | Shotgun | STH | x | | Age, Sex, BMI | 86 | 89 | 17 255 200 |
| [41] | Shotgun | T2D | | x | Age, Sex, BMI | 18 | 19 | 44 929 627 |
| [42] | Shotgun | IBD | x | x | Age, Sex | 27 | 103 | 19 628 426 |
| [43] | Shotgun | Adenoma | x | x | Age, Sex, BMI | 24 | 27 | 89 916 338 |
| [43] | Shotgun | CRC | x | x | Age, Sex, BMI | 28 | 32 | 39 538 771 |
| [43] | Shotgun | CRC | x | x | Age, Sex, BMI | 40 | 40 | 42 904 588 |
| [43] | Shotgun | CRC | x | x | Age, Sex, BMI | 24 | 29 | 95 517 110 |
| [44] | Shotgun | CRC | x | x | Age, Sex, BMI | 52 | 52 | 63 755 391 |
| [45] | Shotgun | CRC | x | x | Age, Sex, BMI | 65 | 60 | 48 601 165 |
| [46] | Shotgun | Asthma | x | | Age, BMI | 177 | 24 | 72 321 657 |
| [46] | Shotgun | Migraine | x | | Age, BMI | 177 | 49 | 72 195 837 |
| [47] | Shotgun | Adenoma | x | x | Age, Sex, BMI | 251 | 67 | 42 606 059 |
| [47] | Shotgun | CRC | x | x | Age, Sex, BMI | 251 | 258 | 41 899 750 |
| [48] | Shotgun | BD | x | | Age, Sex, BMI | 45 | 20 | 41 731 548 |
| [49] | Shotgun | CRC | x | x | Age, Sex, BMI | 54 | 74 | 55 154 620 |
| [20] | Shotgun | Adenoma | x | x | Age, Sex, BMI | 61 | 42 | 54 303 740 |
| [20] | Shotgun | CRC | x | x | Age, Sex, BMI | 61 | 53 | 55 276 139 |
| [50] | Shotgun | Schizophrenia | x | | Age, Sex, BMI | 81 | 90 | 73 408 097 |

**Table A1.2** The shotgun datasets used in this benchmarking study

**Additional details on the analysis frameworks**

*Construction of the pairs of datasets in the split-data analyses*

In the split-data analyses, each exploratory-validation pair of datasets was constructed by randomly splitting an original dataset into two equal sized halves. Within each pair, one of the halves was randomly chosen as the exploratory dataset and the other one as the validation dataset. The splitting was done stratified by the case/control status (see Figure 1b). Only original datasets with at least 20 samples per group were used. This criterion was fulfilled by 57 datasets (from 43 studies). To increase the number of dataset pairs and thus to decrease the randomness in the results, we performed the splitting five times for each dataset. This resulted in 285 (57 x 5) pairs of datasets.

*Construction of the pairs of datasets in the separate study analyses*

In the separate study analyses, each exploratory-validation pair was made up from datasets from different studies which examined the same condition and had the same sequencing type. Within each pair, the dataset with the smaller sample size was set as the exploratory dataset because the results from a smaller study are more likely to replicate in the larger study than vice versa. We included only datasets with at least 10 samples per group (and with at least one taxon detected by at least one method with FDR level .05). This yielded the inclusion of 50 datasets (23 16S and 27 shotgun datasets), of which 37 (15 16S and 22 shotgun datasets) were used as exploratory datasets. In the cases of several pairs of datasets having the same exploratory dataset, the overall values of Conflict% and Replication% were calculated so that each taxon in each exploratory dataset received an equal weight in the calculations (see below for details).

*Analyses with covariates*

We carried out the split-data analyses also by including covariates in DAA. As age, BMI, and sex are typical covariates in human microbiome studies and at least some of them were provided with most of the original datasets, we included the available ones of these three covariates in DAA. There were 45 original datasets that provided age, sex or BMI and were eligible to the split-data analyses. Covariates included in each dataset are given in Tables A1.1 and A1.2. As metganomeSeq did not provide the possibility to include covariates, it was excluded from this analysis.

Age and BMI were treated as continuous variables and standardized before DAA. If at most 10% of the values of any covariate were missing, they were imputed by the group-wise median (age and BMI) or mode (sex) of the covariate. If >10% of the values were missing, the covariate was not included in DAA.

As the covariates were not chosen by subject matter considerations but merely by their availability and by their common use in microbiome studies, the goal of these analyses was not to evaluate how accurately the DAA methods can control for the effect of confounding variables. Instead, the goal was to evaluate how robustly different DAA methods perform when a more complex analysis than a mere two-group comparison is performed. Especially, we compared how the replication performance of each DAA method changed when covariates were included.

**Derivation of the acceptable level for the percentage of conflicting results in the split-data analyses**

For simplicity, we assume below that significance level $\alpha = .05$ is used in the exploratory datasets.

As statistical significance was defined in the exploratory datasets as (FDR adjusted p =) $q < .05$, a properly performing method should control the false discovery rate (FDR) at level .05. Consequently, at most 5% of the *candidate* taxa can be allowed to have incorrectly estimated the sign of DA in an exploratory dataset. Assuming that for most taxa the "true" DA is approximately zero, the "true" DA is likely close to zero also for the findings with incorrectly estimated sign. Therefore, as significance is based on unadjusted p-values < .05 in the validation datasets, a proper DAA method should provide a significant result for at most 5% of these findings in the validation dataset. Furthermore, half of these findings should be estimated to have the opposite direction to that in the exploratory dataset. Consequently, we obtain an approximation for an upper limit of acceptable Conflict%: $.05 \times .05 \times .50 = .00125$ or .125%. Thus, Conflict% < .125% can be interpreted to indicate the method providing proper p and q values.

**Calculation of 83.4% confidence intervals**

If a method provided confidence intervals in its output, they were used in our analyses. If a method provided only standard errors (SE), the 83.4% confidence intervals (CI) were calculated as follows. $CI = \hat{\beta} +/- t(.917)_{df} \times SE$, where $\hat{\beta}$ is the DA estimate and $t(.917)_{df}$ is the $1 - (1 - .834) / 2 = .917$ quantile of the t-statistic with df degrees of freedom. The degrees of freedom were chosen so that the confidence intervals matched the p-values provided by the method.

**Details on the calculation of Conflict%, Replication% and CI% in the separate study analyses**

In the separate study analyses, in case there were multiple pairs of datasets with the same exploratory dataset, the results in the validation datasets were weighted so that each candidate taxon in each exploratory dataset received an equal weight in the calculations. The calculation is best illustrated with an example. We use Replication% here as an example but values for Conflict% and CI% were calculated in a similar manner.

Assume an exploratory dataset E1 had three validation datasets V1a, V1b and V1c and assume that four taxa (Taxon 1, Taxon 2, Taxon 3, and Taxon 4) were significant in E1. Let us further assume that Taxon 1 was absent in all validation datasets, Taxon 2 was present in V1a and V1b but replicated only in V1a, Taxon 3 was present in V1b and V1c but replicated in neither of them, and Taxon 4 was present in all validation datasets but replicated only in V1a and V1c. Finally, assume that an exploratory dataset E2 had validation datasets V2a and V2b, and that only Taxon 4 was significant in E2 and present only in V2a where it replicated.

There were thus three candidate taxa from E1 (Taxon 2, Taxon 3 and Taxon 4; note that Taxon 1 was absent in V1a, V1b and V1c and was therefore not considered as a candidate taxon) and one candidate taxon from E2 (Taxon 4). In E1 Taxon 2 was 1/2 replicated, Taxon 3 was not replicated, and Taxon 4 was 2/3 replicated. In E2 Taxon 4 was fully replicated as it was replicated in the datasets (only V2a) where it was present. Now Replication% = $([1/2 + 0 + 2/3] + [1]) / (3 + 1) = 54.2\%$.

**Details on running the DAA methods**

The DAA methods were mostly run with the default settings. However, if a method provided a filter for taxa with too low prevalence, filtering was not used as we had removed taxa with prevalence < 10%. All other exceptions from the default settings are detailed below.

*ALDEx2*
Functions in the R package *ALDEx2* were used. First, CLR transformed Monte Carlo samples from the Dirichlet distribution were generated using the *aldex.clr* function. The default number (n = 128) of Monte Carlo samples were generated. The function *aldex.glm* was then run to perform DAA. Confidence intervals were calculated based on the standard errors provided by *aldex.glm*. Additionally, we ran ALDEx2 using the default *aldex* function (where covariates cannot be included). The t-test-based p-values were very similar to those provided by *aldex.glm* (data not shown). Instead, the results based on Wilcoxon test (of ALDEx2) (ALDEx2-Wilcox in Figures A14.1-A14.4) were slightly different from those based on the *aldex.glm*. Furthermore, by setting the gamma parameter to a non-null value (gamma = 0.5) in the *aldex* function, we were able to incorporate scale uncertainty into ALDEx2 [51]. This version of ALDEx2 was even less sensitive than the standard version (ALDEx2-scale in Figures A14.1-A14.4).

*ANCOM-BC2*
The function *ancombc2* from the R package *ANCOMBC* was used to perform DAA. Importantly, *the taxa that did not pass the sensitivity analysis for zeros were never considered significant*, as suggested by the authors of ANCOM-BC2. Furthermore, *ancombc2* function was run separately for the p values (with p_adj_method = 'none' and alpha = .05) and for the FDR adjusted p values on each FDR level (by setting p_adj_method = 'BH' and alpha = .01, .05, .10 or .20). This was done to the sensitivity analysis for zeros to work properly. Otherwise the *ancombc2* function was used with the default settings, e.g. the detection of structural zeros was not implemented (struc_zero = FALSE).

*corncob*
The function *differentialTest* from the R package *corncob* was used to perform DAA. The dispersion parameter was *not* allowed to vary between groups (phi.formula = ~ 1 and phi.formula_null = ~ 1). This choice was made as we observed that the performance of *corncob* dropped drastically when dispersion was allowed vary between the groups (corncob-UEV in Figures A14.1-A14.4). We used p-values based on the likelihood ratio test (test = “LRT”) as they were found to give a little better results (higher sensitivity) compared to results based on Wald p-values (data not shown). The confidence intervals were based on the Wald’s approximation, however, and they did not therefore match exactly the p-values based on LRT.

*DESeq2*
Functions in the R package *DESeq2* were used. First, a *DESeqDataSet* object was created from the count matrix by using *DEseqDatasetFromMatrix.* Next, size factors were estimated using *EstimateSizeFactorsForMatrix* (with type = “poscounts”). The results based on the Wald test were then calculated with the function *DESeq*. In the exploratory datasets, we used the adjusted p-values provided by DESeq (instead of performing the Benjamini-Hochberg correction for the raw p-values). Additionally, we calculated results based on the likelihood ratio test (*DESeq* function with test = “LRT” and the model without the group variable as the reduced model). With the likelihood ratio test, DESeq2 had lower error rates (without covariates) but was also clearly less sensitive (DESeq2-LRT in Figures A14.1-A14.4). Furthermore, we tried DESeq2 with GMPR normalized counts (*GMPR* function from *GUniFrac* package). The results, however, were not better than with the default size factors. (Data not shown).

*edgeR*
Functions in the R package *edgeR* were used. First, *DGEList* object was created using the *DGEList* function. Next, the default TMM normalization factors were calculated with function *calcNormFactors*. Then dispersions were estimated with the *estimateDisp* function. Finally, the results based on the quasi-likelihood test were calculated using the *glmQLFTest function*.

*fastANCOM*
DAA was performed using the *fastANCOM* function (from package *fastANCOM*) with the default settings (pseudo = 0.5, sig = 0.05 and ref.rate = .05, struc_zero = FALSE). Statistical significance was based on the p-values and not on the fraction of rejected log-ratio tests (*REJECT* value) provided by *fastANCOM*. This choice was made so that the performance of fastANCOM could be compared to that of the other methods.

*LDM*
DAA was performed using function *ldm* (from package *LDM*) with the default settings. The q values provided by LDM were used as adjusted p-values in the exploratory datasets. The p and q values based on the omnibus test combing the results with non-transformed and arcsine-root-transformed abundances were used (p.otu.omni and q.otu.omni). We also run LDM with CLR transformed counts (comp.anal = TRUE) as suggested recently [52] but this did not improve the performance of LDM (LDM-CLR in Figures A14.1-A14.4).

*limma-voom*
Functions from packages *edgeR* and *limma* were used. First, a *DGEList* object was created from count matrix using the *DGEList* function. Then TMM normalization factors were calculated using function *calcNormFactors*. Next, the mean-variance relationship and thus the weights for the observations were calculated using the *voom* function. A linear model was then fitted for each taxon using the lmFit function. Lastly, the final results based on the empirical Bayes moderated standard errors were obtained by applying the function *eBayes*.

*LinDA*
DAA was performed using the function *linda* from package *LinDA* with the default settings (adaptive = TRUE, pseudo.cnt = 0.5). As the zero-handling approach may have varied between exploratory and validation datasets when the default “Adaptive” setting was used, we tried LinDA also with the pseudo-count and imputation approaches available in the *linda* function. Using these alternative options had very little effect on the results, however (Data not shown).

*LogR (Logistic regression for presence/absence of taxa)*
The count data were first transformed so that non-zero counts (present taxa) were replaced by 1 and zero counts (absent taxa) were left as zeros. The DAA for each taxon was then performed using Firth type logistic regression [53]. This was implemented using the *logistf* function in the *logistf* R package. We used this robust version of logistic regression as the sample size was rather small (N = 20) in some cases and, furthermore, in some cases the prevalence of some taxa was zero in one of the groups.

*MaAsLin2/t-test*
DAA was performed using function *Maaslin2* in the R package *Maaslin2*. We used the default approach, i.e. a linear model for log transformed TSS normalized counts (analysis_method = “LM”, transform = “LOG” and normalization = “TSS”). Additionally, we tried MaAsLin2 with arcsine-square root transformed relative counts (transform = AST) as in [54], but this only dropped the number of replicated taxa detected by it (MaAsLin2-AST in Figures A14.1 - A14.4). We also tried MaAsLin2 with CLR, CSS and TMM normalized counts (normalization = “CLR”, “CSS” or “TMM”, respectively) but using these alternative normalizations did not improve its performance (MaAsLin2-CLR, MaAsLin2-CSS and MaAsLin2-TMM in Figures A14.1 – A14.4).

*metagenomeSeq*
Functions in R package *metagenomeSeq* were used. First, an *MRexperiment* object was created from the count matrix by using the function *newMRexperiment*. Next, CSS normalization factors were calculated using the function *cumNorm* (with parameter p = 0.5). DAA was then performed using the *fitFeatureModel* function.

*NegBin (Negative binomial regression)*
We included in the additional analyses also pure negative binomial regression (Figures A14.1 – A14.4). The analyses were performed using function *glmmTMB* (with family = nbinom2) in the R package *glmmTMB*. We used log(library size) as an offset term to effectively achieve TSS normalization.

*ORM/Wilcoxon (Ordinal regression model)*
An ordinal regression model was used to analyze TSS normalized counts for each taxon. The analyses were performed using the function *orm* in the R package *rms*. We used p-values based on the score test as they were closest to p-values from the Wilcoxon test. We also tried p-values based on the likelihood ratio and Wald tests. The former were slightly anti-conservative when the prevalence of a taxon was zero in one group and the latter were not sensible in such cases (Data not shown). The confidence intervals were based on the Wald's approximation, however, and did not therefore match exactly the p-values based on the score test. Additionally, we used ORM/Wilcoxon with GMPR and Wrench normalized counts (Figures A14.1 – A14.4).

*radEmu*
DAA was performed using the *emuFit* function from the *radEmu* package. We extracted p values based on Wald tests instead of the default score tests (return_wald_p = T, run_score_tests = F) as the score tests took an excessively long time to converge. We did not include radEmu in the main text of this study because it was introduced in a paper that was available only as a pre-print [55].

*ZicoSeq*
DAA was performed using the *ZicoSeq* function (from the *GUniFrac* package) with the default settings (e.g. is.winsor = TRUE, is.post.sample = TRUE). Especially, we employed the default square root link function [link.func = list(function(x) sign(x) * (abs(x))^0.5)]. The number of permutations was, however, set to 999 (perm.no = 999), instead of the default 99, to obtain more reliable p values.

There were also two very recent DAA methods that we originally planned to be included in this study, i.e. DACOMP [56] and LOCOM [57]. They were eventually excluded, however, as they did not always provide all the required quantities (p-value and DA estimate) and their performance was thus difficult to compare to the other methods.

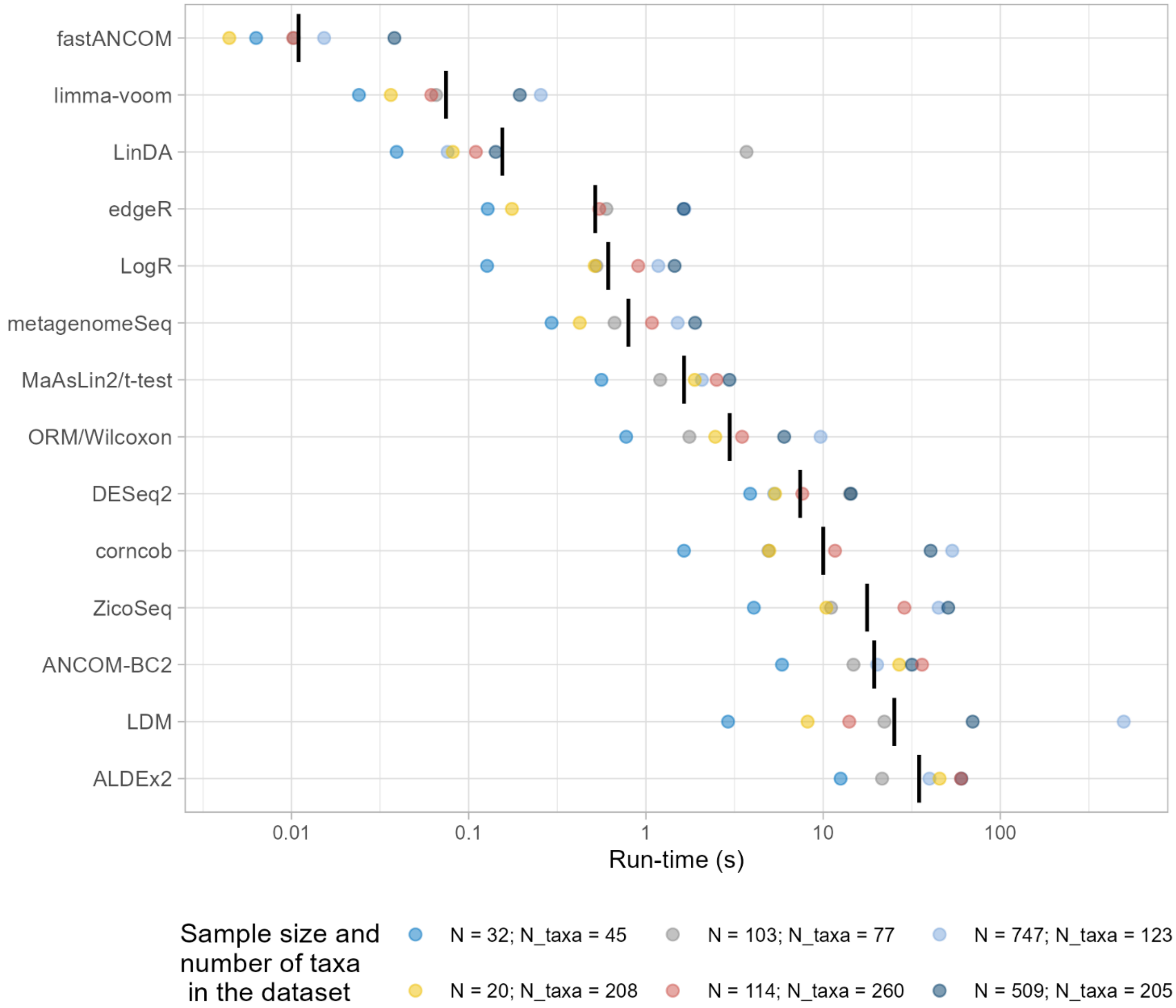

**Figure A2** The run-times of the 14 compared DAA methods on six microbiome datasets with varying numbers of samples and taxa (shown in the legend). The black vertical lines are the geometric means of the run-times (given in Table 2). The methods were run on a standard laptop (Intel(R) Core(TM) i7-8565U CPU @ 1.80GHz 1.99 GHz, 16GB RAM, 64bit, Windows 11).

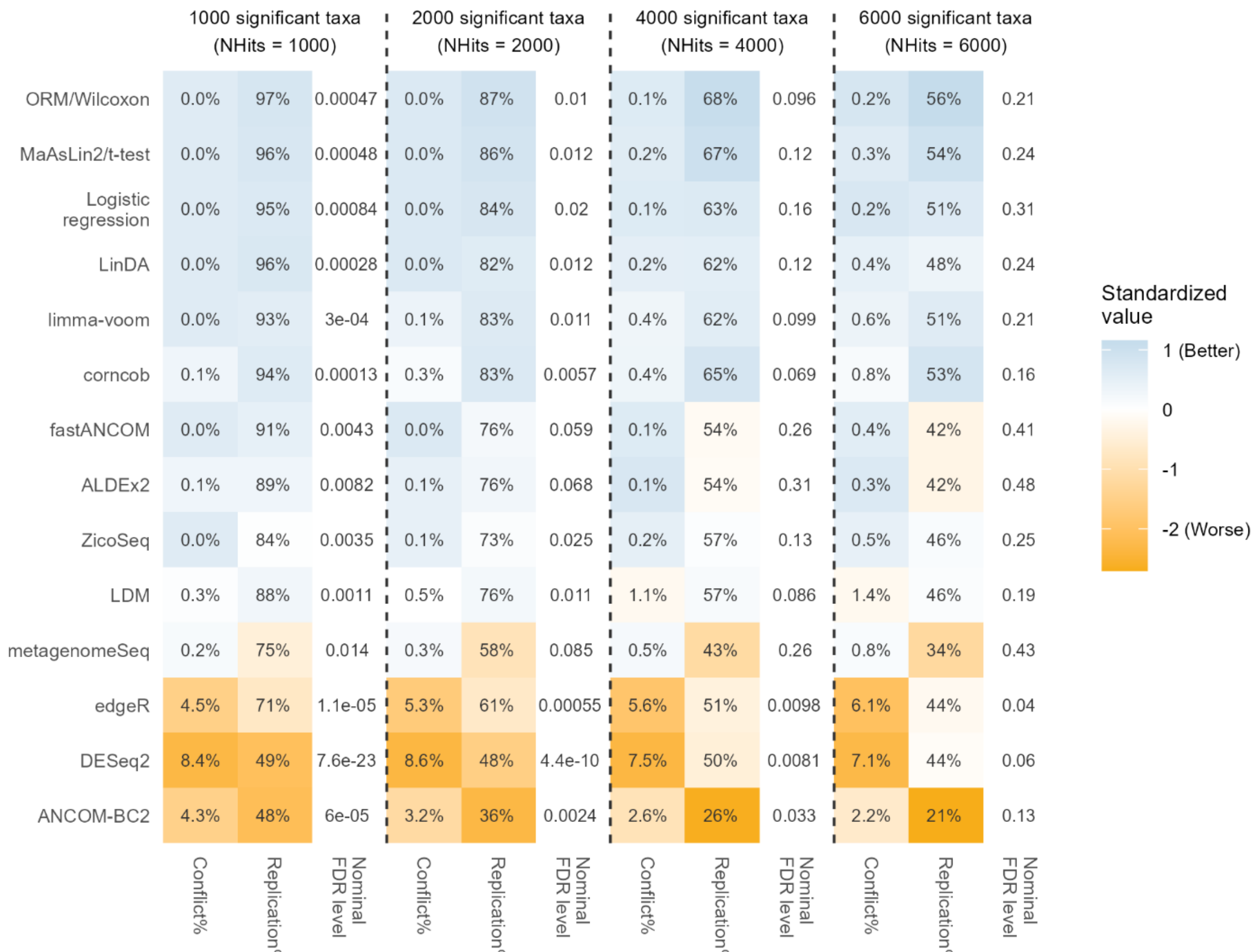

**Figure A3** The consistency of the 14 DAA methods when the nominal FDR levels were chosen so that each method detected a total of 1000, 2000, 4000 or 6000 significant taxa in the 285 exploratory datasets in the split-data analyses. The methods are in rank order based on the mean of the standardized values of the percentage of conflicting results (Conflict%) and replication percentage (Replication%) on all the four values of total number of significant taxa. (Conflict% was square root transformed before the standardization.) The number 6000 is approximately the number of significant taxa that the most sensitive methods (DESeq2 and edgeR) detected with nominal FDR level .05. (For ANCOM-BC2, the sensitivity filter was employed with FDR level .20. See Details on running the DAA methods.)

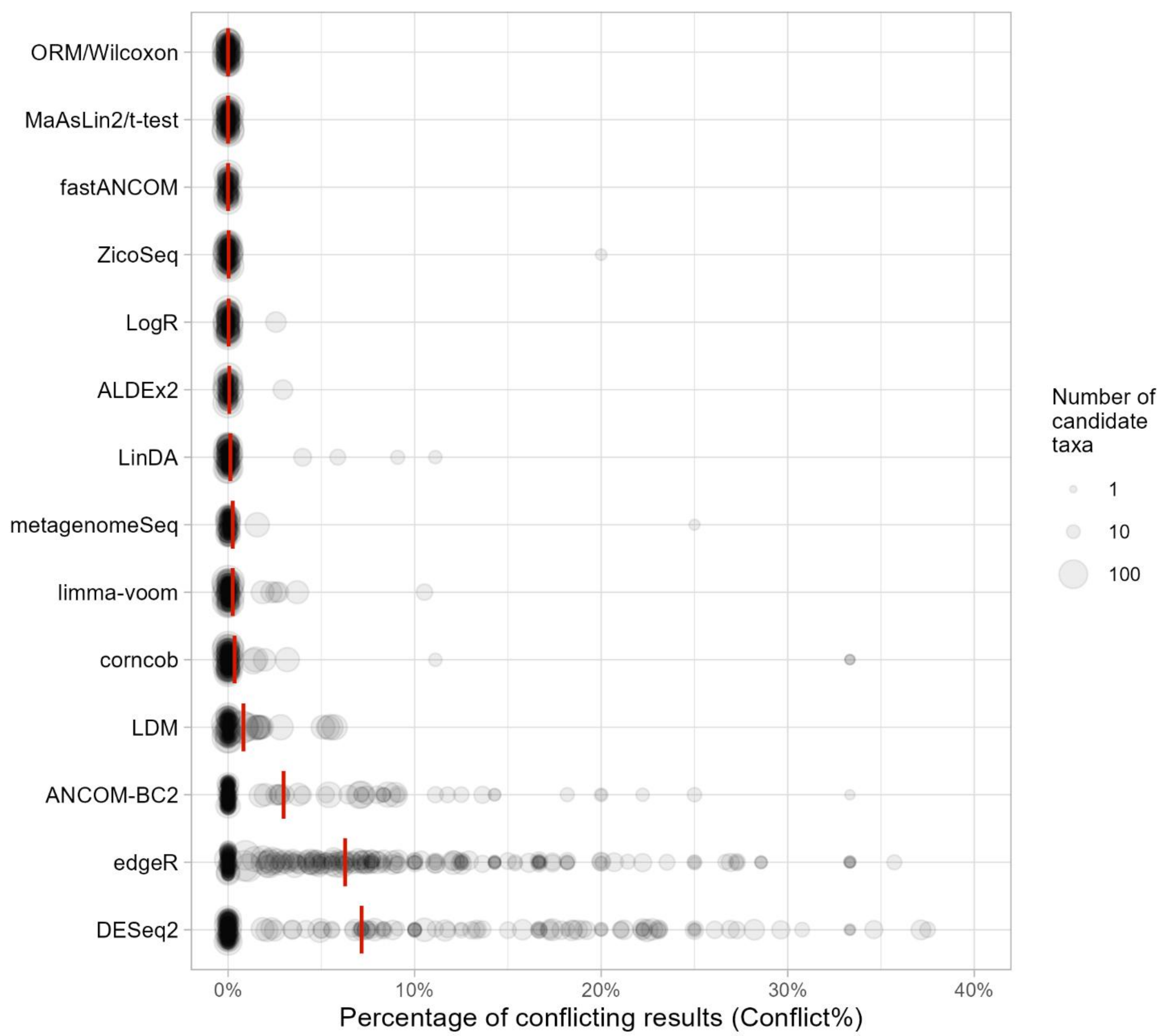

**Figure A4.1** The percentage of conflicting results (Conflict%) in the 285 pairs of datasets (with at least one candidate taxon) in the split-data analyses. The red lines indicate the overall Conflict%. The methods are ordered according to the overall values. The Figure is cut at 40%. Nominal FDR level = .05 was employed in the exploratory datasets.

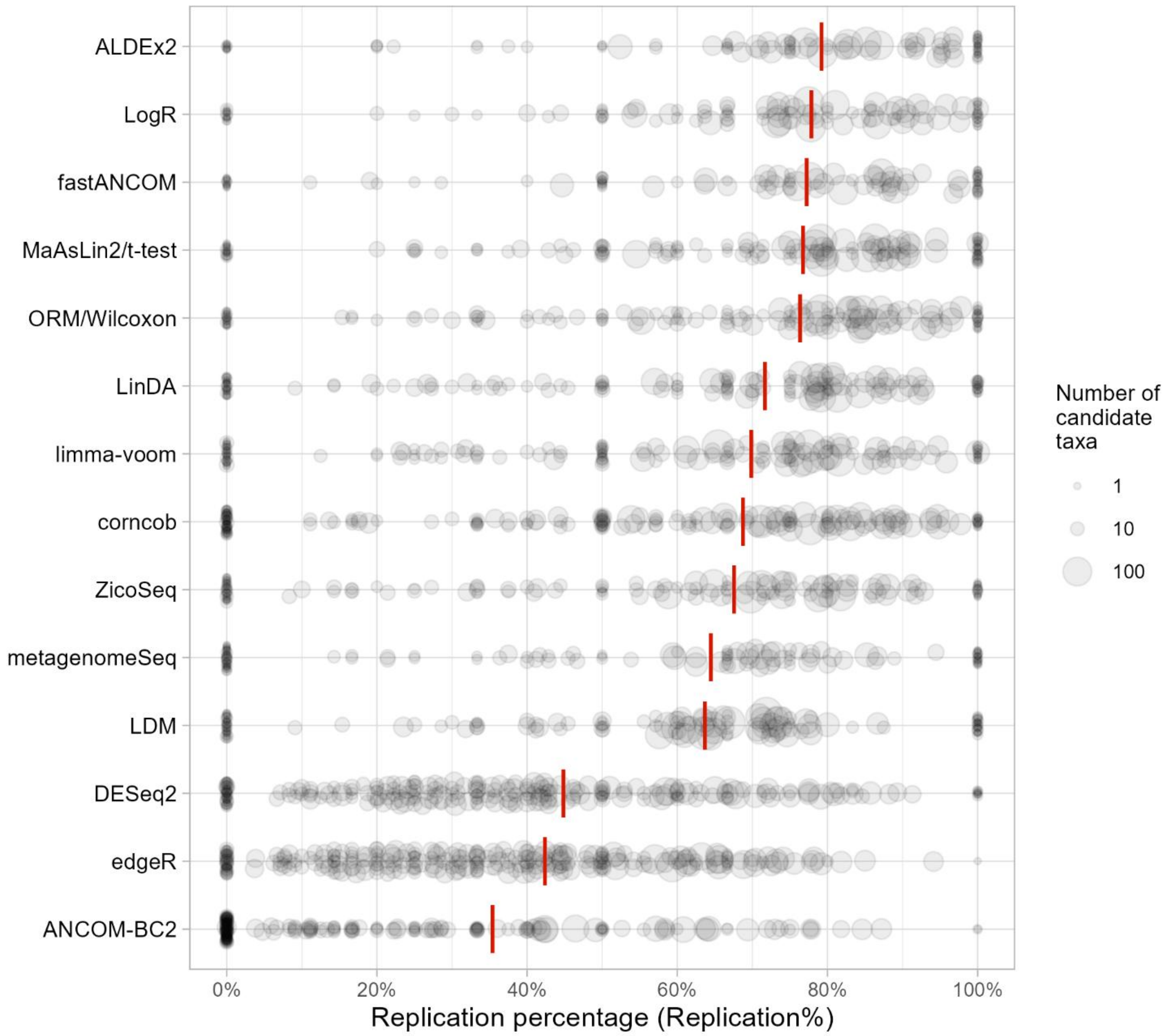

**Figure A4.2** The replication percentages (Replication%) in the 285 pairs of datasets (with at least one candidate taxon) in the split-data analyses. The red lines indicate the overall Replication%. The methods are ordered according to the overall values. Nominal FDR level = .05 was employed in the exploratory datasets.

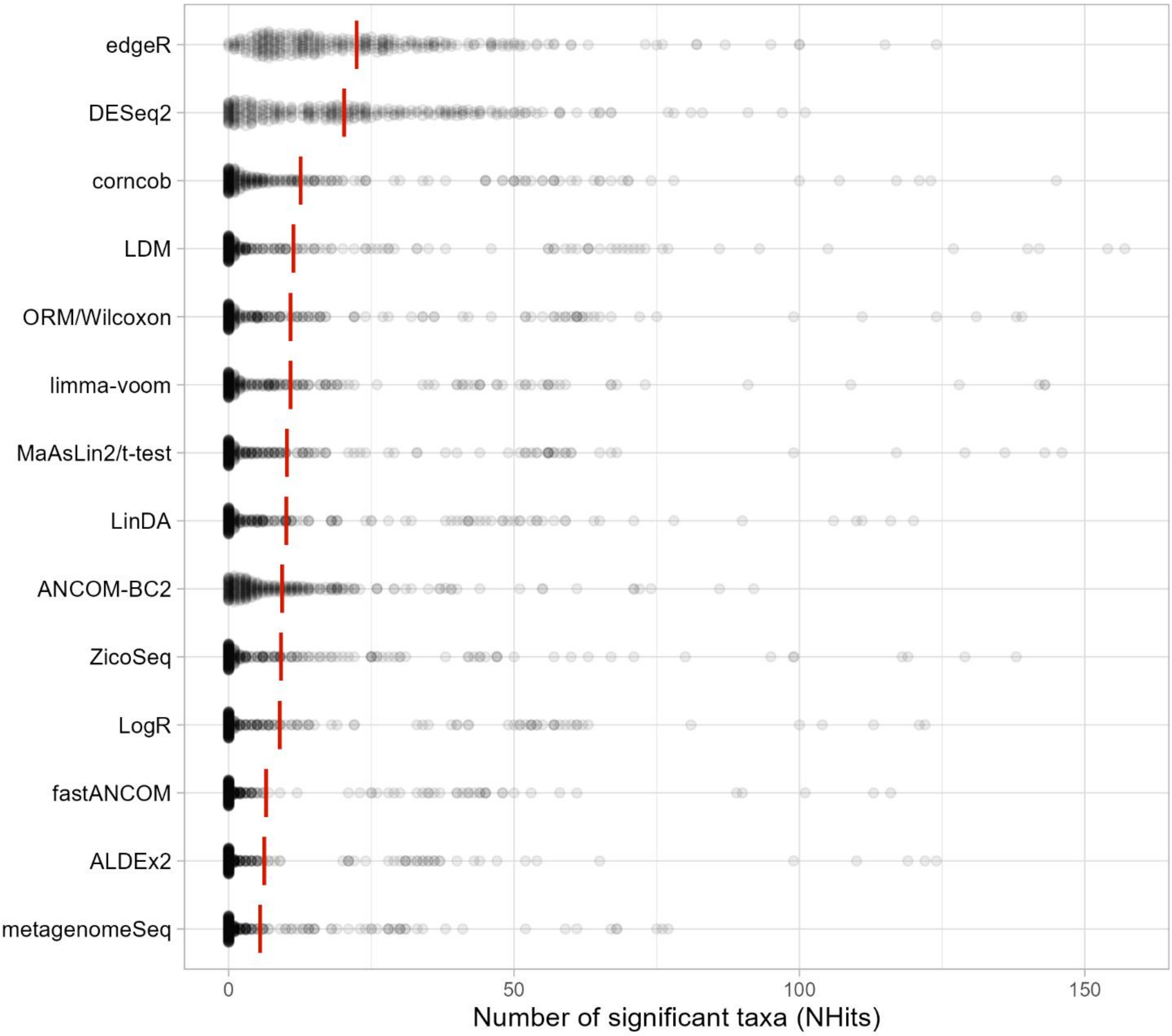

**Figure A4.3** The number of significant taxa (FDR adjusted p < .05) in the 285 exploratory datasets in the split-data analyses. The red lines indicate the mean numbers of significant taxa detected in the exploratory datasets. The methods are ordered according to these mean numbers.

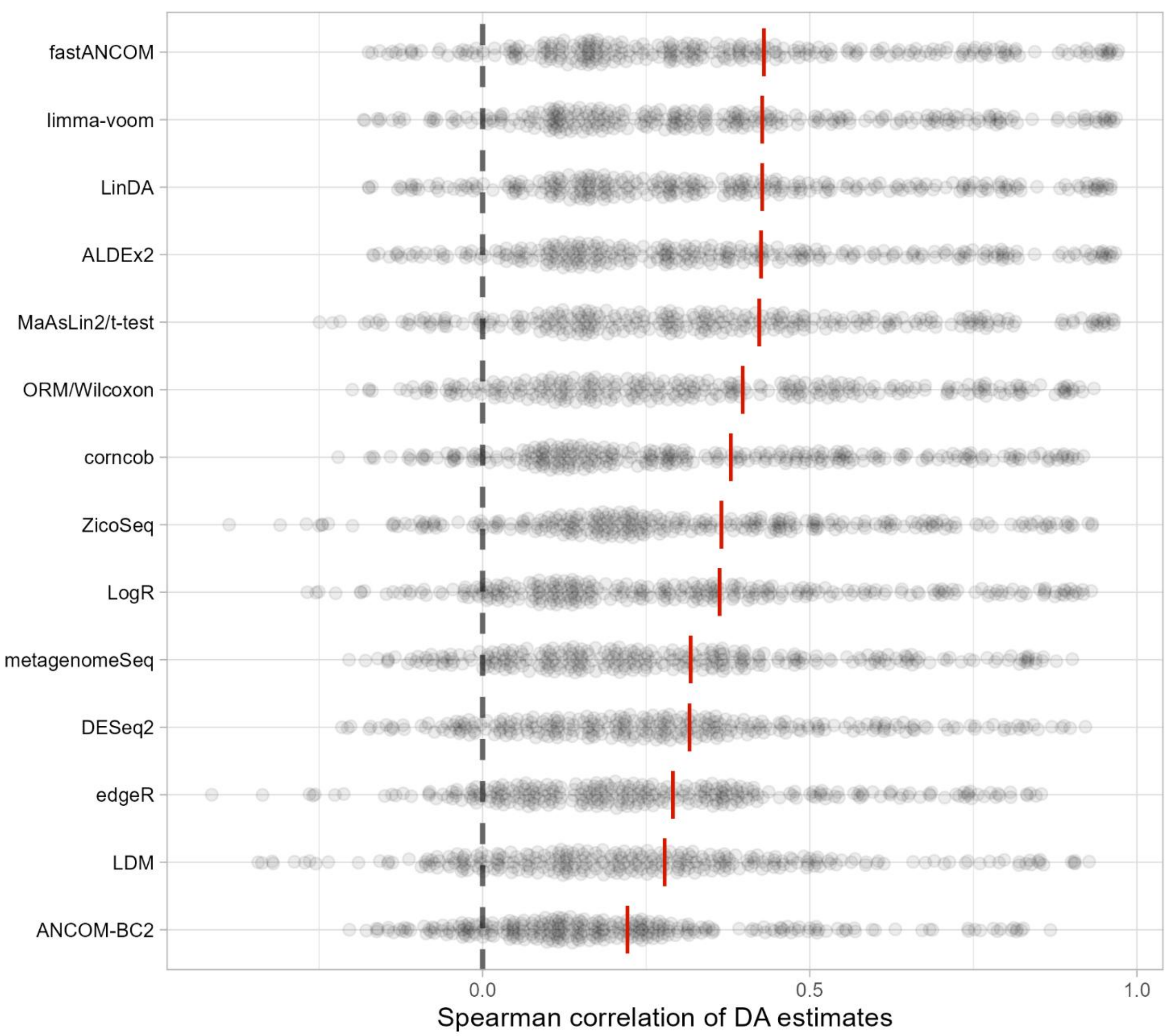

**Figure A5** The Spearman correlation coefficients of the DA estimates between exploratory and validation datasets in the split-data analyses. Values on all 285 exploratory-validation pairs of datasets are shown. The red lines indicate the average correlations, namely, the hyperbolic tangent transformed means of the inverse hyperbolic tangent transformed correlation coefficients. The methods are ordered according to the average correlation.

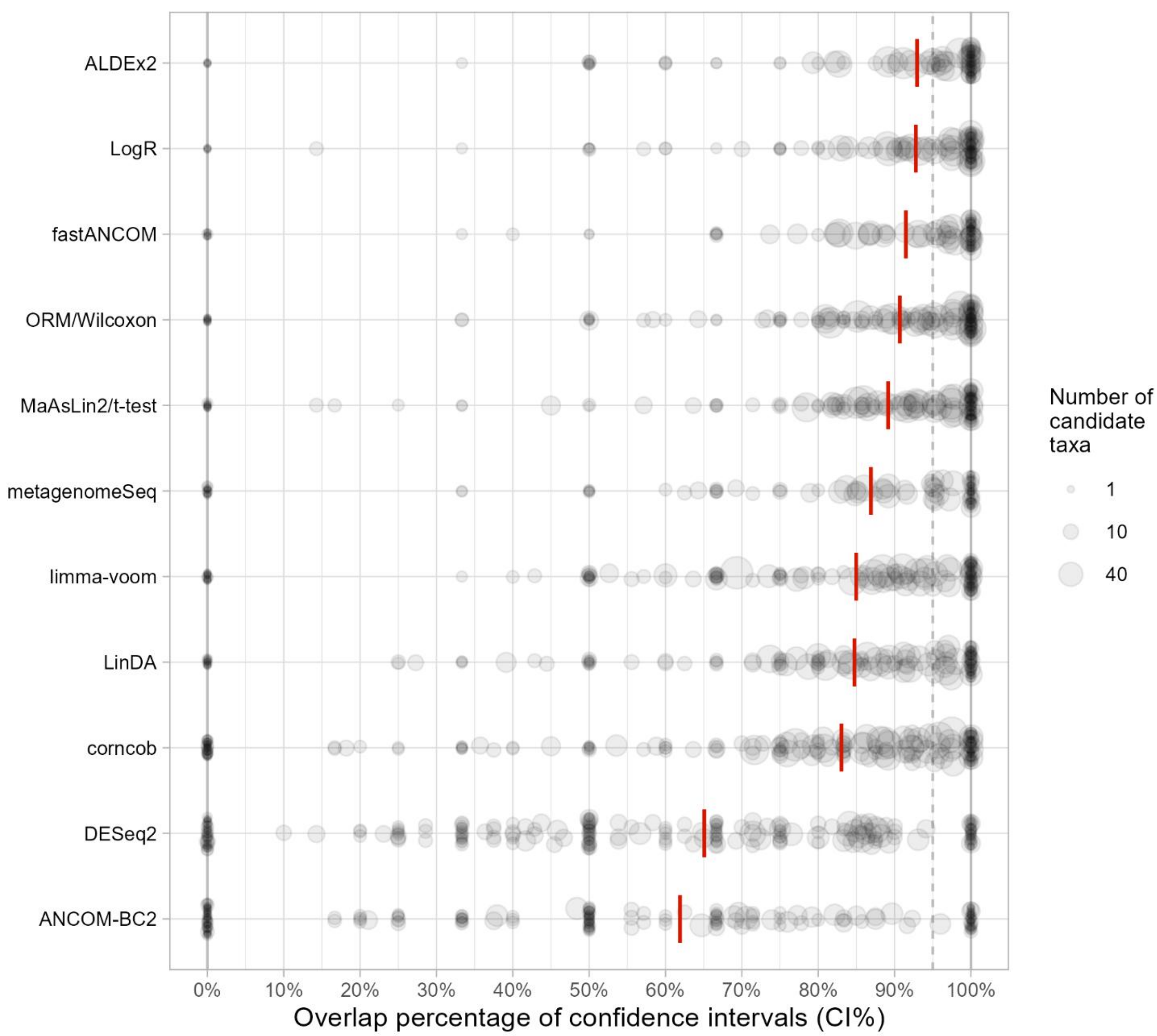

**Figure A6** The overlap percentage of 83.4% confidence intervals in the 285 pairs of datasets (with at least one candidate taxon) in the split-data analyses. Each grey point indicates the overlap percentage on a pair of datasets. The red lines indicate the overlap percentages (CI%) calculated over all candidate taxa in all pairs of datasets. The methods are ordered according to the overall value. The grey dashed line at 95% indicates a theoretical overlap percentage if DA estimates were normally distributed.

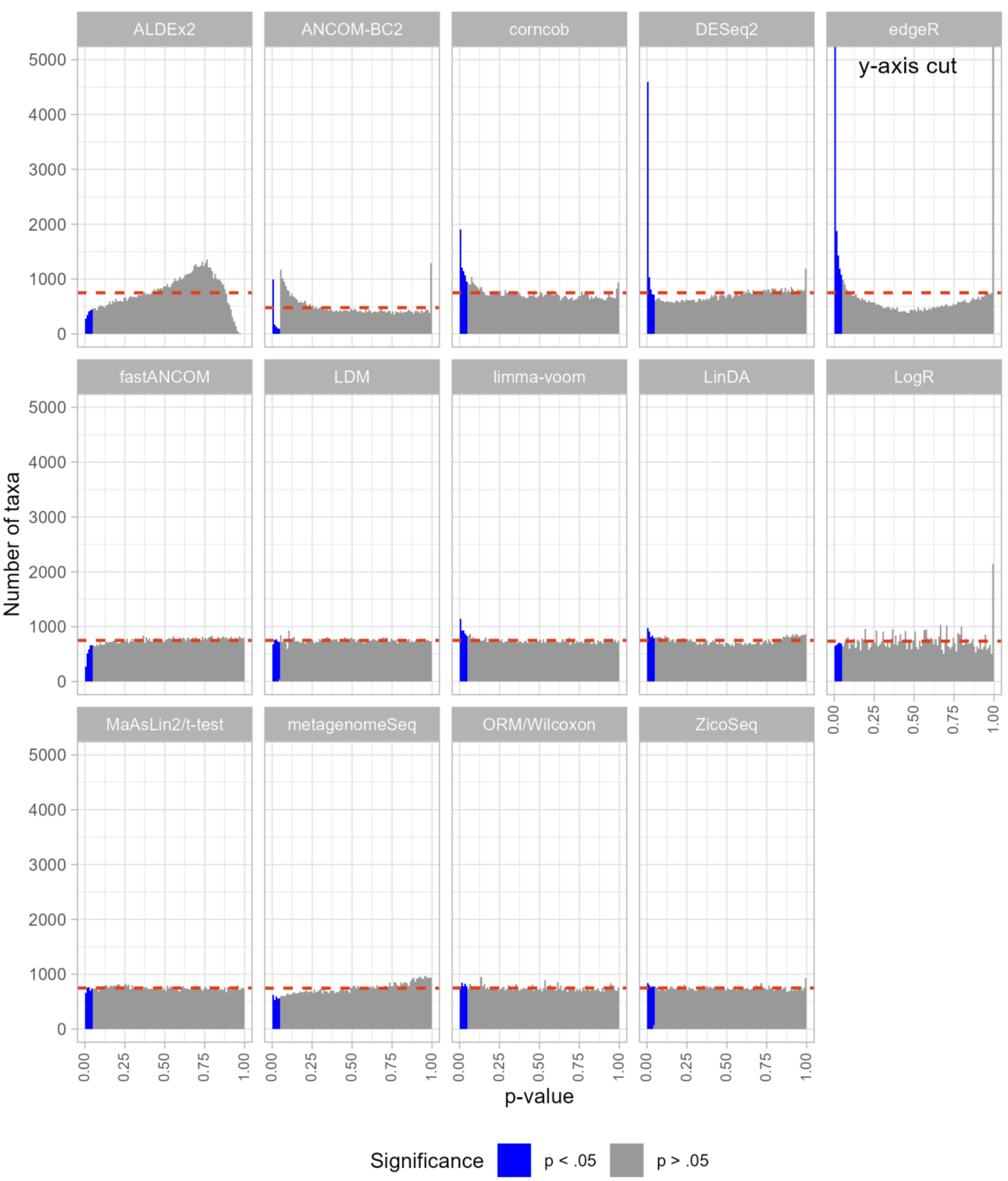

**Figure A7** The distribution of p-values (not adjusted for multiple testing) provided by the 14 DAA methods when run on 500 microbiome datasets with *randomly permuted* group labels (case/control). The width of each bin is .01. In this scenario, where there are no truly differentially abundant taxa, an ideally performing method should provide uniformly distributed p-values (indicated by the dashed red lines). Especially, the number of p-values < .05 (in blue color) should be at (or below) the red dashed line. For visualization purposes the y-axis is cut at 5000 taxa. For edgeR, the number of taxa with p < .05 is 9241. For ANCOM-BC2, the p-values that do not pass the sensitivity analysis for zeros (at p < .05) are filtered out. The 500 datasets were constructed by randomly permuting the group labels 10 times on the 50 real datasets included in the separate study analyses. The percentage of p < .05 for each method: ALDEx2 (2.5%), ANCOM-BC2 (3.1%), corncob (8.4%), DESeq2 (10.5%), edgeR (18.4%), fastANCOM (3.6%), LDM (4.8%), limma-voom (6.3%), LinDA (5.8%), LogR (4.6%), MaAsLin2/t-test (4.8%), metagenomeSeq (3.9%), ORM/Wilcoxon (5.3%), ZicoSeq (5.2%).

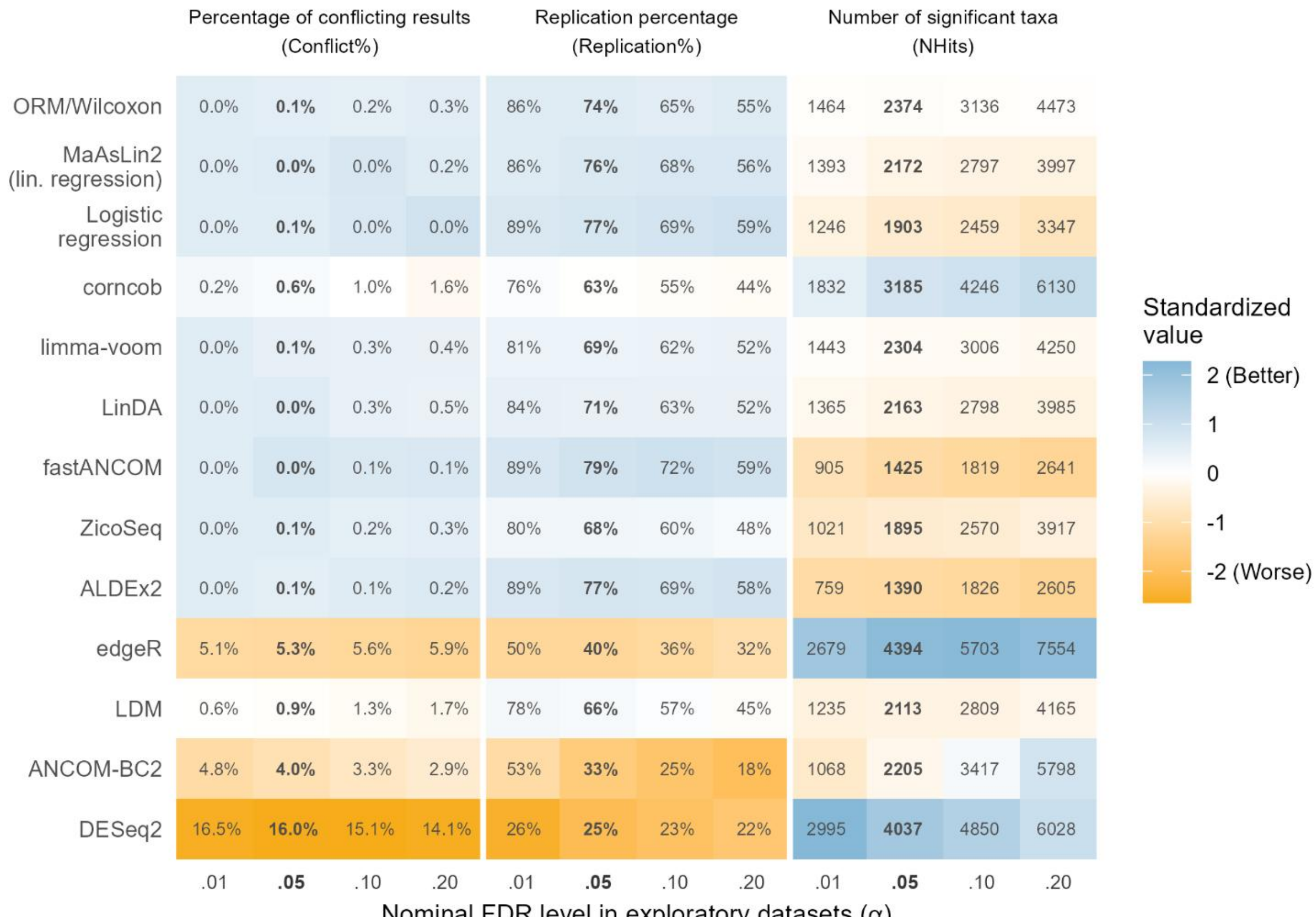

**Figure A8** The results of the split-data analyses when covariates (age, sex or BMI) are included in DAA. The methods are in rank order based on the mean of the standardized values of the metrics. Values based on the nominal FDR level $\alpha = .05$ are shown in bold. Altogether 45 original datasets for which at least one of the covariates age, sex or BMI was available were used in the analysis. Each original dataset was split five times to form pairs consisting of an exploratory and a validation dataset, thus totaling 225 (= 45 × 5) pairs of datasets. Candidate taxon = a taxon that was significant (FDR adjusted $p < \alpha$) in an exploratory dataset and present in the validation dataset. Conflict% = the percentage of candidate taxa that were significant ($p < .05$) in the validation dataset, but in the opposite direction to that in the exploratory dataset. Values below 0.025%, 0.125%, 0.25% and 0.50% were considered ideal for $\alpha = .01$, .05, .10 and .20, respectively. Replication% = the percentage of candidate taxa that were significant ($p < .05$) in the validation dataset in the same direction as in the exploratory dataset. NHits = the total number of significant taxa found in the 225 (45 × 5) exploratory datasets. A higher value of NHits can be considered better when it is accompanied by a low Conflict% and a high Replication%.

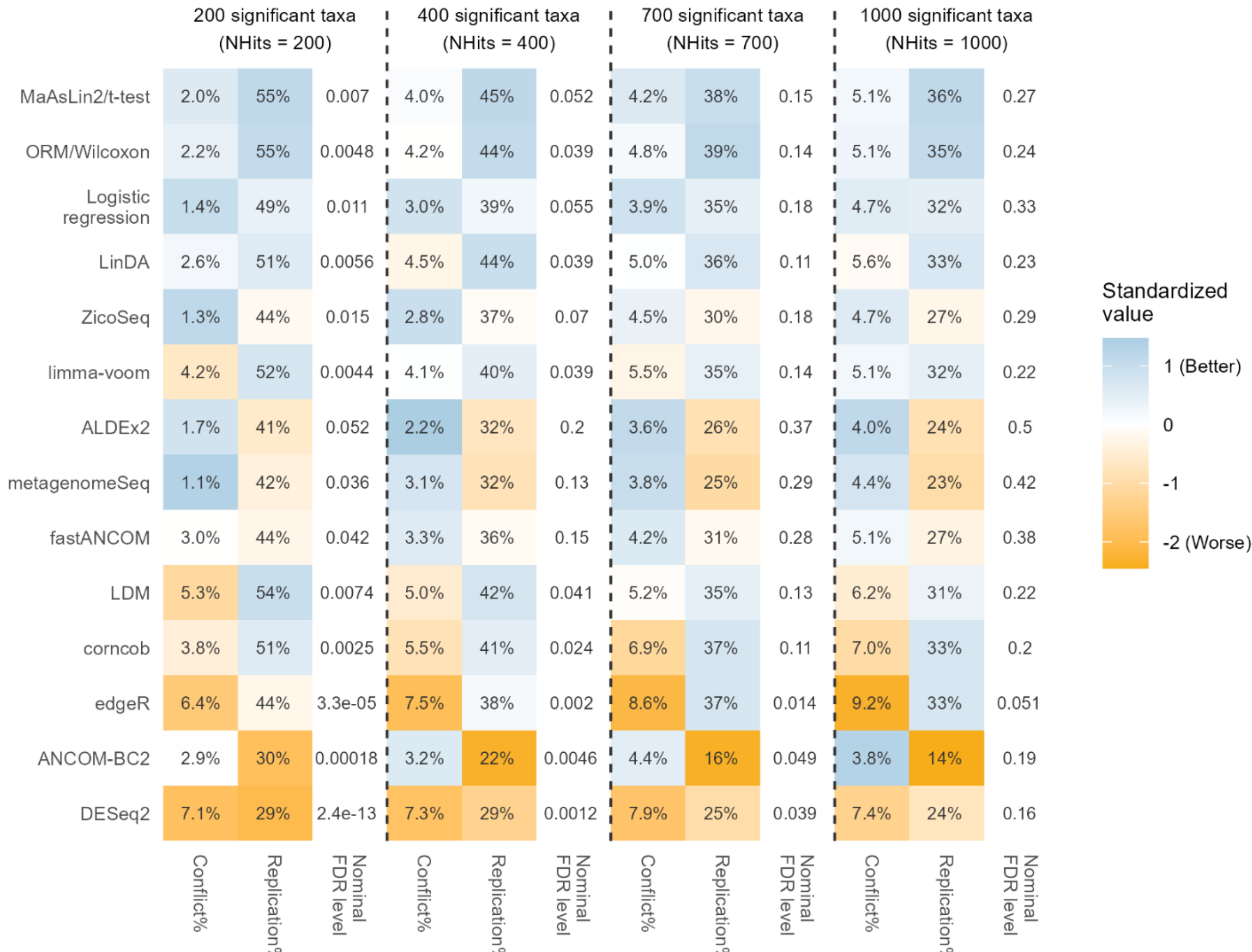

**Figure A9** The consistency of the 14 DAA methods when the nominal FDR levels were chosen so that each method detected a total of 200, 400, 700 or 1000 significant taxa in the 37 exploratory datasets in the separate study analyses. The methods are in rank order based on the mean of the standardized values of the percentage of conflicting results (Conflict%) and replication percentage (Replication%) on all the four values of total number of significant taxa. (Conflict% was square root transformed before the standardization.) The number 1000 is approximately the number of significant taxa that the most sensitive method (edgeR) detected with nominal FDR level .05. (For ANCOM-BC2, the sensitivity filter was employed with FDR level .20. See Details on running the DAA methods.)

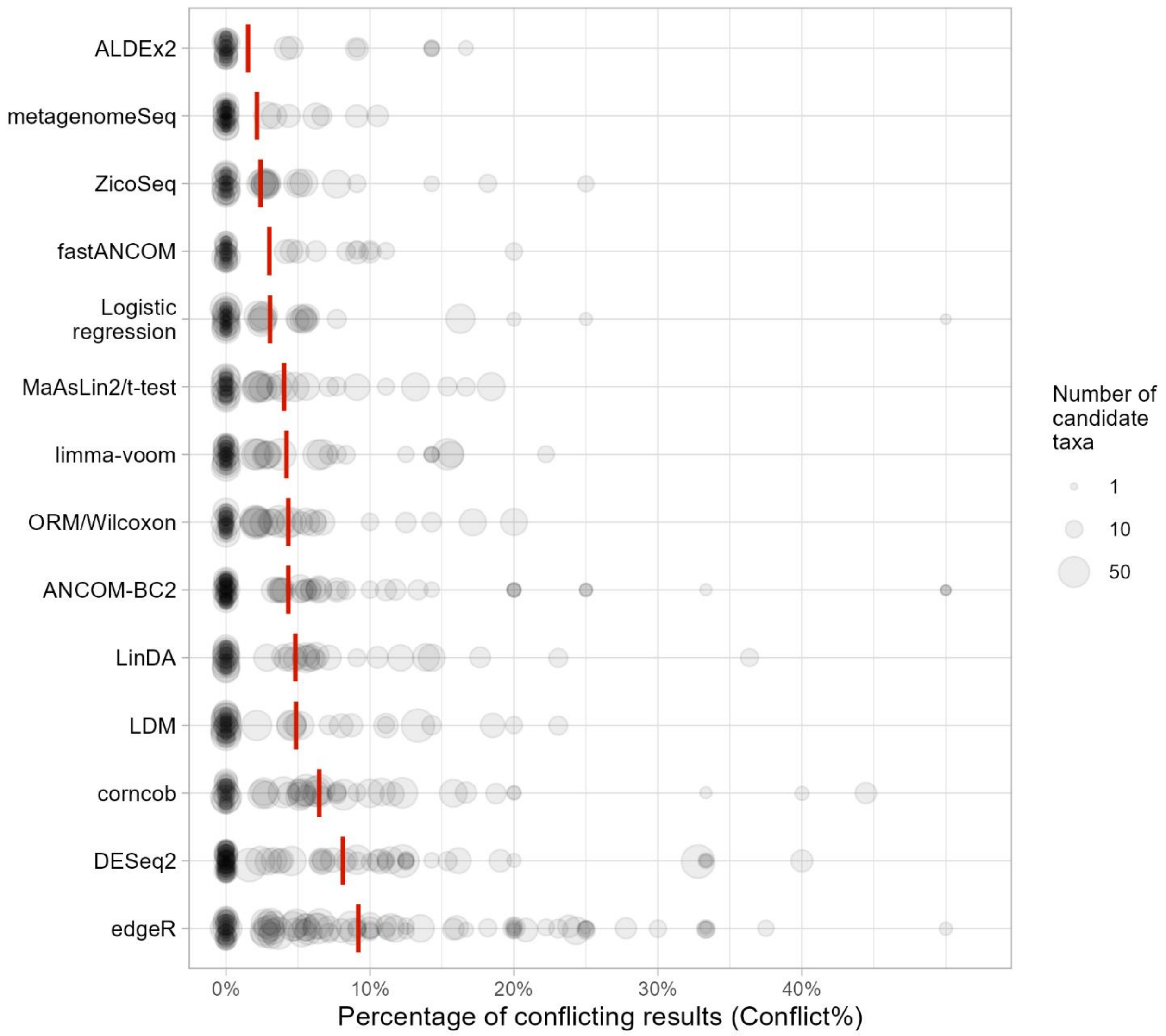

**Figure A10.1** The percentage of conflicting results (Conflict%) in all pairs of datasets (with at least one candidate taxon) in the separate study analyses. The red lines indicate the overall Conflict%. The methods are ordered according to the overall values. The figure is cut at 52%. Nominal FDR level = .05 was employed in the exploratory datasets.

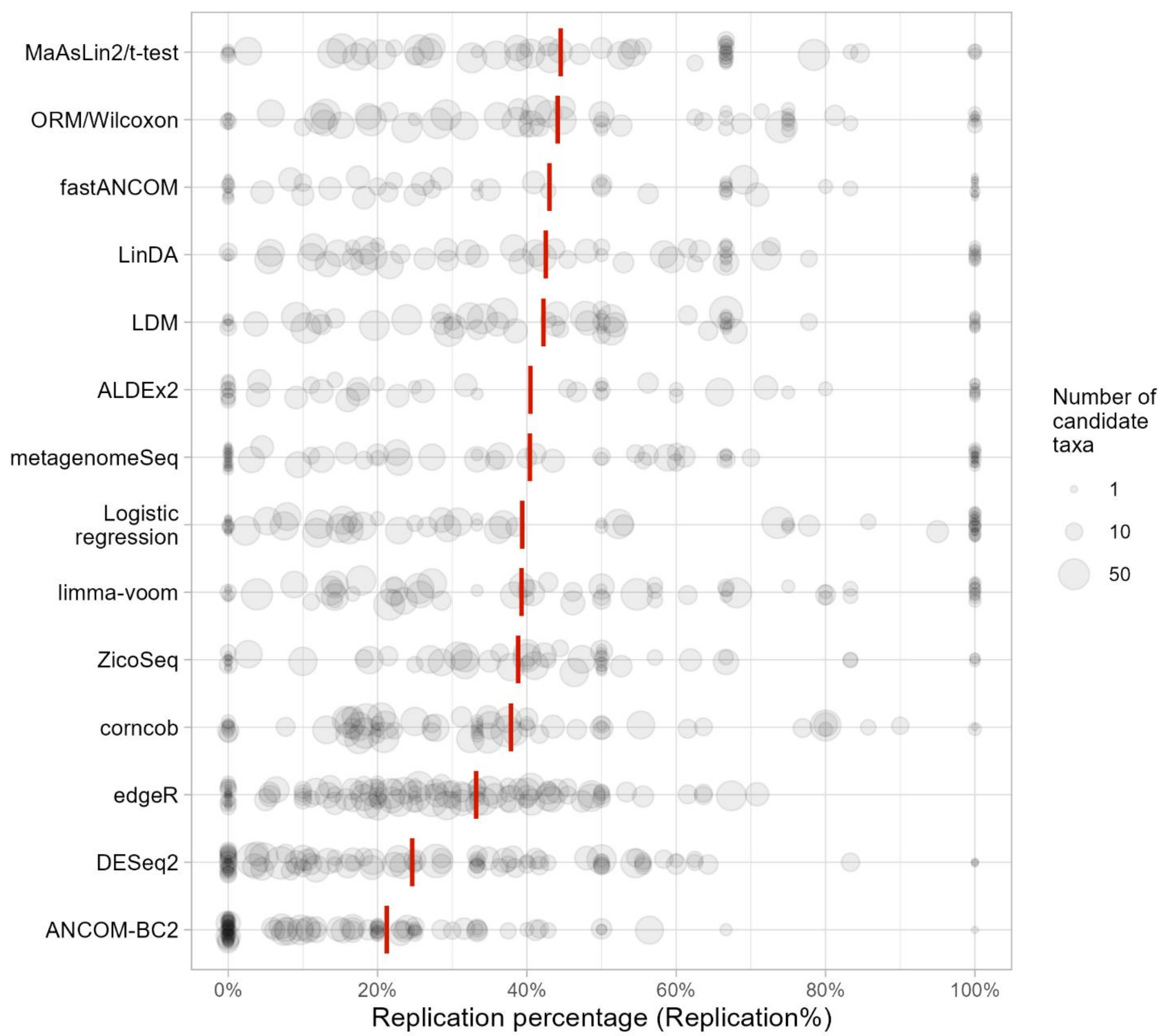

**Figure A10.2** The replication percentages (Replication%) in all pairs of datasets (with at least one candidate taxon) in the separate study analyses. The red lines indicate the overall Replication%. The methods are ordered according to the overall values. Nominal FDR level = .05 was employed in the exploratory datasets.

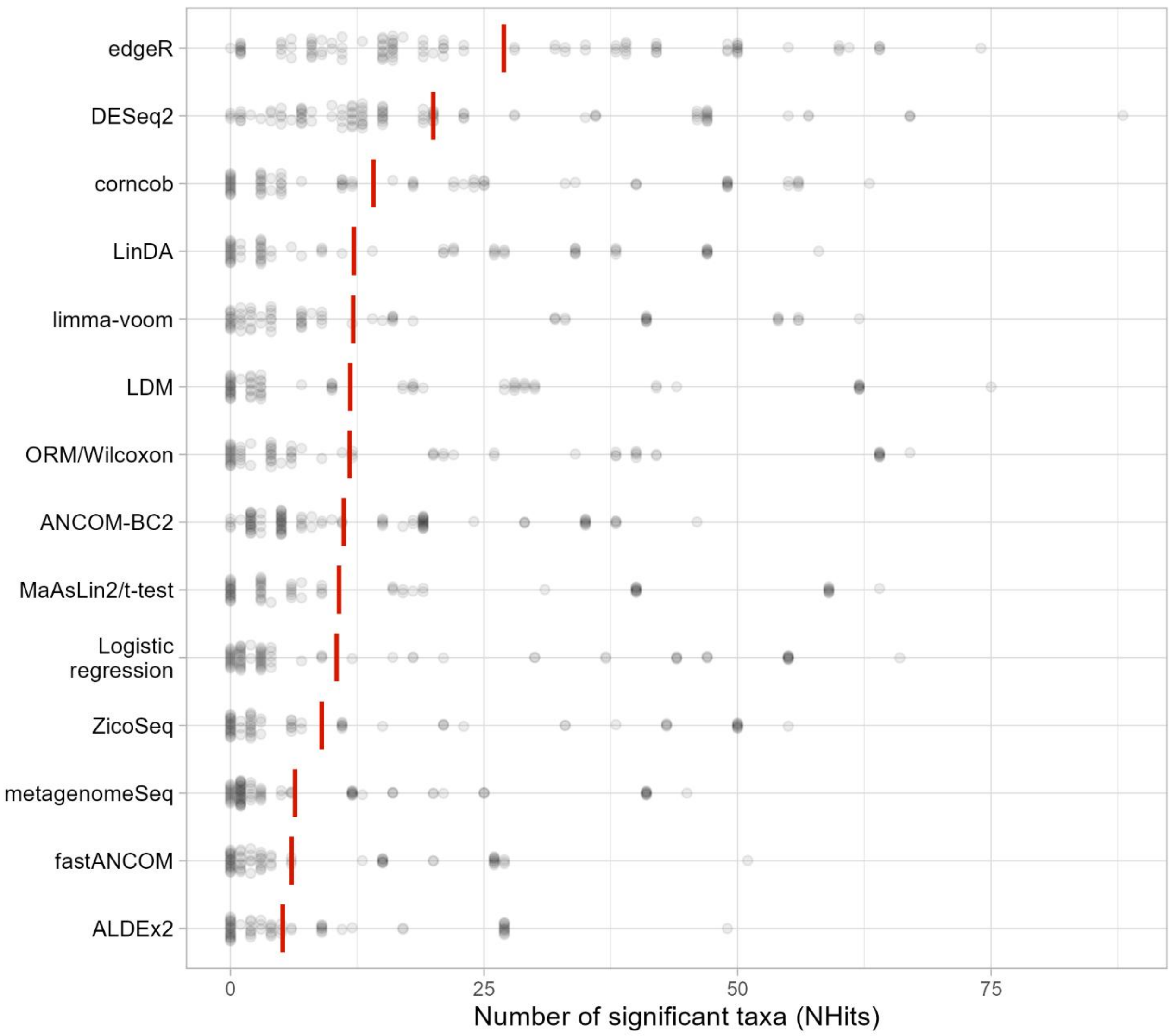

**Figure A10.3** The number of significant taxa (FDR adjusted p < .05) in the 37 exploratory datasets in the separate study analyses. The red lines indicate the mean number of significant taxa detected in the exploratory datasets. The methods are ordered according to these mean numbers.

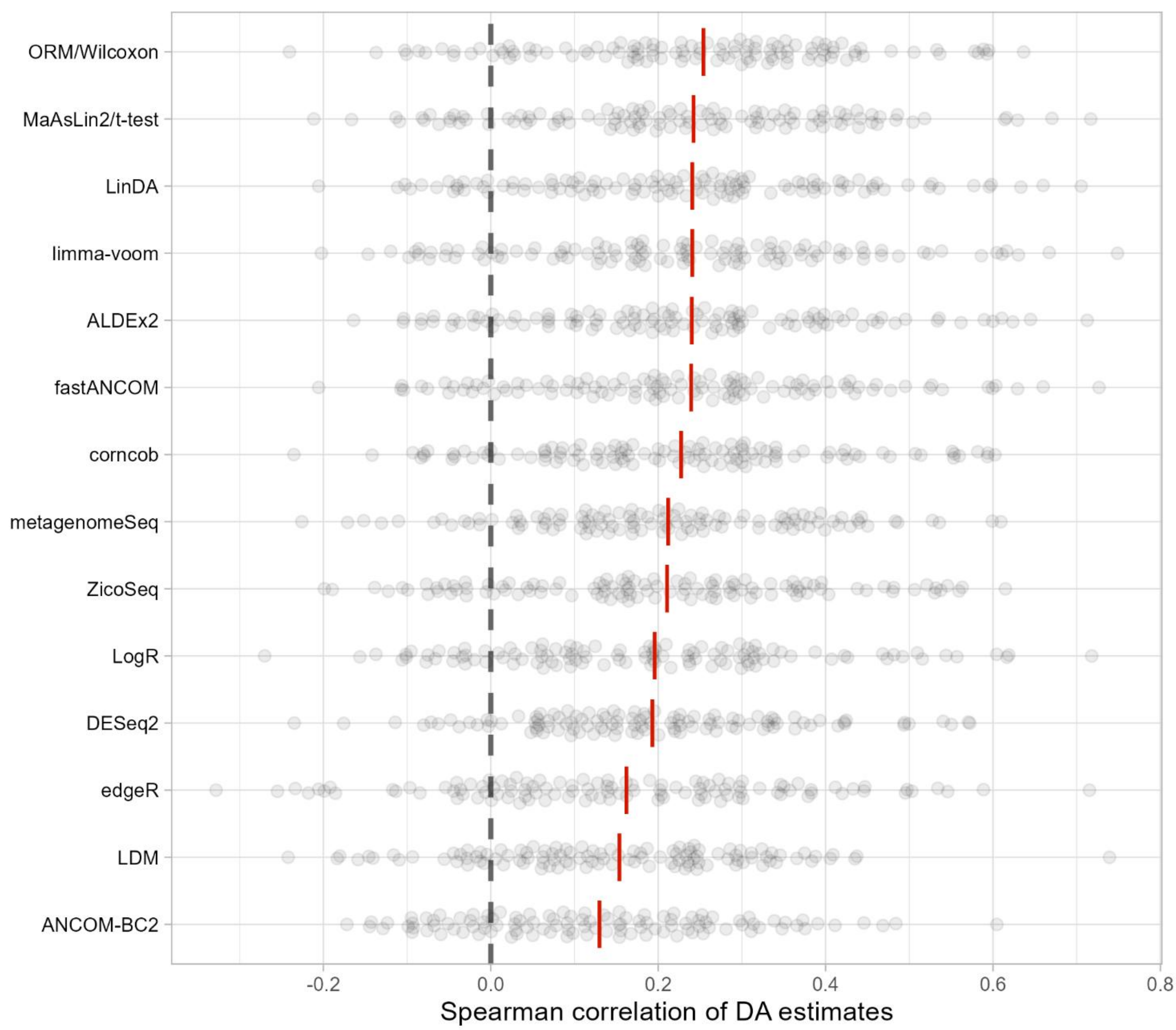

**Figure A11** The Spearman correlation coefficients of the DA estimates between the exploratory and validation datasets in the separate study analyses. The values in all exploratory-validation pairs of datasets are shown. The red lines indicate the average correlations, i.e., the hyperbolic tangent transformed mean of the inverse hyperbolic tangent transformed correlation coefficients. The methods are ordered according to the average correlation.

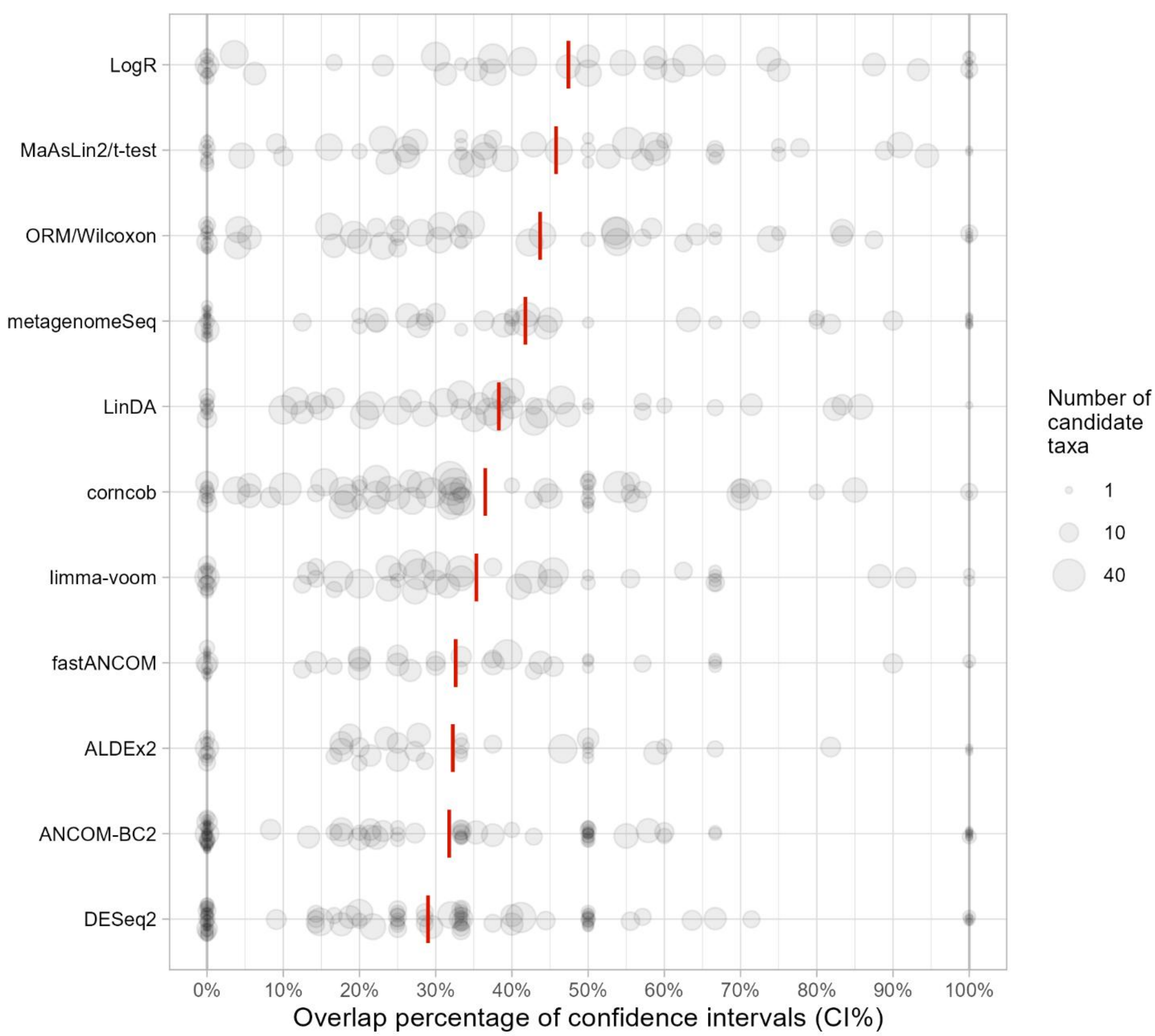

**Figure A12** The overlap percentage of 83.4% confidence intervals in all pairs of datasets (with at least one candidate taxon) in the separate study analyses. Each grey point indicates the overlap percentage in a pair of datasets. The red lines indicate the overall overlap percentages (CI%) calculated over all candidate taxa in all pairs of datasets. The methods are ordered according to the overall value.

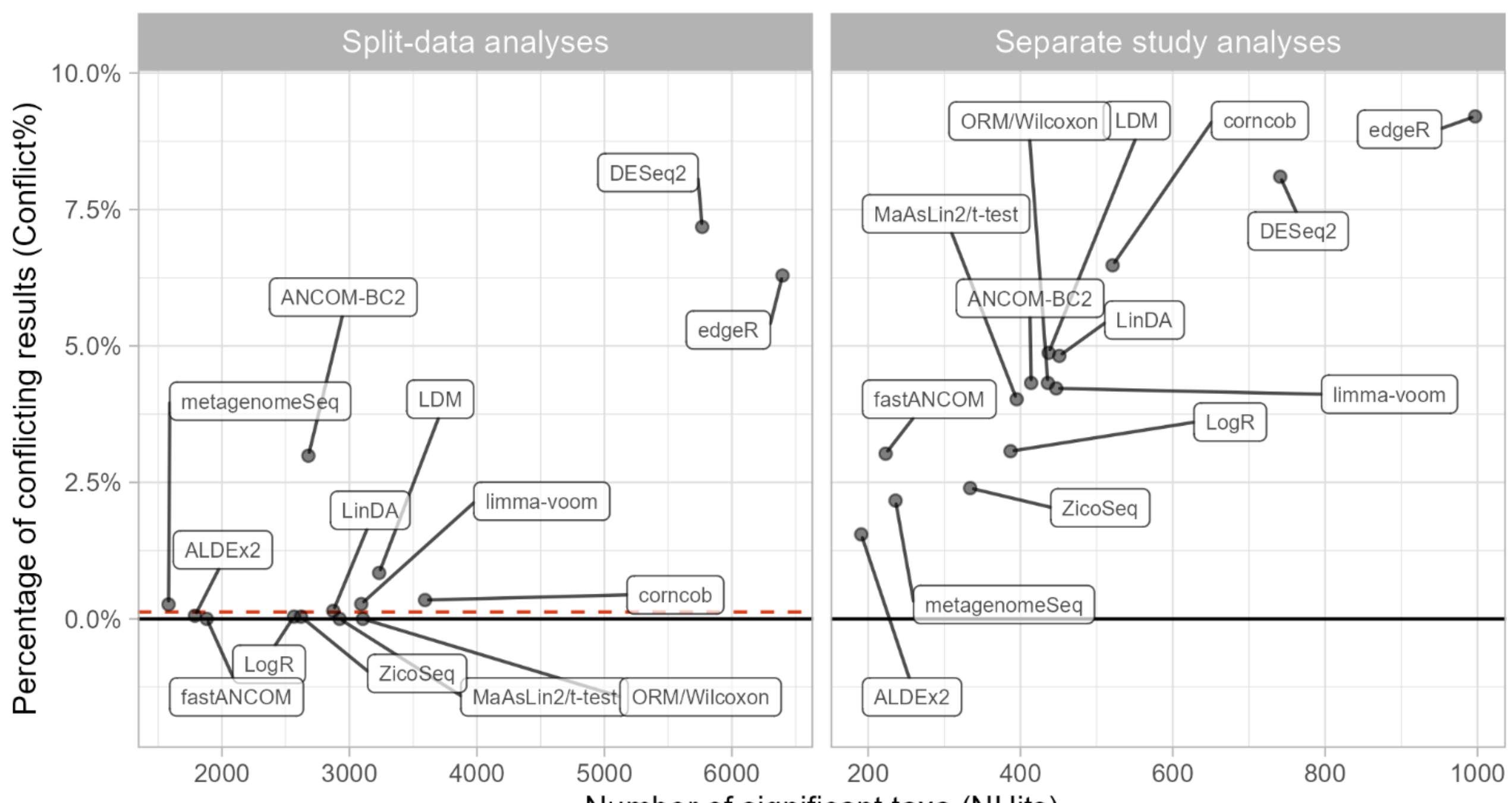

**Figure A13** This figure illustrates the correlation between the sensitivity and inconsistency of DAA methods. The sensitivity is measured by the total number of significant taxa found in all exploratory datasets (NHits). The inconsistency is measured by the percentage of conflicting results (Conflict%). Nominal FDR level α = .05 is used in both subfigures. The red dashed line on the left indicates the level of acceptable Conflict% (.125%).

**Results for additional methods and alternative versions of the 14 DAA methods**

The figures A14.1 – A14.4 below correspond Figures 2 - 5 in the main text. For details on the shown methods see the subsection "Details on running the DAA methods" above. The additional methods or versions are

- ALDEx2 with scale uncertainty (ALDEx2-scale) [51]
- ALDEx2 based on Wilcoxon test (ALDEx2-Wilcox)
- corncob with unequal variances allowed (corncob-UEV)
- DESeq2 with p values based on likelihood ratio test (DESeq2-LRT)
- LDM with CLR normalization (LDM-CLR)
- MaAsLin2 with arcsine square root transformation (MaAsLin2-AST)
- MaAsLin2 with CLR normalization (MaAsLin2-CLR)
- MaAsLin2 with CSS normalization (MaAsLin2-CSS)
- MaAsLin2 with TMM normalization (MaAsLin2-TMM)
- Negative binomial regression with TSS normalization (NegBin)
- ORM/Wilcoxon with GMPR normalization (ORM/W-GMPR)
- ORM/Wilcoxon with Wrench normalization (ORM/W-Wrench)
- radEmu [55]

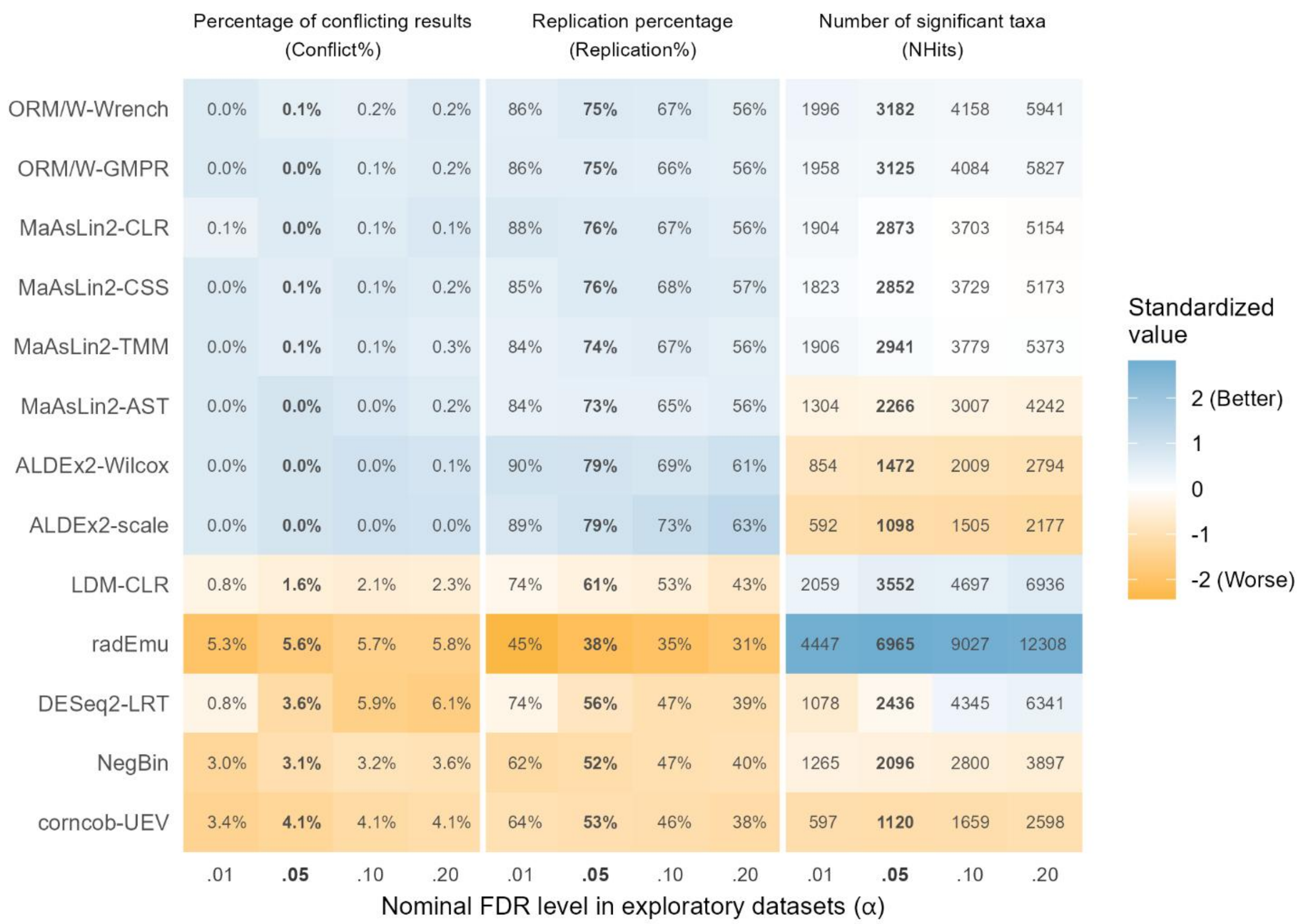

**Figure A14.1** The results for alternative versions of the methods in the split-data analyses. The figure corresponds to Figure 2 in the main text.

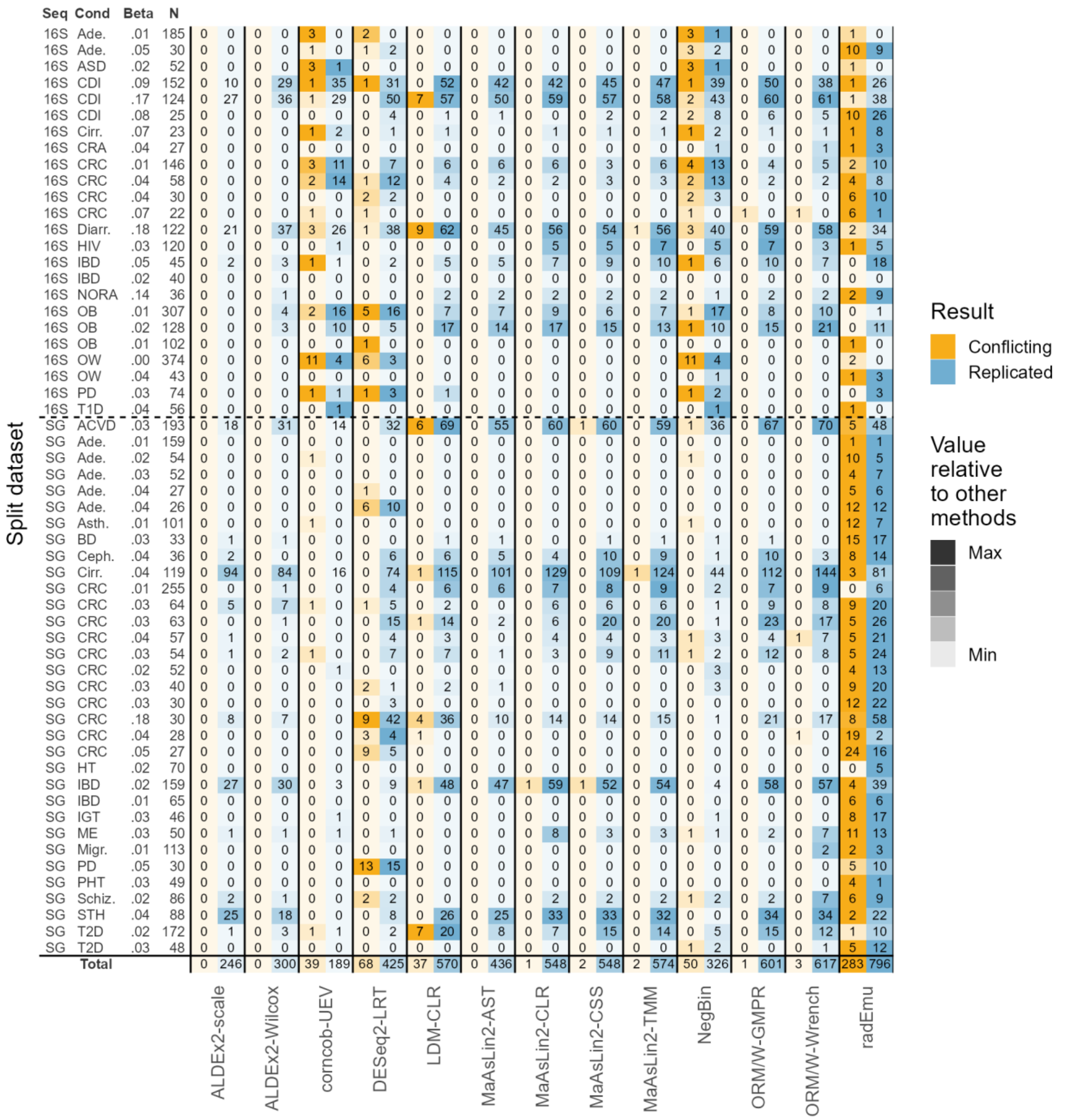

**Figure A14.2** The number of conflicting and replicated results in the split-data analyses for alternative versions of the methods. This figure corresponds to Figure 3 in the main text.

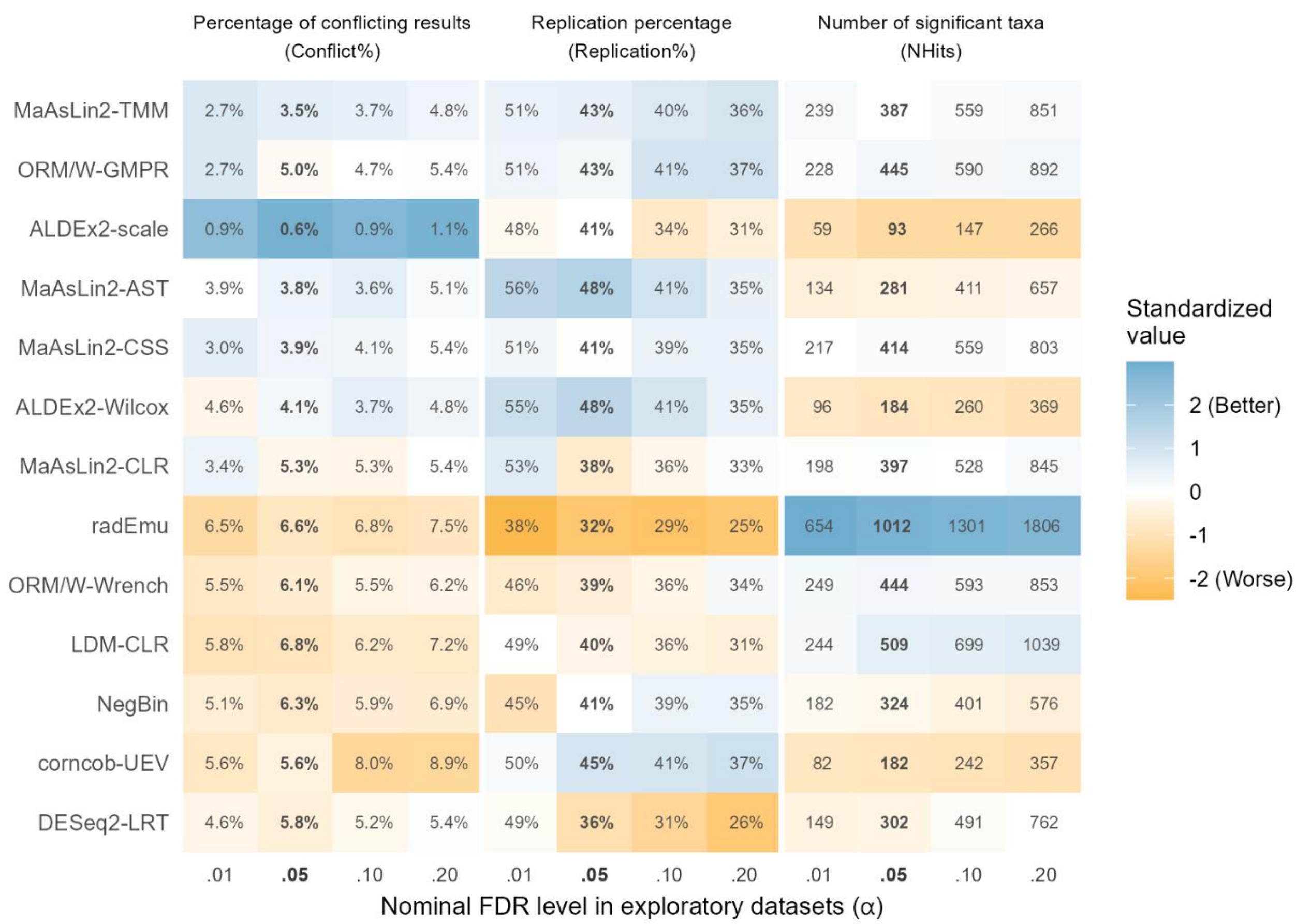

**Figure A14.3** The results for alternative versions of the methods in the separate study analyses. The figure corresponds to Figure 4 in the main text.

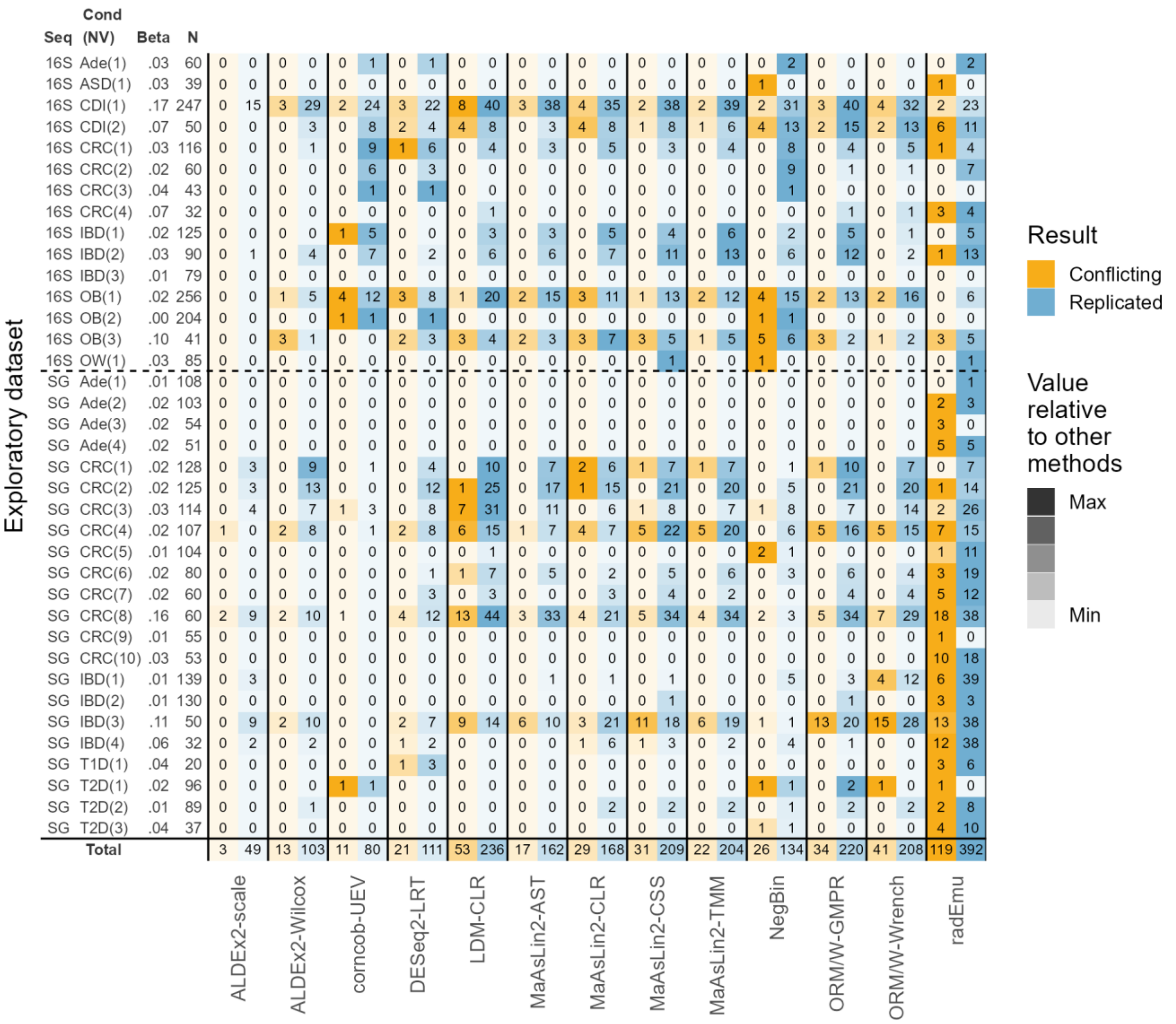

**Figure A14.4** The number of conflicting and replicated results in the separate study analyses for alternative versions of the methods. The figure corresponds to Figure 5 in the main text.

**Additional analysis to evaluate how different normalization strategies perform under large systematic differences in the total absolute abundances**

We here describe the additional analysis of how different methods can address the case of large systematic differences in total absolute abundances in practice. We used a real dataset from a gut microbiome study on 12 (6 + 6) mice measured at four time points [58]. In this study the absolute microbial abundances were measured to be clearly systematically higher in one group. We performed DAA with each method on this dataset in the standard way using only the observed counts. We then compared the direction of DAA estimates provided by the methods to the "true" directions. For each taxon, the "true" direction was defined as the sign of the difference of the arithmetic means of the measured absolute abundances. If the mean absolute abundance was greater in the case group, the sign was positive (otherwise it was negative). For each method, we then calculated the accuracy of estimating the sign correctly (accuracy = correct signs / number of taxa).

The results are shown in Figure A15. The methods employing TSS normalization perform generally well. Furthermore, methods employing CLR-transformation *in some phase* of DAA perform generally below average.

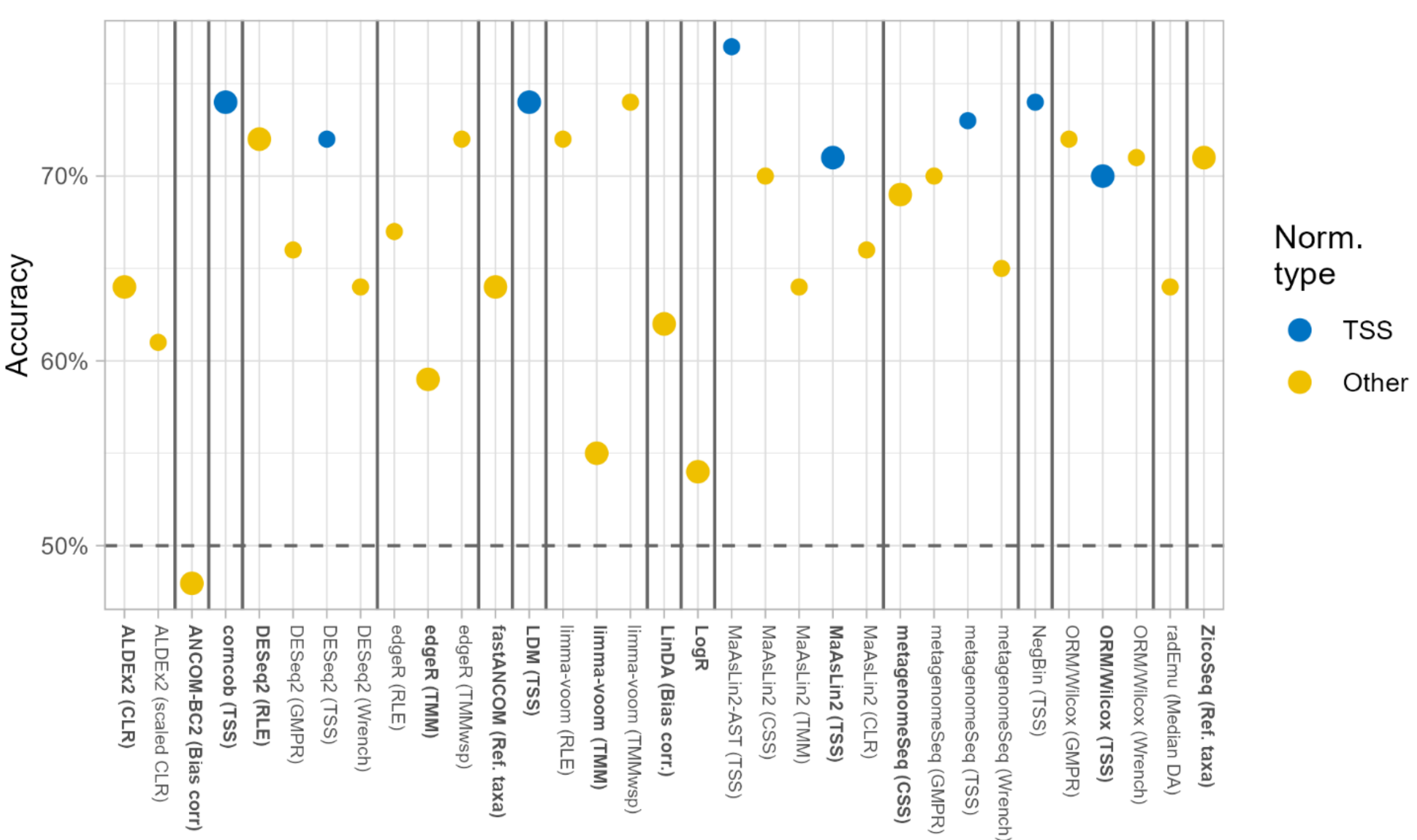

**Figure A15** The figure illustrates the ability of DAA methods to estimate the direction of "true" DA based on measured absolute abundances on a real dataset from a microbiome study on mice. The bolded names and larger points indicate the versions of the methods that are included in the main text. The normalization method/strategy is given in parenthesis. Norm. type indicates whether TSS normalization (or its equivalent) was used or whether some other type of normalization strategy was employed.

**Additional analyses to evaluate how different normalization strategies affect the performance of DAA methods**

Here we simply evaluated the methods based on the statistical significance of the results (q < .05) they provided on 50 datasets used in the separate study analyses. Jaccard distances based on significance of the results were calculated and Multidimensional Scaling (MDS) analysis based on those distances was performed. The results for the two most important MDS coordinates are shown below (Figure A16a). We only show results for the more appropriately performing methods (e.g. DESeq and edgeR would show rather far away from other methods). In Figure A16b we show the results only for datasets from studies investigating CDI (Clostridium difficile infection) as we consider it likely that there may occur systematic differences in the total absolute abundances.

The results show that, generally, other factors than normalization method (apart from CLR normalization) affect mostly the findings made by a method, especially for MaAsLin2 and ORM/Wilcoxon.

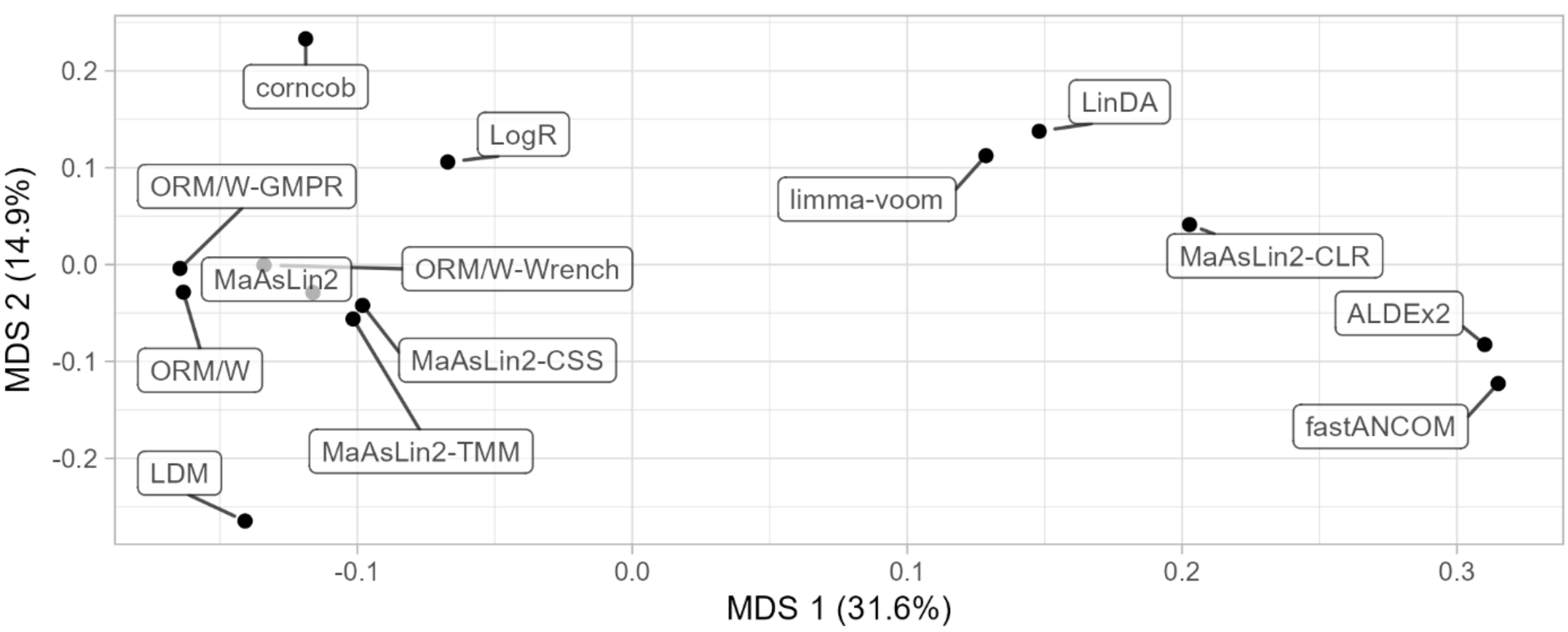

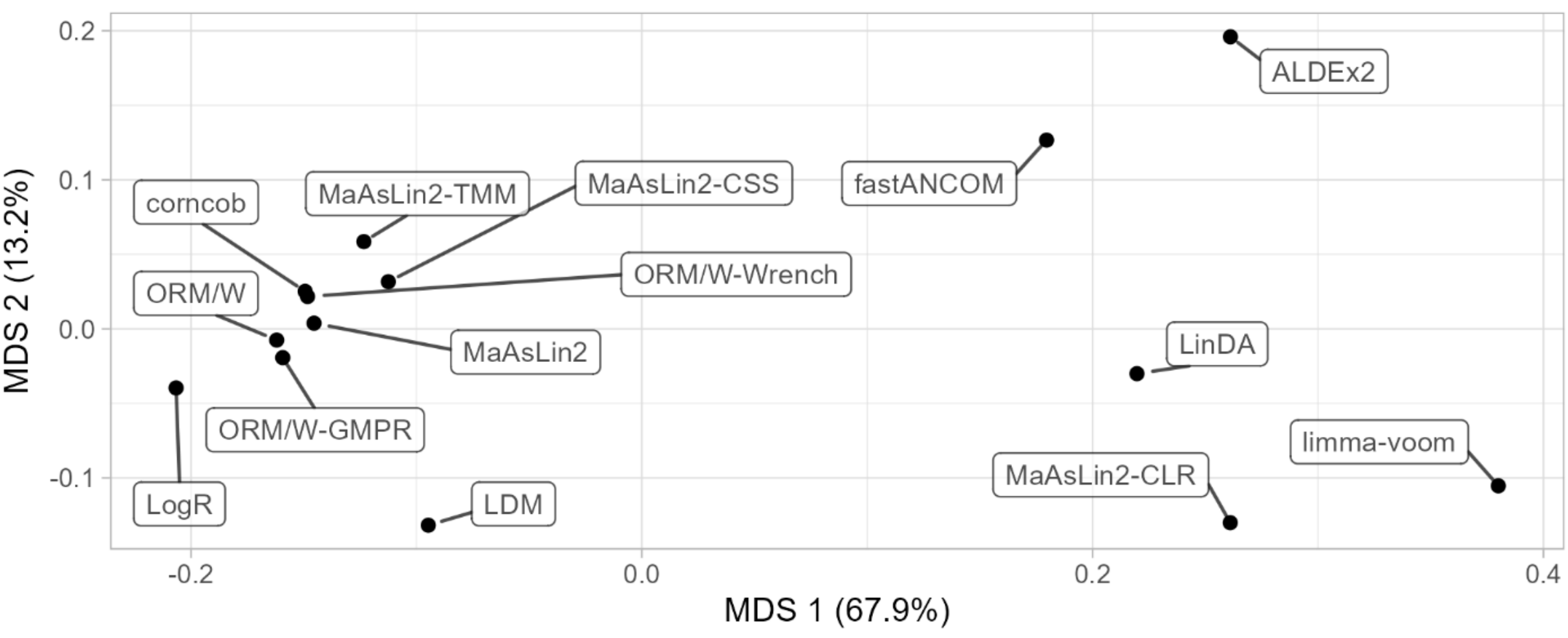

**Figure A16** DAA methods clustered based on Jaccard distances which are calculated based on statistical significance of the results provided by the methods on 50 datasets (a) and on three 3 16S datasets from studies investigating CDI (Clostridioides difficile infection) (b). The variance explained by the principal coordinate is shown in the parentheses. ORM/W = ORM/Wilcoxon. If some non-default normalization method was employed, it is shown after the name of the method.

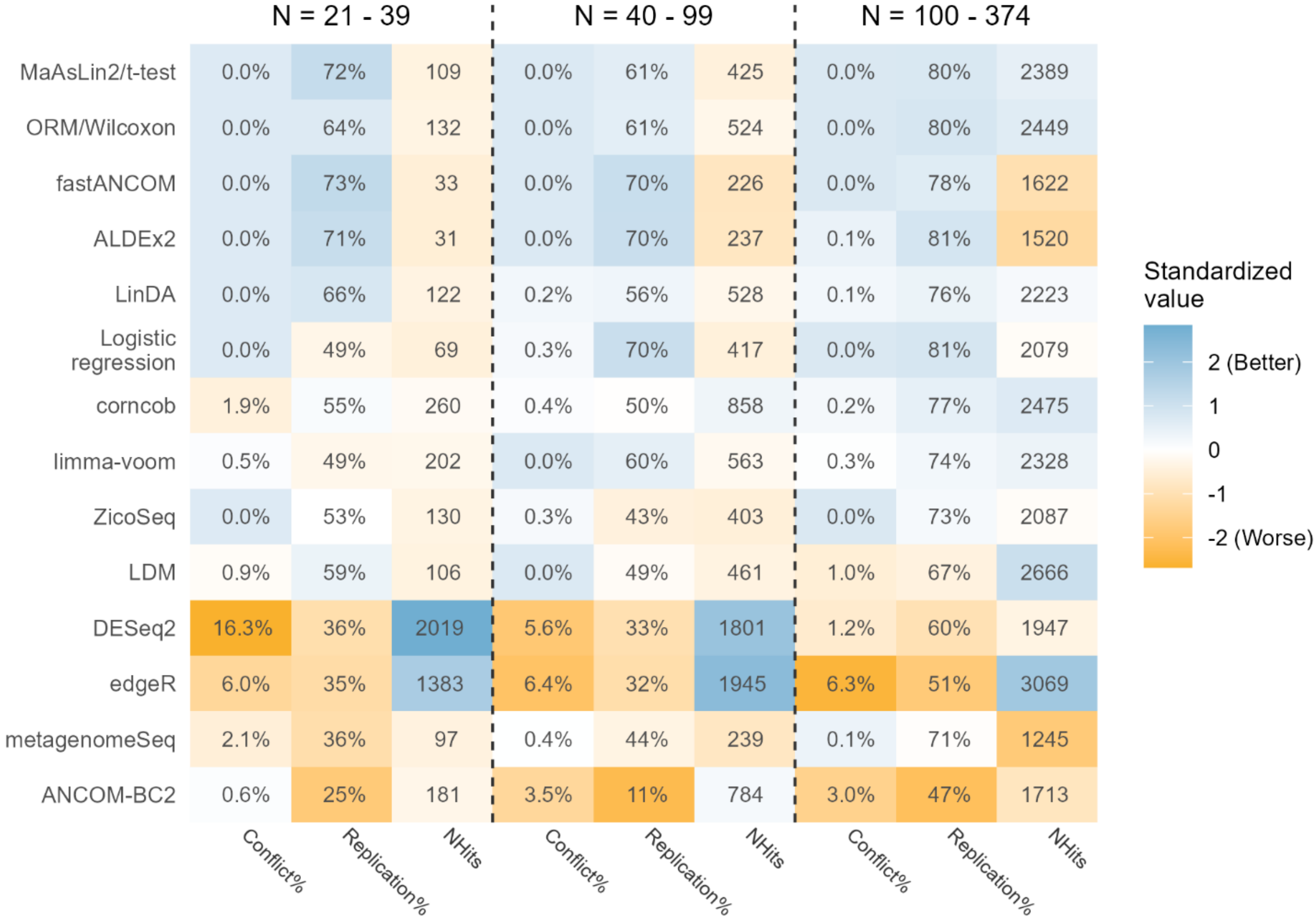

**Figure A17.1** The main results of the split-data analyses, stratified by sample size (N). The sample size refers to the number of subjects in a single exploratory or validation dataset. There were 5 x 18, 5 x 22 and 5 x 17 exploratory and validation datasets with sample sizes between 21 and 39, between 40 and 99, and between 100 and 374, respectively. FDR level α = .05 was employed in the exploratory datasets.

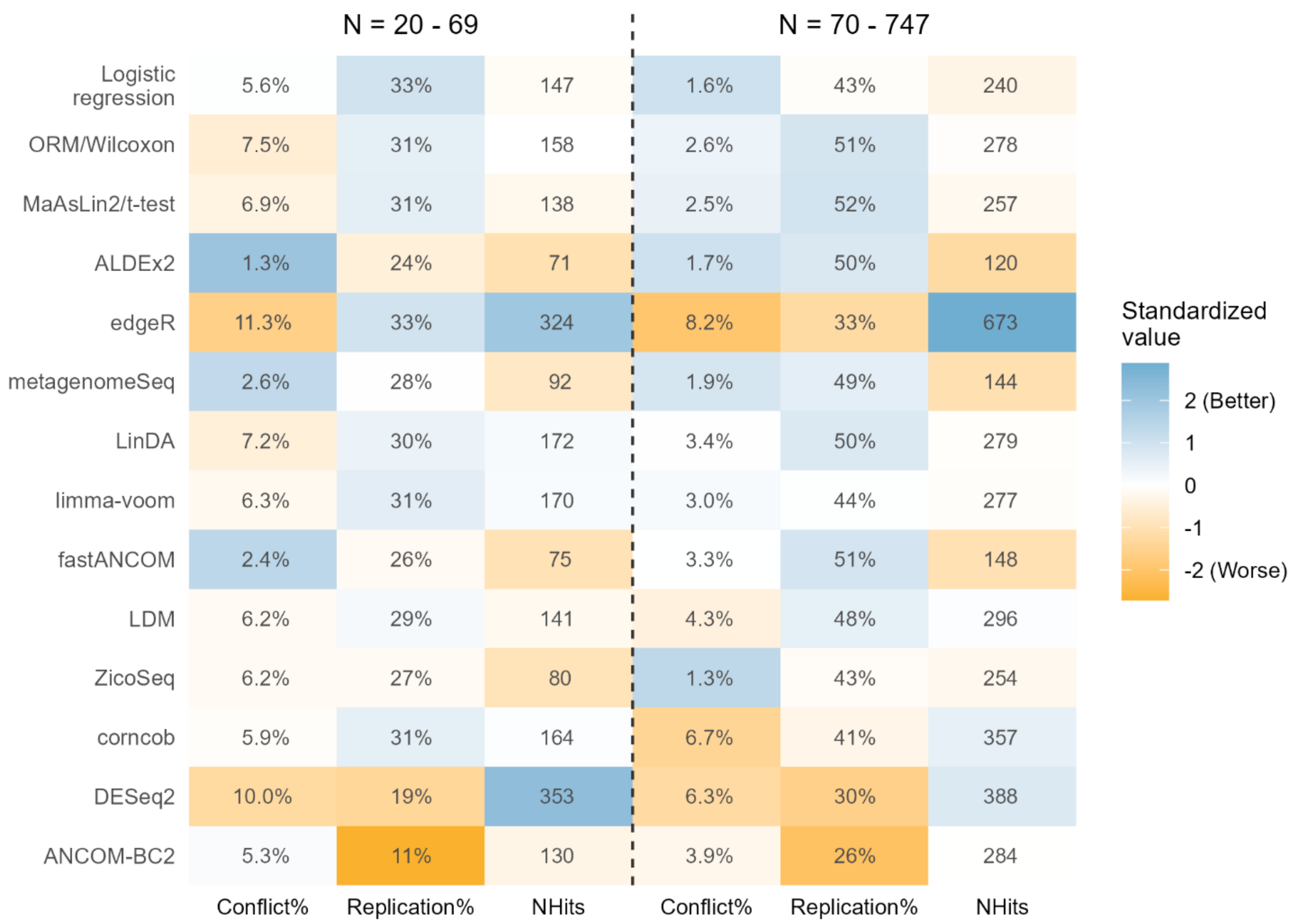

**Figure A17.2** The main results of the separate study analyses, stratified by sample size (N). The sample size refers to the number of subjects in the *exploratory* datasets. The larger datasets (N > 70) were used as the validation datasets also for the smaller datasets (20 <= N < 70). There were 17 and 20 exploratory datasets with N = 20 – 69 and N = 70 - 747, respectively. FDR level α = .05 was employed in the exploratory datasets.

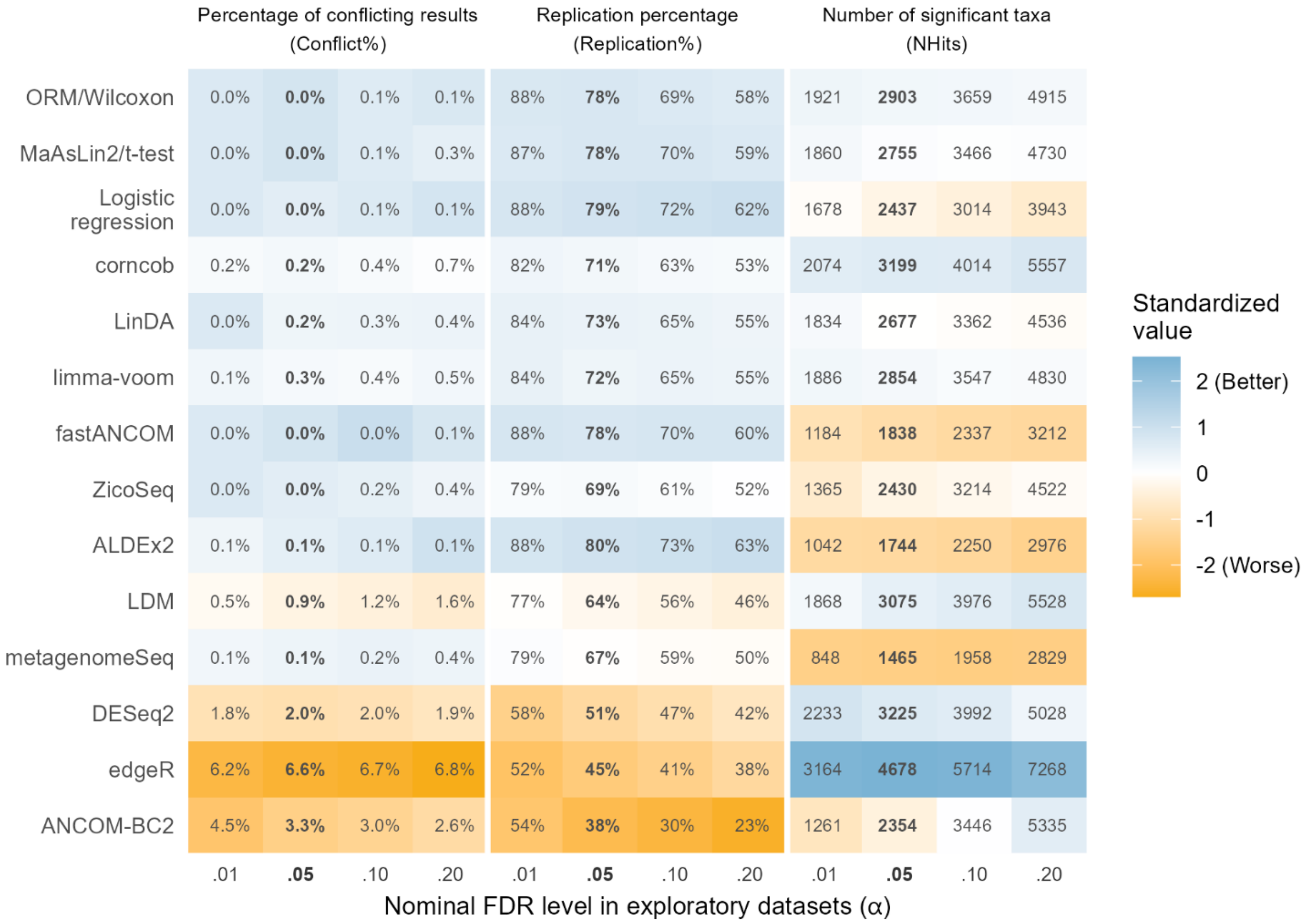

**Figure A18.1** The main results of the split-data analyses when exploratory and validation datasets with sample size N < 50 were filtered out (N referring to the number of subjects in a single exploratory or validation dataset). This corresponds to Figure 2 in the main text.

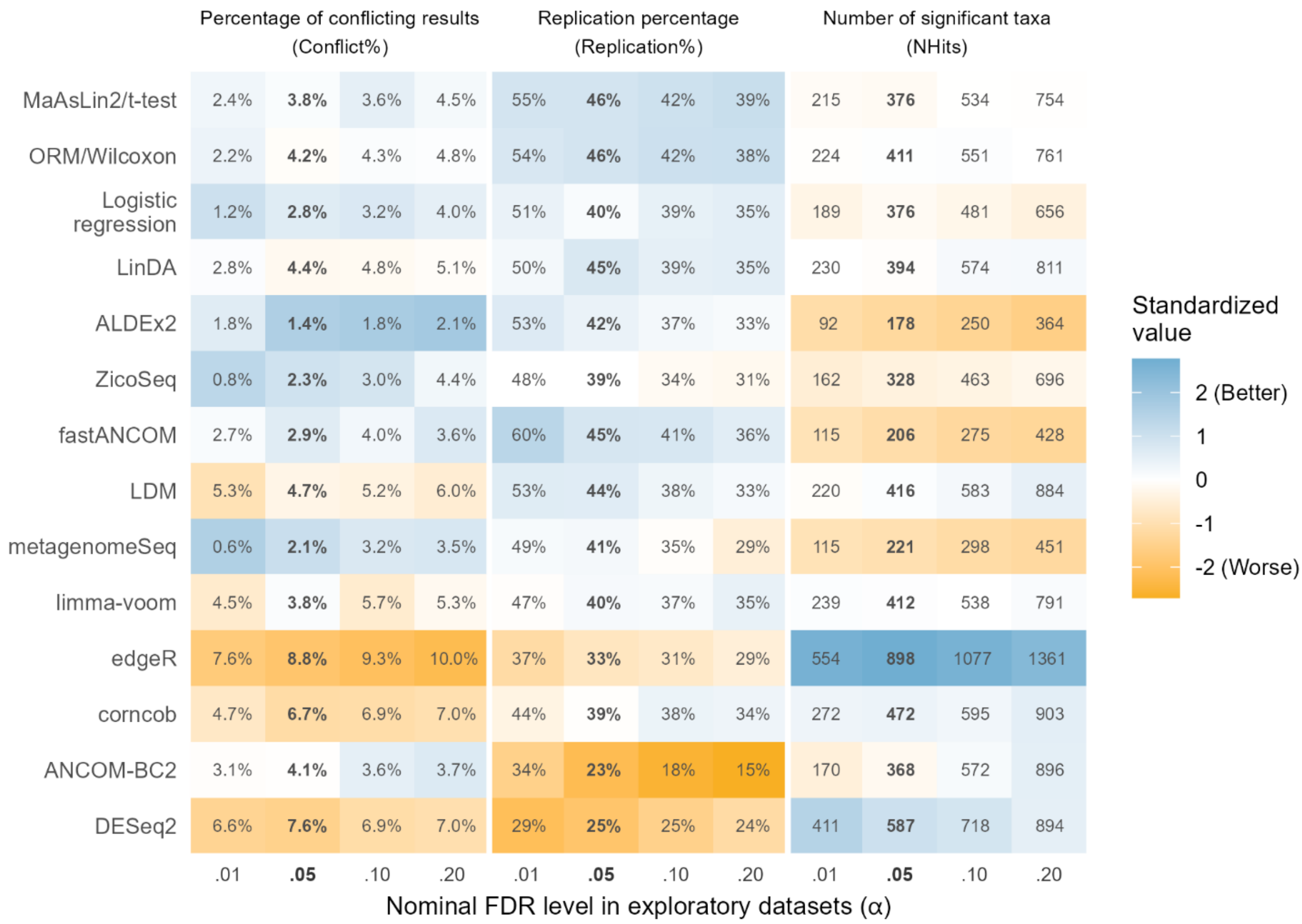

**Figure A18.2** The main results of the separate study analyses when exploratory and validation datasets with sample size N < 50 were filtered out. This corresponds to Figure 4 in the main text.

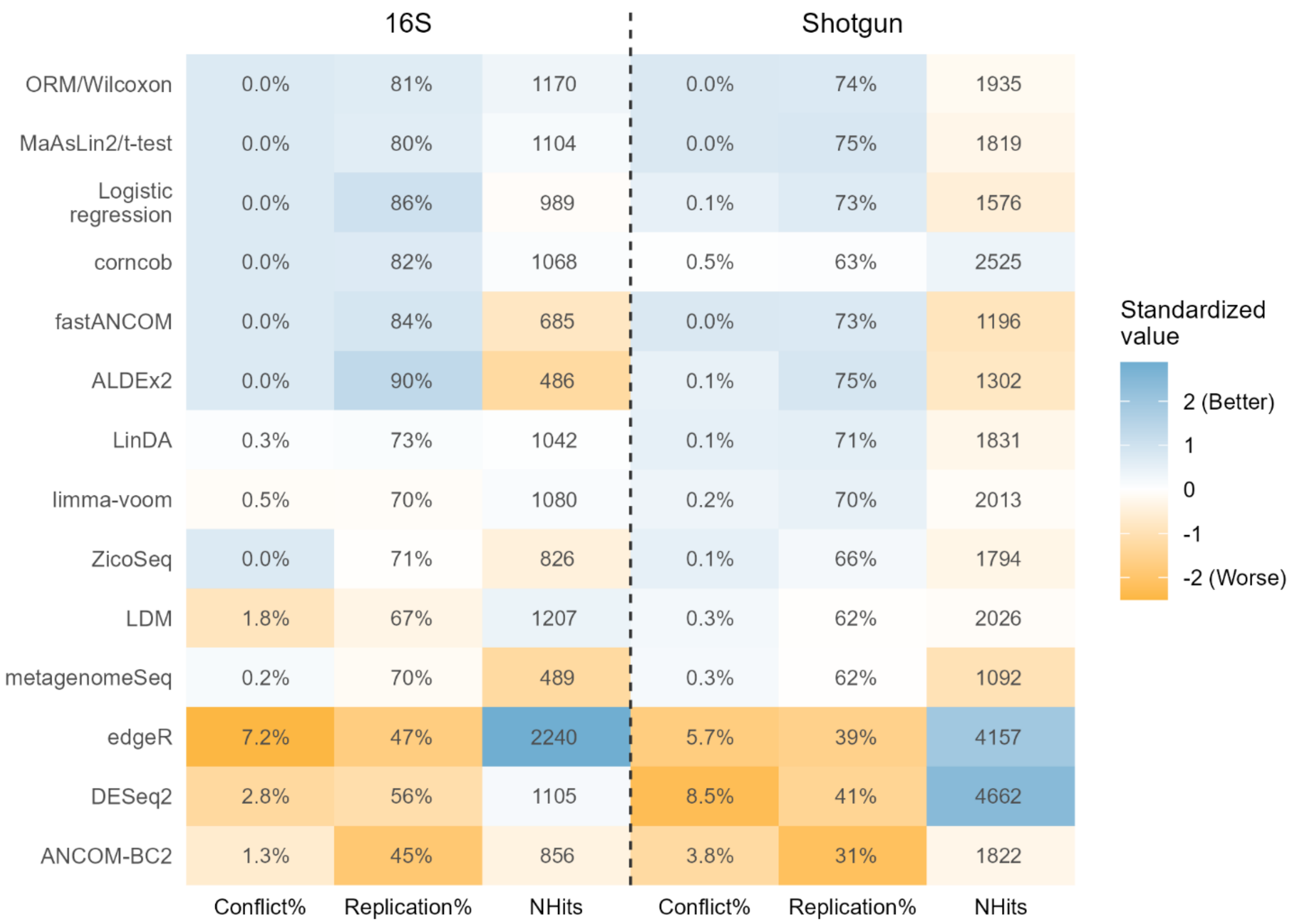

**Figure A19.1** The main results of the split-data analyses, stratified by sequencing type (16S or Shotgun). FDR level α = .05 was employed in the exploratory datasets.

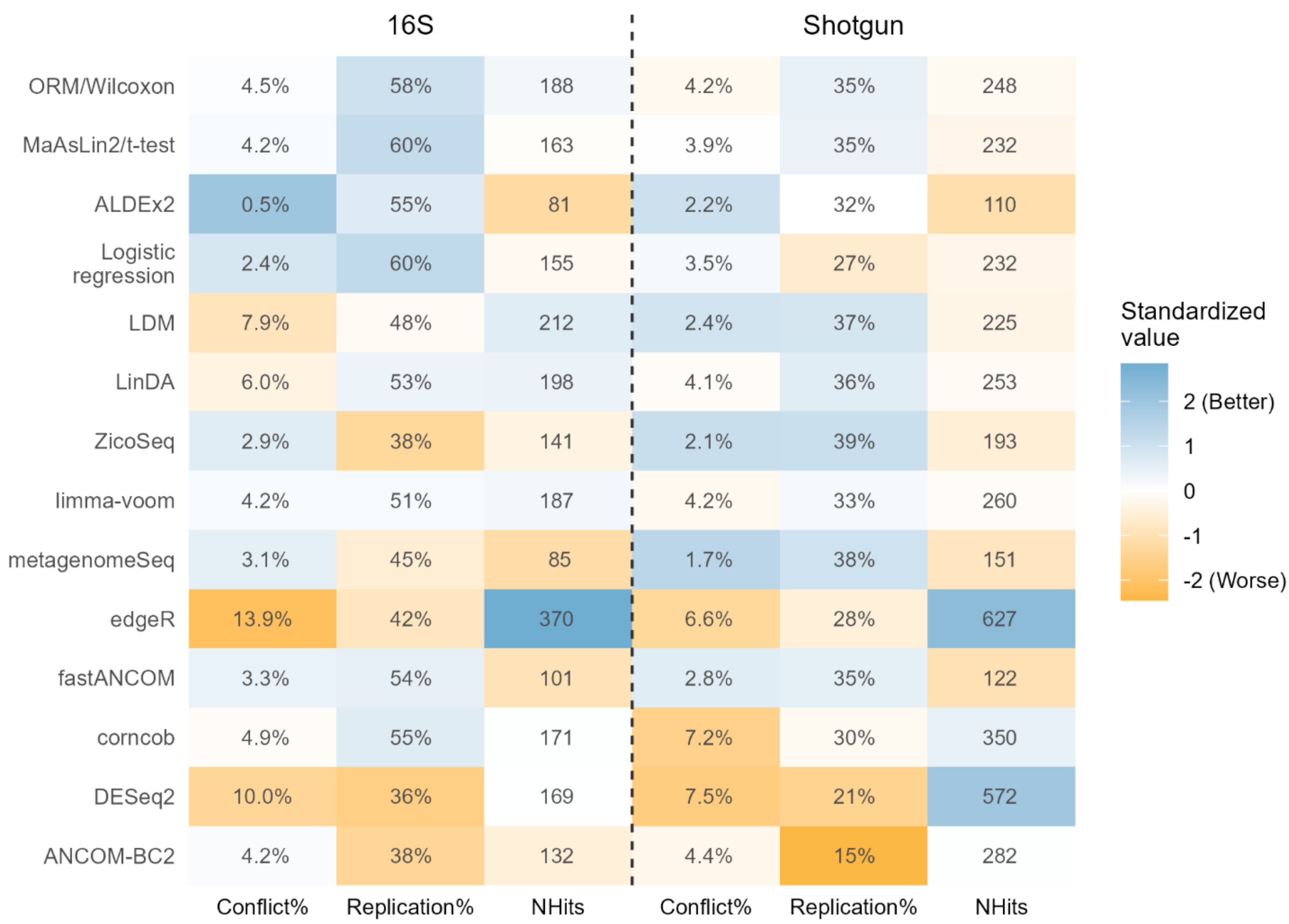

**Figure A19.2** The main results of the separate study analyses, stratified by sequencing type (16S or Shotgun). FDR level α = .05 was employed in the exploratory datasets.

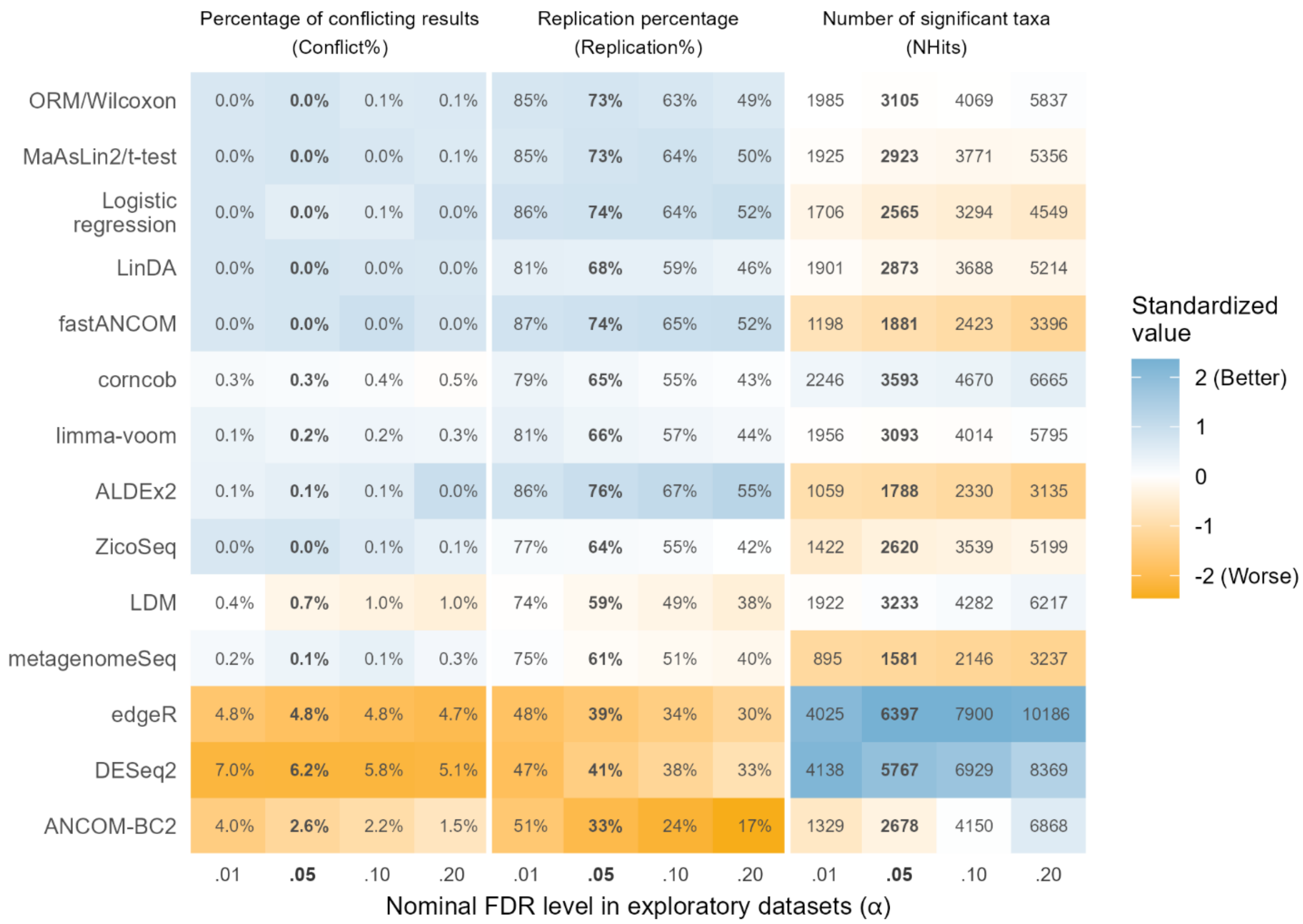

**Figure A20.1** The main results of the split-data analyses when p-values that were FDR adjusted over *candidate* taxa in the validation datasets were employed (instead of unadjusted p-values). That is, for each candidate taxon, the FDR adjusted p-value in the validation dataset was calculated by employing Benjamini-Hochberg method over the p-values of the candidate taxa. The taxon was then considered as statistically significant in the validation dataset if this FDR adjusted p-value was below .05.

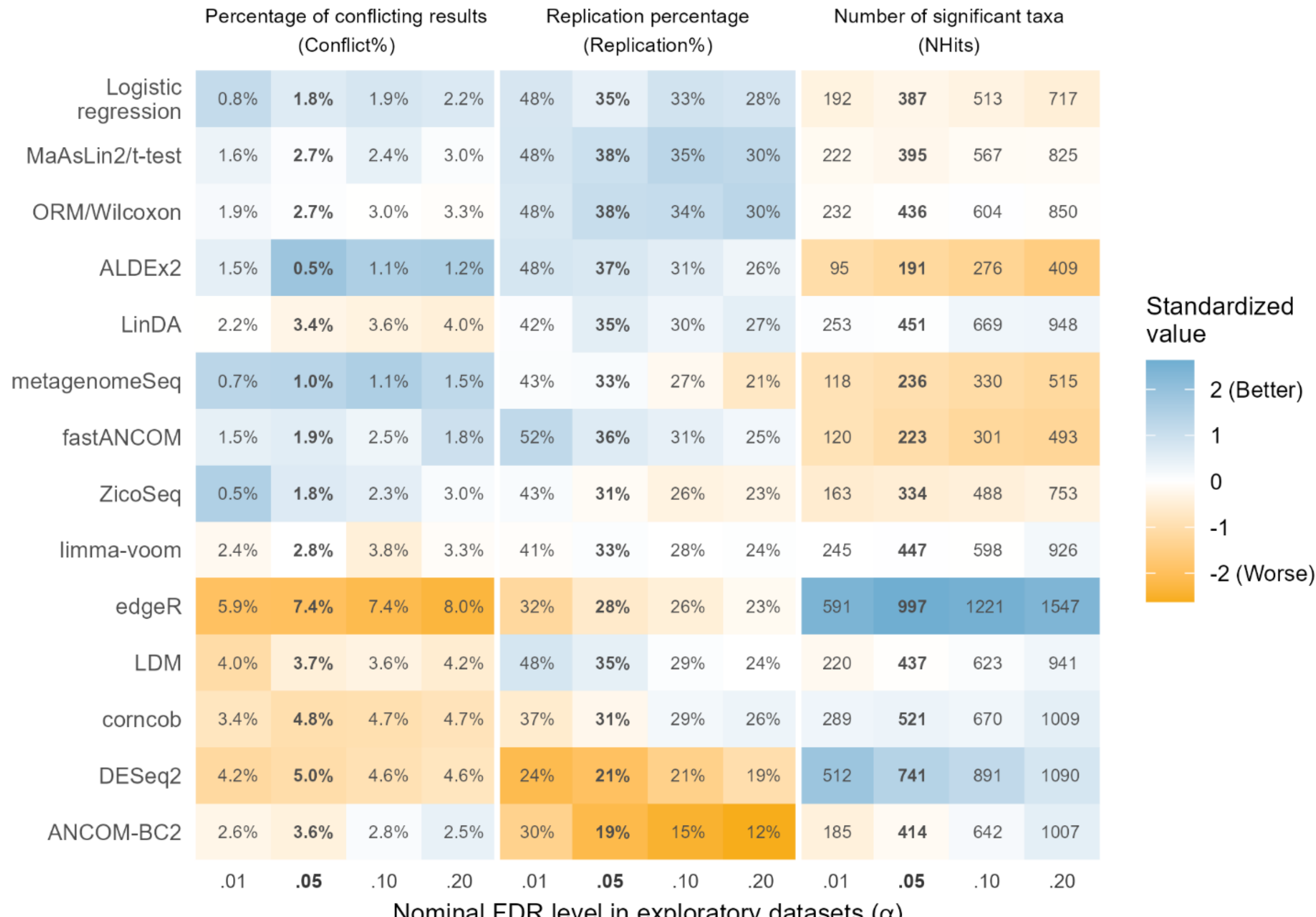

**Figure A20.2** The main results of the separate study analyses when p-values that were FDR adjusted over *candidate* taxa in the validation datasets were employed (instead of unadjusted p-values). That is, for each candidate taxon, the FDR adjusted p-value in the validation dataset was calculated by employing Benjamini-Hochberg method over the p-values of the candidate taxa. The taxon was then considered as statistically significant in the validation dataset if this FDR adjusted p-value was below .05.

**References in the Appendix**

[1] A. K. Alkanani *et al.*, ‘Alterations in Intestinal Microbiota Correlate With Susceptibility to Type 1 Diabetes’, *Diabetes*, vol. 64, no. 10, pp. 3510–3520, Oct. 2015, doi: 10.2337/db14-1847.

[2] N. T. Baxter, M. T. Ruffin, M. A. M. Rogers, and P. D. Schloss, ‘Microbiota-based model improves the sensitivity of fecal immunochemical test for detecting colonic lesions’, *Genome Med*, vol. 8, no. 1, p. 37, Dec. 2016, doi: 10.1186/s13073-016-0290-3.

[3] W. Chen, F. Liu, Z. Ling, X. Tong, and C. Xiang, ‘Human Intestinal Lumen and Mucosa-Associated Microbiota in Patients with Colorectal Cancer’, *PLoS One*, vol. 7, no. 6, p. e39743, Jun. 2012, doi: 10.1371/journal.pone.0039743.

[4] D. Gevers *et al.*, ‘The Treatment-Naive Microbiome in New-Onset Crohn’s Disease’, *Cell Host Microbe*, vol. 15, no. 3, pp. 382–392, Mar. 2014, doi: 10.1016/j.chom.2014.02.005.

[5] J. K. Goodrich *et al.*, ‘Human Genetics Shape the Gut Microbiome’, *Cell*, vol. 159, no. 4, pp. 789–799, Nov. 2014, doi: 10.1016/j.cell.2014.09.053.

[6] D.-W. Kang *et al.*, ‘Reduced Incidence of Prevotella and Other Fermenters in Intestinal Microflora of Autistic Children’, *PLoS One*, vol. 8, no. 7, p. e68322, Jul. 2013, doi: 10.1371/journal.pone.0068322.

[7] X. C. Morgan *et al.*, ‘Dysfunction of the intestinal microbiome in inflammatory bowel disease and treatment’, *Genome Biol*, vol. 13, no. 9, p. R79, 2012, doi: 10.1186/gb-2012-13-9-r79.

[8] M. Noguera-Julian *et al.*, ‘Gut Microbiota Linked to Sexual Preference and HIV Infection’, *EBioMedicine*, vol. 5, pp. 135–146, Mar. 2016, doi: 10.1016/j.ebiom.2016.01.032.

[9] E. Papa *et al.*, ‘Non-Invasive Mapping of the Gastrointestinal Microbiota Identifies Children with Inflammatory Bowel Disease’, *PLoS One*, vol. 7, no. 6, p. e39242, Jun. 2012, doi: 10.1371/journal.pone.0039242.

[10] F. Scheperjans *et al.*, ‘Gut microbiota are related to Parkinson’s disease and clinical phenotype’, *Movement Disorders*, vol. 30, no. 3, pp. 350–358, Mar. 2015, doi: 10.1002/mds.26069.

[11] J. U. Scher *et al.*, ‘Expansion of intestinal Prevotella copri correlates with enhanced susceptibility to arthritis’, *Elife*, vol. 2, Nov. 2013, doi: 10.7554/eLife.01202.

[12] A. M. Schubert *et al.*, ‘Microbiome Data Distinguish Patients with Clostridium difficile Infection and Non-C. difficile-Associated Diarrhea from Healthy Controls’, *mBio*, vol. 5, no. 3, Jul. 2014, doi: 10.1128/mBio.01021-14.

[13] P. Singh *et al.*, ‘Intestinal microbial communities associated with acute enteric infections and disease recovery’, *Microbiome*, vol. 3, no. 1, p. 45, Dec. 2015, doi: 10.1186/s40168-015-0109-2.

[14] J. S. Son *et al.*, ‘Comparison of Fecal Microbiota in Children with Autism Spectrum Disorders and Neurotypical Siblings in the Simons Simplex Collection’, *PLoS One*, vol. 10, no. 10, p. e0137725, Oct. 2015, doi: 10.1371/journal.pone.0137725.

[15] P. J. Turnbaugh *et al.*, ‘A core gut microbiome in obese and lean twins’, *Nature*, vol. 457, no. 7228, pp. 480–484, Jan. 2009, doi: 10.1038/nature07540.

[16] C. Vincent *et al.*, ‘Reductions in intestinal Clostridiales precede the development of nosocomial Clostridium difficile infection’, *Microbiome*, vol. 1, no. 1, p. 18, Dec. 2013, doi: 10.1186/2049-2618-1-18.

[17] T. Wang *et al.*, 'Structural segregation of gut microbiota between colorectal cancer patients and healthy volunteers', *ISME J*, vol. 6, no. 2, pp. 320–329, Feb. 2012, doi: 10.1038/ismej.2011.109.

[18] B. P. Willing *et al.*, 'A Pyrosequencing Study in Twins Shows That Gastrointestinal Microbial Profiles Vary With Inflammatory Bowel Disease Phenotypes', *Gastroenterology*, vol. 139, no. 6, pp. 1844-1854.e1, Dec. 2010, doi: 10.1053/j.gastro.2010.08.049.

[19] J. P. Zackular, M. A. M. Rogers, M. T. Ruffin, and P. D. Schloss, 'The Human Gut Microbiome as a Screening Tool for Colorectal Cancer', *Cancer Prevention Research*, vol. 7, no. 11, pp. 1112–1121, Nov. 2014, doi: 10.1158/1940-6207.CAPR-14-0129.

[20] G. Zeller *et al.*, 'Potential of fecal microbiota for early-stage detection of colorectal cancer.', *Mol Syst Biol*, vol. 10, no. 11, p. 766, Nov. 2014, doi: 10.15252/msb.20145645.

[21] Z. Zhang *et al.*, 'Large-Scale Survey of Gut Microbiota Associated With MHE Via 16S rRNA-Based Pyrosequencing', *American Journal of Gastroenterology*, vol. 108, no. 10, pp. 1601–1611, Oct. 2013, doi: 10.1038/ajg.2013.221.

[22] L. Zhu *et al.*, 'Characterization of gut microbiomes in nonalcoholic steatohepatitis (NASH) patients: A connection between endogenous alcohol and NASH', *Hepatology*, vol. 57, no. 2, pp. 601–609, Feb. 2013, doi: 10.1002/hep.26093.

[23] M. L. Zupancic *et al.*, 'Analysis of the Gut Microbiota in the Old Order Amish and Its Relation to the Metabolic Syndrome', *PLoS One*, vol. 7, no. 8, p. e43052, Aug. 2012, doi: 10.1371/journal.pone.0043052.

[24] J. R. Bedarf *et al.*, 'Functional implications of microbial and viral gut metagenome changes in early stage L-DOPA-naïve Parkinson's disease patients.', *Genome Med*, vol. 9, no. 1, p. 39, Apr. 2017, doi: 10.1186/s13073-017-0428-y.

[25] Q. Feng *et al.*, 'Gut microbiome development along the colorectal adenoma-carcinoma sequence.', *Nat Commun*, vol. 6, p. 6528, Mar. 2015, doi: 10.1038/ncomms7528.

[26] A. Gupta *et al.*, 'Association of Flavonifractor plautii, a Flavonoid-Degrading Bacterium, with the Gut Microbiome of Colorectal Cancer Patients in India', *mSystems*, vol. 4, no. 6, Dec. 2019, doi: 10.1128/MSYSTEMS.00438-19.

[27] A. B. Hall *et al.*, 'A novel Ruminococcus gnavus clade enriched in inflammatory bowel disease patients.', *Genome Med*, vol. 9, no. 1, p. 103, Nov. 2017, doi: 10.1186/s13073-017-0490-5.

[28] G. D. Hannigan, M. B. Duhaime, M. T. Ruffin, C. C. Koumpouras, and P. D. Schloss, 'Diagnostic Potential and Interactive Dynamics of the Colorectal Cancer Virome.', *mBio*, vol. 9, no. 6, Nov. 2018, doi: 10.1128/mBio.02248-18.

[29] A. Heintz-Buschart *et al.*, 'Integrated multi-omics of the human gut microbiome in a case study of familial type 1 diabetes.', *Nat Microbiol*, vol. 2, p. 16180, Oct. 2016, doi: 10.1038/nmicrobiol.2016.180.

[30] U. Z. Ijaz *et al.*, 'The distinct features of microbial "dysbiosis" of Crohn's disease do not occur to the same extent in their unaffected, genetically-linked kindred.', *PLoS One*, vol. 12, no. 2, p. e0172605, 2017, doi: 10.1371/journal.pone.0172605.

[31] Z. Jie *et al.*, 'The gut microbiome in atherosclerotic cardiovascular disease.', *Nat Commun*, vol. 8, no. 1, p. 845, Oct. 2017, doi: 10.1038/s41467-017-00900-1.

[32] F. H. Karlsson *et al.*, 'Gut metagenome in European women with normal, impaired and diabetic glucose control.', *Nature*, vol. 498, no. 7452, pp. 99–103, Jun. 2013, doi: 10.1038/nature12198.

[33] J. Li *et al.*, 'An integrated catalog of reference genes in the human gut microbiome.', *Nat Biotechnol*, vol. 32, no. 8, pp. 834–41, Aug. 2014, doi: 10.1038/nbt.2942.

[34] J. Li *et al.*, 'Gut microbiota dysbiosis contributes to the development of hypertension.', *Microbiome*, vol. 5, no. 1, p. 14, Feb. 2017, doi: 10.1186/s40168-016-0222-x.

[35] D. Nagy-Szakal *et al.*, 'Fecal metagenomic profiles in subgroups of patients with myalgic encephalomyelitis/chronic fatigue syndrome.', *Microbiome*, vol. 5, no. 1, p. 44, Apr. 2017, doi: 10.1186/s40168-017-0261-y.

[36] H. B. Nielsen *et al.*, 'Identification and assembly of genomes and genetic elements in complex metagenomic samples without using reference genomes.', *Nat Biotechnol*, vol. 32, no. 8, pp. 822–8, Aug. 2014, doi: 10.1038/nbt.2939.

[37] J. Qin *et al.*, 'A metagenome-wide association study of gut microbiota in type 2 diabetes.', *Nature*, vol. 490, no. 7418, pp. 55–60, Oct. 2012, doi: 10.1038/nature11450.

[38] N. Qin *et al.*, 'Alterations of the human gut microbiome in liver cirrhosis.', *Nature*, vol. 513, no. 7516, pp. 59–64, Sep. 2014, doi: 10.1038/nature13568.

[39] F. Raymond *et al.*, 'The initial state of the human gut microbiome determines its reshaping by antibiotics.', *ISME J*, vol. 10, no. 3, pp. 707–20, Mar. 2016, doi: 10.1038/ismej.2015.148.

[40] M. A. Rubel *et al.*, 'Lifestyle and the presence of helminths is associated with gut microbiome composition in Cameroonians.', *Genome Biol*, vol. 21, no. 1, p. 122, May 2020, doi: 10.1186/s13059-020-02020-4.

[41] K. Sankaranarayanan *et al.*, 'Gut Microbiome Diversity among Cheyenne and Arapaho Individuals from Western Oklahoma.', *Curr Biol*, vol. 25, no. 24, pp. 3161–9, Dec. 2015, doi: 10.1016/j.cub.2015.10.060.

[42] M. Schirmer *et al.*, 'Dynamics of metatranscription in the inflammatory bowel disease gut microbiome.', *Nat Microbiol*, vol. 3, no. 3, pp. 337–346, Mar. 2018, doi: 10.1038/s41564-017-0089-z.

[43] A. M. Thomas *et al.*, 'Metagenomic analysis of colorectal cancer datasets identifies cross-cohort microbial diagnostic signatures and a link with choline degradation.', *Nat Med*, vol. 25, no. 4, pp. 667–678, Apr. 2019, doi: 10.1038/s41591-019-0405-7.

[44] E. Vogtmann *et al.*, 'Colorectal Cancer and the Human Gut Microbiome: Reproducibility with Whole-Genome Shotgun Sequencing.', *PLoS One*, vol. 11, no. 5, p. e0155362, 2016, doi: 10.1371/journal.pone.0155362.

[45] J. Wirbel *et al.*, 'Meta-analysis of fecal metagenomes reveals global microbial signatures that are specific for colorectal cancer.', *Nat Med*, vol. 25, no. 4, pp. 679–689, Apr. 2019, doi: 10.1038/s41591-019-0406-6.

[46] H. Xie *et al.*, 'Shotgun Metagenomics of 250 Adult Twins Reveals Genetic and Environmental Impacts on the Gut Microbiome.', *Cell Syst*, vol. 3, no. 6, pp. 572-584.e3, Dec. 2016, doi: 10.1016/j.cels.2016.10.004.

[47] S. Yachida *et al.*, 'Metagenomic and metabolomic analyses reveal distinct stage-specific phenotypes of the gut microbiota in colorectal cancer.', *Nat Med*, vol. 25, no. 6, pp. 968–976, Jun. 2019, doi: 10.1038/s41591-019-0458-7.

[48] Z. Ye *et al.*, 'A metagenomic study of the gut microbiome in Behcet's disease.', *Microbiome*, vol. 6, no. 1, p. 135, Aug. 2018, doi: 10.1186/s40168-018-0520-6.

[49] J. Yu *et al.*, 'Metagenomic analysis of faecal microbiome as a tool towards targeted non-invasive biomarkers for colorectal cancer.', *Gut*, vol. 66, no. 1, pp. 70–78, Jan. 2017, doi: 10.1136/gutjnl-2015-309800.

[50] F. Zhu *et al.*, 'Metagenome-wide association of gut microbiome features for schizophrenia.', *Nat Commun*, vol. 11, no. 1, p. 1612, Mar. 2020, doi: 10.1038/s41467-020-15457-9.

[51] M. P. Nixon, G. B. Gloor, and J. D. Silverman, 'Beyond Normalization: Incorporating Scale Uncertainty in Microbiome and Gene Expression Analysis', *bioRxiv*, p. 2024.04.01.587602, Apr. 2024, doi: 10.1101/2024.04.01.587602.

[52] Y.-J. Hu and G. A. Satten, 'Compositional analysis of microbiome data using the linear decomposition model (LDM)', *bioRxiv*, p. 2023.05.26.542540, May 2023, doi: 10.1101/2023.05.26.542540.

[53] D. Firth, 'Bias Reduction of Maximum Likelihood Estimates', *Biometrika*, vol. 80, no. 1, p. 27, Mar. 1993, doi: 10.2307/2336755.

[54] J. T. Nearing *et al.*, 'Microbiome differential abundance methods produce different results across 38 datasets', *Nat Commun*, vol. 13, no. 1, p. 342, Jan. 2022, doi: 10.1038/s41467-022-28034-z.

[55] D. S. Clausen and A. D. Willis, 'Estimating Fold Changes from Partially Observed Outcomes with Applications in Microbial Metagenomics', *ArXiv*, Feb. 2024, Accessed: Mar. 05, 2025. [Online]. Available: https://arxiv.org/abs/2402.05231v1

[56] B. Brill, A. Amir, and R. Heller, 'Testing for differential abundance in compositional counts data, with application to microbiome studies', *Ann Appl Stat*, vol. 16, no. 4, pp. 2648–2671, 2022.

[57] Y. Hu, G. A. Satten, and Y. J. Hu, 'LOCOM: A logistic regression model for testing differential abundance in compositional microbiome data with false discovery rate control', *Proc Natl Acad Sci U S A*, vol. 119, no. 30, p. e2122788119, Jul. 2022, doi: 10.1073/PNAS.2122788119/SUPPL_FILE/PNAS.2122788119.SAPP.PDF.

[58] J. T. Barlow, S. R. Bogatyrev, and R. F. Ismagilov, 'A quantitative sequencing framework for absolute abundance measurements of mucosal and lumenal microbial communities', *Nature Communications 2020 11:1*, vol. 11, no. 1, pp. 1–13, May 2020, doi: 10.1038/s41467-020-16224-6.